\documentclass[11pt]{article}

\usepackage{graphics,latexsym}

\begin{document}
\title{\bf  A Full Review of the Theory of Electromagnetism}

\author{Daniele Funaro\\
\small Department of Mathematics, University of Modena \\
\small Via Campi 213/B, 41100 Modena (Italy)\\
\small E-mail: funaro@unimo.it}
\vskip1.truecm

\maketitle
\begin{abstract} \sl We will provide detailed arguments showing that the set of Maxwell
equations, and the corresponding wave equations, do not properly describe the evolution of
electromagnetic wave-fronts. We propose a nonlinear corrected version  that is proven to be far more
appropriate for the modellization of electromagnetic phenomena. The suitability of this approach will
soon be evident to the reader, through a sequence of astonishing congruences, making the model as
elegant as Maxwell's, but with increased chances of development. Actually, the new set of equations will
allow us to explain many open questions, and  find links between electromagnetism and other theories
that have been searched for a long time, or not even imagined.

\end{abstract}

\vskip1.truecm

\section{Short introduction}

The theory of electromagnetism, in the form conceived by J.C. Maxwell, can boast 130 years of honored
service. It survived the severest tests, proving itself to be, for  completeness and elegance, among the
most solid theories. Very few would doubt its validity, to the extent that they may be more inclined to
modify the point of view of other theories, rather than question the Maxwell equations. The trust in the
model has been strong enough to obscure a certain number of ``minor'' incongruities and to incite the
search for justifications in the development of other theories.
\par\smallskip

Nevertheless, even if the time-honored equations excellently solve complex problems, they are not able
to simulate the simplest things. They are not capable for instance of describing what a solitary
signal-packet is, one of the most elementary electromagnetic phenomena. Alternative models have been
proposed with the aim of including solitons, but they did not succeed in gaining a long-lasting
relevance, because they were based on deliberate adjustments, that, accommodating specific aspects on
one hand, were causing the model to lose general properties on the other.
\par\smallskip

The development of  modern field theory, which was very prosperous in the years 1930-1960, has magnified
the role of the equations, giving them a universal validation in the relativistic framework. This
progress came to a stop, leaving however the impression of being not too far from the goal of
compenetrating electromagnetism and gravitation theory.
\par\smallskip

We are going to make some statements that many readers will certainly consider heretic. We think that
the various anomalies, which are present in the model, are not incidental, but consequences of a still
insufficient theoretical description of electromagnetic phenomena. Actually, it is our opinion that the
flaws are more severe than expected, and therefore, such a fundamental ``brick'' of Physics needs
extensive revision. The review process must be so deep that the entire setting necessitates  re-planning
from the beginning.  On the other hand, if it were just a matter of small adaptations, this revision
would have already been made a long time ago.
\par\smallskip

We shall start to analyse some substantial facts, that at a practical level may be considered marginal,
with the aim to evidentiate contradictions. We solve these problems by suitably redesigning the Maxwell
equations. This will allow for the construction of a new model, solving all the inconsistencies and
achieving the scope of a better understanding of electromagnetic phenomena. In a very natural way, the
new approach also leaves the door more than open, to those links and generalizations that were expected
to come from the Maxwell equations, but which, although vaguely insinuated, could never be realized in
practice.
\par\smallskip

None of the gracefulness that characterizes the Maxwell model will be lost. The reader who has the
patience to follow our arguments through to the end, will discover that all the pieces find their exact
place in a global scheme, with due elegance and harmony. We do not wish to say more in this short
introduction. The model will be developed step by step, up to its final form, in order to let the reader
appreciate the phases of its maturation. The mathematical tools used are classical, and maybe dated. On
the other hand, our intention is to examine what would have happened to the evolution of Physics, if our
model was taken into consideration, in place of the Maxwell equations. We will elaborate and clarify
many important concepts, leaving the path well clear for future developments, not considered here due to
lack of time.

\par\bigskip

\setcounter{equation}{0}
\section{Criticism of the theory of electromagnetism}

In this section, we  make some fine considerations regarding the evolution of electromagnetic waves, and
the way they are modelled by the Maxwell equations. We  start by pointing out deficiencies mainly at the
level of mathematical elegance. These will reveal other more severe incoherences. In the end, even
taking into account the correctness, up to a certain degree of approximation, of the physical approach,
our judgement will be rather negative. As a matter of fact, in section 3, with the aim of finding a
remedy to the problems that have emerged, substantial revision will be proposed.

\par\smallskip
From now on, until section 11, we  assume that we are in void three-dimensional space. As usual, the
constant $c$ indicates the speed of light. In this case, the classical Maxwell equations are:
\begin{equation}\label{eq:rotb}
 {\partial {\bf E}\over \partial t}~=~ c^2 {\rm curl} {\bf B}
\end{equation}
\begin{equation}\label{eq:dive}
{\rm div}{\bf E} ~=~0
\end{equation}
\begin{equation}\label{eq:rote}
{\partial {\bf B}\over \partial t}~=~ -{\rm curl} {\bf E}
\end{equation}
\begin{equation}\label{eq:divb}
{\rm div}{\bf B} ~=~0
\end{equation}
where the vector field ${\bf E}$ is dimensionally equivalent to an acceleration
multiplied by a mass and divided by an electrical charge; while ${\bf B}$ is a
frequency multiplied by a mass and divided by a charge.
\par\smallskip

The above equations are supposed to be satisfied point-wise at any instant
of time. Their solutions are assumed to be smooth enough to
allow differential calculus. Therefore, discontinuous or singular solutions
are not allowed. The equations (\ref{eq:dive}) and  (\ref{eq:divb}) could
be considered unnecessary,
since they are easily deduced from (\ref{eq:rotb}) and (\ref{eq:rote})
respectively, after applying the divergence operator. Later on, for
the reasons that we are going to explain, we will question the
validity of  (\ref{eq:dive}) and  (\ref{eq:divb}). As a consequence, the entire
formulation will lose its credibility.
\par\smallskip

As far as the evolution of an electromagnetic plane wave (with infinite extent and linearly polarized)
is concerned, we have no objections to make. In Cartesian coordinates, a monocromatic wave of this type,
moving along the direction of the $z$-axis, is written as:
\begin{equation}\label{eq:campip}
{\bf E}~=~(c\sin \omega(t-z/c) ,~0,~0)~~~~~~~{\bf B}~=~(0,~\sin
\omega(t-z/c),~0)
\end{equation}
In this case, the Maxwell equations are all satisfied point-wise.
\par\smallskip

The next step is to examine the case of a spherical wave, which is far more delicate. The wave could be
generated by an oscillating dipole of negligible size. However, the way the wave is produced and
supplied is not of interest to us at the moment, being more concerned with analyzing the geometrical
aspects of its evolution at a distance from the source.

\par\smallskip
Let us denote by  ${\bf P}={\bf E}\times {\bf B}$ the Poynting vector. It is customary to assume that
${\bf E}$ and ${\bf B}$ are orthogonal, and that the wave-front propagates at constant speed $c$,
through spherical concentric surfaces. One may argue that perfect spherical waves do not exist in
nature. Nevertheless, for the sake of simplicity, we maintain this hypothesis which can be removed
later, without modifying the essence of our reasoning.

\par\smallskip
We are basically confronted with two possibilities. In the first one, the Poynting vector follows
exactly the radial direction. This means that ${\bf E}$ and ${\bf B}$ locally belong to the tangent
plane to the wave-front. In such a circumstance, as detailed below, we are able to show that
(\ref{eq:rotb}) and (\ref{eq:rote}) cannot be both satisfied everywhere. More precisely, it is known
that (\ref{eq:rotb}) and (\ref{eq:rote}) are true up to an error that decays quadratically with the
distance from the source. Since the intensity of a spherical electromagnetic wave only decays linearly
in amplitude, the above mentioned inaccuracy has no influence on practical applications. However, we
record a first negative mark.
\par\smallskip

The second possibility is that, in order to satisfy all the set of Maxwell equations, we loose the
orthogonality of the Poynting vector with respect to the wave-front surface. This is a more unpleasant
situation, considering that the Poynting vector represents the direction of propagation of the energy
flow. The lack of orthogonality between the wave-front tangent plane and the direction of propagation
violates the Huygens principle (recall that we are in vacuum), leading to a deformation of the front
itself. As we will check later, this results in relevant defects in the development of the wave-shape.

\par\smallskip
Let us study the problem more in detail, by taking into account the transformation
in spherical coordinates:
\begin{equation}\label{eq:coor}
(x,~y,~z)~=~(r\sin\phi \cos\theta, ~r\sin \phi\sin\theta,~ r \cos\phi)
\end{equation}
with  $0\leq \theta <2\pi$, $0\leq \phi\leq \pi$ and $r$ large enough.
We look for vector fields having the following form:
\begin{equation}\label{eq:campi}
{\bf B}~=~(0,~0,~ u)~~~~~~~~~~~~~{\bf E}~=~(v,~ w,~0)
\end{equation}
where $u$, $v$, $w$ are functions of the variables  $t$, $r$ and $\phi$ (no dependency on  $\theta$ is
assumed). In (\ref{eq:campi}), the first component of the vectors is referred to the variable  $r$, the
second one to $\phi$, and the third one to $\theta$. The unknowns in the system of Maxwell equations
reduce from six to three. Choosing
 a more general form for the fields only complicates
the computations, without adding anything to the substance.
\par\smallskip

We start by observing that equation (\ref{eq:divb}) is immediately satisfied. Moreover:
\begin{equation}\label{eq:curlb}
 {\rm curl}{\bf B}=\Big({u\cos\phi\over r\sin \phi}+{u_\phi\over r},~
 -({u\over r}+u_r),~ 0\Big)
\end{equation}

\begin{equation}\label{eq:divec}
{\rm div}{\bf E}~=~v_r+{2v\over r}~+~{w\cos\phi\over r\sin\phi}
 ~+~{w_\phi\over r}
\end{equation}

\begin{equation}\label{eq:curle}
 {\rm curl}{\bf E}=\Big(0,~0,~w_r+{w\over r}-{v_\phi\over r} \Big)
\end{equation}
\par\smallskip
\noindent Therefore, the equations in spherical coordinates become:

\begin{equation}\label{eq:ut}
u_t~=~-\Big(w_r+{w\over r}\Big)~+~{v_\phi\over r}
\end{equation}

\begin{equation}\label{eq:vt}
v_t~=~{c^2\over r}\Big(u ~{\cos\phi\over\sin\phi}~+~u_\phi \Big)
\end{equation}

\begin{equation}\label{eq:wt}
w_t~=~-c^2\Big({u\over r}~+~u_r\Big)
\end{equation}

\noindent To avoid discontinuities, we must introduce the following boundary constraints:
\begin{equation}\label{eq:concon1}
u(t,r,0)~=~u(t,r,\pi )~=~0~~~~~~~~w(t,r,0)~=~w(t,r,\pi )~=~0
\end{equation}
\begin{equation}\label{eq:concon2}
{\partial v\over\partial \phi}(t,r,0)~=~{\partial v\over\partial \phi}(t,r,
\pi )~=~0
\end{equation}
\par\smallskip
In the case of the pure radiation field of an oscillating dipole, when $r$ is sufficiently large, one
usually sets $v=0$ and $w=cu$. Within this hypothesis, the two equations (\ref{eq:ut}) and (\ref{eq:wt})
are equivalent. They bring us to the general solution:
\begin{equation}\label{eq:solgen}
w(t, r, \phi )~ =~ c~ u(t, r, \phi )~=~ {c\over r} f(\phi) ~g(t-r/c)
\end{equation}
where $f$ (with $f(0)=f(\pi )=0$) and $g$ are arbitrary functions (the only restrictions apply to their
regularity). Among these solutions there is the one corresponding to $f(\phi )=\sin\phi$, which is often
present in classical texts (see for instance \cite{bleaney}, p.284), being the one with more physical
relevance. Nevertheless, we unfortunately note that equation (\ref{eq:vt}) is compatible with  $v=0$
only when:
\begin{equation}\label{eq:bsing}
{\bf B}~=~\Big(0, ~0, ~{1\over r\sin\phi}~g(t-r/c)\Big)
\end{equation}
\begin{equation}\label{eq:esing}
{\bf E}~=~\Big(0,  ~{c\over r\sin\phi}~g(t-r/c), ~0\Big)
\end{equation}
which manifest singularities at the points corresponding to $\phi =0$ and $\phi =\pi$.
In general, we have the following statement:
\begin{equation}\label{eq:divsing}
{\rm div}{\bf E}={1\over r^2} g(t-r/c)\left ({\cos \phi\over\sin\phi} f(\phi )
+f^\prime (\phi)\right)=0~~~~\Leftrightarrow~~~~~f(\phi )={1\over \sin\phi}
\end{equation}
Note that such a strong singularity at the poles cannot be removed only by requiring the wave-front
 not to be perfectly spherical.
\par\smallskip

We observe that $f$ can be taken in such a way that ${\rm div}{\bf E}$ is vanishing at the poles (for
example $f(\phi )=(\sin\phi)^2$), but not in the proximity of them. In addition, we observe that, if $f$
is regular with $f(0)=f(\pi )=0$, for any fixed $r$, the points in which the divergence of ${\bf E}$
does not vanish belong to a bidimensional set whose measure is different from zero. For instance, if
$f(\phi )=\sin\phi$, we find out that ${\rm div}{\bf E}$ is proportional to $\cos\phi$, so that this set
consists of all points of the sphere of radius $r$, with the exception of the equator. It is certainly
true that even if the divergence is not zero, it is  negligible when designing, for instance, a device
like an antenna. This argument, however, is not going to be valid here, since we would like to carry out
an in depth analysis of what is really happening in the evolution of an electromagnetic wave, compared
to what the Maxwell theory is able to predict.

\par\smallskip
Let us now follow a different path  and try to find other solutions, of the form
given in (\ref{eq:campi}), satisfying
 the set of all Maxwell equations (including ${\rm div}{\bf E}=0$).
If we do not want $f$ to be singular somewhere, we have to accept that $v$ is different from zero. This
means that ${\bf E}$ has a radial component, so that the Poynting vector cannot be perfectly radial. We
have to better check what happens in this last case.
\par\smallskip

\noindent It is well-known that the the Maxwell equations lead to:
\begin{equation}\label{eq:eonde}
{1\over c^2}{\partial^2 {\bf E}\over \partial t^2}~=~\Delta {\bf E}~~~~~~~{\rm and}
~~~~~~~~{1\over c^2}{\partial^2 {\bf B}\over \partial t^2}~=~\Delta {\bf B}
\end{equation}
The above are usually called ``wave equations'', but, shortly, we will see that this name is not
appropriate. The terminology is correct only if the fields involved are scalar. By deriving
(\ref{eq:ut}) with respect to time and using (\ref{eq:vt}) and (\ref{eq:wt}), we arrive at the equation:
\begin{equation}\label{eq:ondeu}
{1\over c^2}~ u_{tt}~=~{1\over r^2}(r^2 u_r)_r~+~{1\over r^2}\left(
{1\over \sin\phi}(u\sin\phi)_\phi\right)_{\hskip-.1truecm\phi}
\end{equation}
corresponding to the third component of the second equation in (\ref{eq:eonde})
in spherical coordinates.
\par\smallskip

It is worthwhile noting that (\ref{eq:ondeu}) is not the wave equation for the scalar field $u$ in
spherical coordinates, due to the fact that in this framework the Laplacian of a vector field is not the
Laplacian of its coordinates (even if  only one of them is different from zero). The wave equation for
$u$ reads as follows:
\begin{equation}\label{eq:ondeum}
{1\over c^2}~ u_{tt}~=~{1\over r^2}(r^2 u_r)_r~+~{1\over r^2 \sin\phi}
(u_\phi \sin\phi)_\phi~=~\Delta u
\end{equation}
This is not a trivial warning, since many texts in electromagnetism erroneously confuse
(\ref{eq:ondeum}) with (\ref{eq:ondeu}). Implicitly, we made the same mistake before, when looking for
d'Alembert type solutions of the form (\ref{eq:solgen}), generating, for this reason, solutions not
compatible with all the Maxwell equations.

\par\smallskip
By separation of variables, for any $k\geq 1$ and any $n\geq 1$, we discover that (\ref{eq:ondeu})
admits the following basis of solutions:
$$
r^{-{1\over 2}}~\cos (ckt)~J_{n+{1\over 2}}(kr)~\sin\phi ~P_n^\prime (\cos\phi )
$$
$$
r^{-{1\over 2}}~\sin (ckt)~J_{n+{1\over 2}}(kr)~\sin\phi ~P_n^\prime (\cos\phi )
$$
$$
r^{-{1\over 2}}~\cos (ckt)~Y_{n+{1\over 2}}(kr)~\sin\phi ~P_n^\prime (\cos\phi )
$$
\begin{equation}\label{eq:legbes}
r^{-{1\over 2}}~\sin (ckt)~Y_{n+{1\over 2}}(kr)~\sin\phi ~P_n^\prime (\cos\phi )
\end{equation}
where  $J_{n+{1\over 2}}$ and $Y_{n+{1\over 2}}$  are Bessel functions of first
and second kind respectively, while  $P_n$ is the $n$-th Legendre polynomial.
\par\smallskip

A classical reference for Bessel functions is \cite{watson}. It is important to note
that the solutions given in \cite{watson} at page 127, for the scalar wave equation in
spherical coordinates, differ from the ones shown in (\ref{eq:legbes}). The reason is that
the functions in \cite{watson} (having $P_n(\cos\phi )$ in place of $\sin\phi~P^\prime_n
(\cos\phi )$) are those solving (\ref{eq:ondeum}), which is not the vector version of
the wave equation, as we already mentioned.
\par\smallskip

\noindent For example, if $n=1$ we have (see \cite{watson}, p.54):
$$J_{3\over 2}(kr)= \sqrt{2\over \pi k r}\left({\sin kr\over kr}-\cos kr\right)$$
\begin{equation}\label{eq:besn1}
Y_{3\over 2}(kr)= \sqrt{2\over \pi k r}\left({\cos kr\over kr}+\sin kr\right)~~~~~~~~~
P_1^\prime (\cos\phi )=1
\end{equation}
In order to understand what the solutions in (\ref{eq:besn1}) look like, it is standard to introduce
some approximation.  Thus,  for $n=1$ and $r$ large, by taking the combination $r^{-1/2}(\sin
(ckt)J_{3/2}(kr)+\cos (ckt)Y_{3/2}(kr))\sin\phi$, up to multiplicative constants, it is possible to get
asymptotically the monocromatic solution $u=r^{-1}\sin\phi ~\sin k(ct-r)$ (compare to
(\ref{eq:solgen})), up to an error which decays quadratically with  $r$. Once again, one ends up with
something similar to a travelling wave, although some cheating has been necessary (that is equivalent,
in the end, to replacing once again (\ref{eq:ondeum}) by (\ref{eq:ondeu})).

\par\smallskip
On the other hand, suppose that $u$ is evaluated exactly as  linear combination of the functions
in (\ref{eq:legbes}). Then, one  recovers $v$ e $w$ by (\ref{eq:vt}) and
(\ref{eq:wt}), through time integration. Successively, it is possible to compute
the Poynting vector:
\begin{equation}\label{eq:pciam}
{\bf P}~=~-c^2\left( u \int\hskip-.2truecm\left({u\over r}+u_r\right)\hskip-.1truecmdt,~~
{u\over r}
\int\hskip-.2truecm\left(u{\cos\phi\over\sin\phi}+u_\phi \right)\hskip-.1truecmdt,~~0  \right)
\end{equation}
which has, as expected, a non radial component. Now, let us fix $r$ and study the behavior, by varying
$\phi$, of the two components of ${\bf P}$. In particular, we are interested to see what happens near
the poles ($\phi=0$ or $\phi =\pi$). We start by noting that, for any $n\geq 1$, the term
$P^\prime_n(\cos\phi )$ tends towards a finite limit for $\phi \rightarrow 0$ or $\phi \rightarrow\pi$
(recall that $P^\prime_n(\pm 1)={1\over 2}(\pm 1)^{n+1} n(n+1)$). Therefore, according to
(\ref{eq:legbes}), the first component in (\ref{eq:pciam}) behaves as $(\sin\phi)^2$ near the poles. It
is a matter of using known properties of Legendre polynomials, in particular the differential equation:
\begin{equation}\label{eq:legeq}
(\sin\phi)^2P^{\prime\prime}_n(\cos\phi )-2\cos\phi ~P^\prime_n(\cos\phi )
+n(n+1)P_n(\cos\phi )=0
\end{equation}
to check that the second component in (\ref{eq:pciam}) behaves as
$\sin\phi$ near the poles.
\par\smallskip

We are ready to draw some preliminary conclusions. Let us note that finally ${\rm div}{\bf E}=0$, hence
all the Maxwell equations are satisfied. As already remarked, it has been necessary to keep the
nonradial component of ${\bf P}$. Surprisingly, for any fixed $r$, such a nonradial component prevails
on the radial one, when approaching the poles. This implies that the shape of the wave-fronts does not
resemble a sphere, but rather a kind of doughnut with the central hole reduced to a single point. The
parts of the fronts corresponding to  the internal side of the doughnut, progressively stratify along
the $z$-axis. We do not see a chance of recognising any sort of Hyugens principle here. This is not what
we would call a travelling wave. It may be argued that this behavior is due to the influence of the
source located at $r=0$. But, if we stop the source, the wave-fronts already produced continue to
develop. If their motion is ruled by the Huygens principle, the hole should fill up quickly, and each
 front should transform to something rounded which is almost a perfect sphere.
The problem is that, during this smoothing process, the vector fields
${\bf E}$ and  ${\bf B}$ are not compatible with both the constraints
${\rm div}{\bf E}=0$ and ${\rm div}{\bf B}=0$. The clue is that ``wave equations''
in vector form have nothing to do with real waves.
\par\smallskip

Some mild analogy between the Maxwell equations and the eikonal equation, governing the movement of the
fronts, was devised a long time ago.
 The equivalence is valid within the limits
of geometrical optics (see \cite{born}, p.110). In spite of this, examining the behavior of the fronts,
our impression is that their natural evolution is in conflict with all restrictions imposed by Maxwell
equations. This statement will be clearer as we proceed with our study. The right connections with the
eikonal equation will be defined in section 10.
\par\smallskip

The different situations analyzed up to now are summarized in figure 1, which should clarify
our point of view: either we keep the singularity
at the poles (manifested by infinite amplitude of the fields or strong
geometrical distorsion), or we allow the divergence of the electric field to be
different from zero. To sustain this proposition, let us collect other elements.

\par\smallskip

\begin{figure}[t]
\begin{picture}(400,250)
\put(0,10){\vector(0,1){220}}
\put(130,10){\vector(0,1){220}}
\put(260,10){\vector(0,1){220}}
\put(-5,120){\vector(1,0){112}}
\put(125,120){\vector(1,0){112}}
\put(255,120){\vector(1,0){112}}
\put(16,237){\makebox(0,0){$\phi =0$}}
\put(146,237){\makebox(0,0){$\phi =0$}}
\put(276,237){\makebox(0,0){$\phi =0$}}
\put(16,13){\makebox(0,0){$\phi =\pi$}}
\put(146,13){\makebox(0,0){$\phi =\pi$}}
\put(276,13){\makebox(0,0){$\phi =\pi$}}
\put(100,120){\vector(0,1){24}}
\put(230,120){\vector(0,1){14}}
\put(360,120){\vector(0,1){38}}
\put(87,170){\vector(-1,2){9}}
\put(87,70){\vector(1,2){9}}
\put(50,207){\vector(-2,1){10}}
\put(50,33){\vector(2,1){10}}
\put(10,219){\vector(-1,0){7}}
\put(10,21){\vector(1,0){7}}
\put(217,170){\vector(-1,2){10}}
\put(217,70){\vector(1,2){10}}
\put(180,207){\vector(-2,1){40}}
\put(180,33){\vector(2,1){40}}
\put(150,218){\vector(-4,1){60}}
\put(150,22){\vector(4,1){60}}
\put(348,170){\vector(-1,1){20}}
\put(348,70){\vector(1,1){20}}
\put(310,203){\vector(-1,0){21}}
\put(310,37){\vector(1,0){21}}
\put(276,200){\vector(-1,-1){10}}
\put(276,40){\vector(1,-1){10}}
\end{picture}
\vskip.5truecm

\caption{\small{\sl Qualitative behavior of the field ${\bf E}$ as a function
of the angle $\phi$. Case 1: ${\bf E}=(0, r^{-1}(\sin\phi)
g(t-r/c),0)$, the wavefronts are perfect spheres,
but ${\rm div}{\bf E}\not =0$. Case 2:  ${\bf E}=(0, (r\sin\phi)^{-1}
g(t-r/c),0)$ the condition ${\rm div}{\bf E}=0$ is satisfied, but there are singularities at
 $\phi =0$ and $\phi =\pi$; Case 3: the corresponding Poynting vector is given in
(\ref{eq:pciam}), the divergence of the electric field is vanishing, but
the wave-fronts are far from being spherical surfaces.}}
\end{figure}
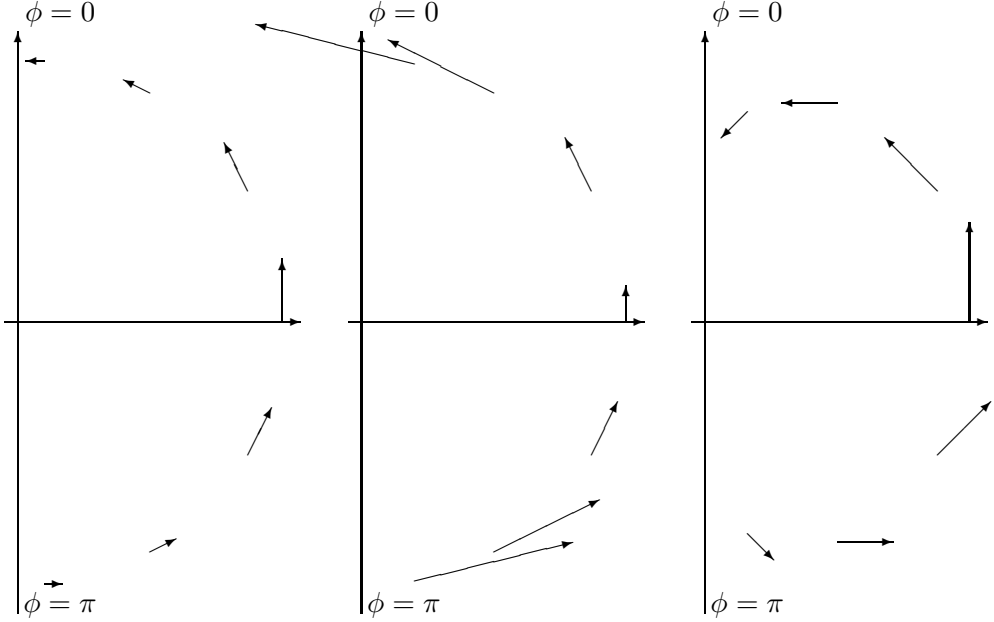

\par\medskip
Some confusion usually arises when one tries to simulate the evolution of
a ``fragment'' of wave. We examine the case of the plane wave given in
(\ref{eq:campip}). For any fixed $z$, we can cut out
a region $\Omega$ in the plane determined by the variables $x$ and $y$, and
follow its evolution in time. For simplicity, $\Omega$ can be the square  $[0,1]\times
[0,1]$. Inside $\Omega$ we assume that the electromagnetic fields evolve
 following (\ref{eq:campip}), in full agreement with Maxwell equations.
Outside $\Omega$, the fields ${\bf E}$ and ${\bf B}$ are supposed to vanish. The question is
understanding what happens at the boundary $\partial\Omega$ of $\Omega$. It is not difficult to realize
that, on the sides $\{ 0\} \times ]0,1[$ and $\{ 1\}\times ]0,1[$, ${\rm curl}{\bf B}$ and ${\rm div}
{\bf E}$ become singular, producing concentrated distributions. Similarly, on the two sides $]0,1[\times
\{ 0\}$ and $]0,1[\times \{ 1\}$, the quantities  ${\rm curl}{\bf E}$ and ${\rm div}{\bf B}$ present
singularities.
\par\smallskip

Some readers may complain because discontinuities of the fields may not exist in nature. Commonly, the
right way to proceed is to consider a thin layer around  $\partial\Omega$, where the solution given by
(\ref{eq:campip}) smoothly decays to zero. Then, one lets the width of the layer tend towards zero. This
in general allows us to determine special relations to be satisfied on $\partial\Omega$ (in place of the
Maxwell equations, which are meaningless there).  Unfortunately, the procedure presents some drawbacks.
Let us first assume that the wave-fronts shift along the $z$-axis maintaining their squared shape. We
also assume that the fields ${\bf E}$ and ${\bf B}$ are orthogonal and smoothly decaying to zero in a
neighbourhood of $\partial\Omega$ (like for instance in figure 2). Our conjecture is that there exists
at least one point where  Maxwell equations are not all satisfied,  because ${\rm div}{\bf E}$ and ${\rm
div}{\bf B}$ cannot both be zero at the same time. Actually, examining figure 2, we discover that there
are infinite points where either ${\rm div}{\bf E}\not =0$ or ${\rm div}{\bf B}\not =0$.  These points
form a set whose area is different from zero. We are free to try other configurations by modifying the
orientation of the vector fields at each point near $\partial\Omega$, but we always arrive at the same
conclusion: some rule of Physics breaks down when  approaching $\partial\Omega$. Now, the question is:
if we do not know what the governing rules are in  the layer around $\partial\Omega$, how can we go to
the limit for the size of the layer tending to zero?

\par\smallskip
Another possibility is that the wave-fronts, due to the strong variation of the fields near the boundary
of $\Omega$, are forced to bend a little. The electromagnetic fields are no longer on a plane, so we
could probably find out the way to enforce all the Maxwell equations. However, this implies that the
Poynting vectors are not parallel to the $z$-axis anymore. Thus, the shape of $\Omega$ is going to be
further modified during the evolution. A little diffusion is bearable, yet our impression is that the
wave-fronts would rapidly change their form. The more they bend, the faster they produce other
distorsion. This is in contrast for instance with the fact that neat electromagnetic signals, of
arbitrary transversal shape, reach our instruments after travelling for years between galaxies. The only
acceptable rule is that all the Poynting vectors must stay orthogonal to the fronts and parallel to the
actual direction of movement; if this does not happen the wave quickly deteriorates, fading completely.
\par\smallskip

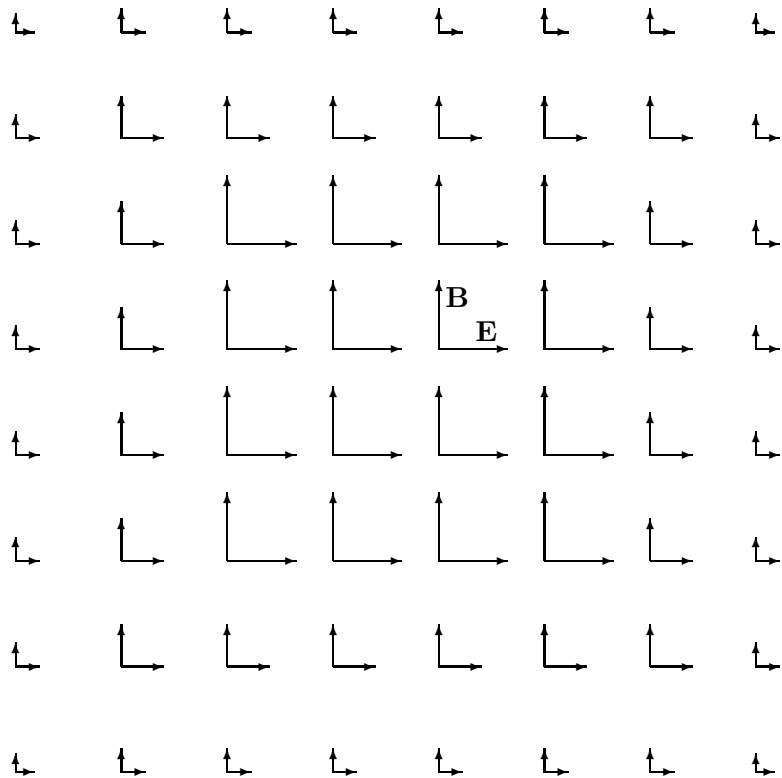
\begin{figure}[t]
\begin{picture}(400,300)
\put(20,10){\vector(0,1){7}}\put(20,10){\vector(1,0){7}}
\put(60,10){\vector(0,1){9}}\put(60,10){\vector(1,0){9}}
\put(100,10){\vector(0,1){9}}\put(100,10){\vector(1,0){9}}
\put(140,10){\vector(0,1){9}}\put(140,10){\vector(1,0){9}}
\put(180,10){\vector(0,1){9}}\put(180,10){\vector(1,0){9}}
\put(220,10){\vector(0,1){9}}\put(220,10){\vector(1,0){9}}
\put(260,10){\vector(0,1){9}}\put(260,10){\vector(1,0){9}}
\put(300,10){\vector(0,1){7}}\put(300,10){\vector(1,0){7}}
\put(20,50){\vector(0,1){9}}\put(20,50){\vector(1,0){9}}
\put(60,50){\vector(0,1){16}}\put(60,50){\vector(1,0){16}}
\put(100,50){\vector(0,1){16}}\put(100,50){\vector(1,0){16}}
\put(140,50){\vector(0,1){16}}\put(140,50){\vector(1,0){16}}
\put(180,50){\vector(0,1){16}}\put(180,50){\vector(1,0){16}}
\put(220,50){\vector(0,1){16}}\put(220,50){\vector(1,0){16}}
\put(260,50){\vector(0,1){16}}\put(260,50){\vector(1,0){16}}
\put(300,50){\vector(0,1){9}}\put(300,50){\vector(1,0){9}}
\put(20,90){\vector(0,1){9}}\put(20,90){\vector(1,0){9}}
\put(60,90){\vector(0,1){16}}\put(60,90){\vector(1,0){16}}
\put(100,90){\vector(0,1){26}}\put(100,90){\vector(1,0){26}}
\put(140,90){\vector(0,1){26}}\put(140,90){\vector(1,0){26}}
\put(180,90){\vector(0,1){26}}\put(180,90){\vector(1,0){26}}
\put(220,90){\vector(0,1){26}}\put(220,90){\vector(1,0){26}}
\put(260,90){\vector(0,1){16}}\put(260,90){\vector(1,0){16}}
\put(300,90){\vector(0,1){9}}\put(300,90){\vector(1,0){9}}
\put(20,130){\vector(0,1){9}}\put(20,130){\vector(1,0){9}}
\put(60,130){\vector(0,1){16}}\put(60,130){\vector(1,0){16}}
\put(100,130){\vector(0,1){26}}\put(100,130){\vector(1,0){26}}
\put(140,130){\vector(0,1){26}}\put(140,130){\vector(1,0){26}}
\put(180,130){\vector(0,1){26}}\put(180,130){\vector(1,0){26}}
\put(220,130){\vector(0,1){26}}\put(220,130){\vector(1,0){26}}
\put(260,130){\vector(0,1){16}}\put(260,130){\vector(1,0){16}}
\put(300,130){\vector(0,1){9}}\put(300,130){\vector(1,0){9}}
\put(20,170){\vector(0,1){9}}\put(20,170){\vector(1,0){9}}
\put(60,170){\vector(0,1){16}}\put(60,170){\vector(1,0){16}}
\put(100,170){\vector(0,1){26}}\put(100,170){\vector(1,0){26}}
\put(140,170){\vector(0,1){26}}\put(140,170){\vector(1,0){26}}
\put(180,170){\vector(0,1){26}}\put(180,170){\vector(1,0){26}}
\put(220,170){\vector(0,1){26}}\put(220,170){\vector(1,0){26}}
\put(260,170){\vector(0,1){16}}\put(260,170){\vector(1,0){16}}
\put(300,170){\vector(0,1){9}}\put(300,170){\vector(1,0){9}}
\put(20,210){\vector(0,1){9}}\put(20,210){\vector(1,0){9}}
\put(60,210){\vector(0,1){16}}\put(60,210){\vector(1,0){16}}
\put(100,210){\vector(0,1){26}}\put(100,210){\vector(1,0){26}}
\put(140,210){\vector(0,1){26}}\put(140,210){\vector(1,0){26}}
\put(180,210){\vector(0,1){26}}\put(180,210){\vector(1,0){26}}
\put(220,210){\vector(0,1){26}}\put(220,210){\vector(1,0){26}}
\put(260,210){\vector(0,1){16}}\put(260,210){\vector(1,0){16}}
\put(300,210){\vector(0,1){9}}\put(300,210){\vector(1,0){9}}
\put(20,250){\vector(0,1){9}}\put(20,250){\vector(1,0){9}}
\put(60,250){\vector(0,1){16}}\put(60,250){\vector(1,0){16}}
\put(100,250){\vector(0,1){16}}\put(100,250){\vector(1,0){16}}
\put(140,250){\vector(0,1){16}}\put(140,250){\vector(1,0){16}}
\put(180,250){\vector(0,1){16}}\put(180,250){\vector(1,0){16}}
\put(220,250){\vector(0,1){16}}\put(220,250){\vector(1,0){16}}
\put(260,250){\vector(0,1){16}}\put(260,250){\vector(1,0){16}}
\put(300,250){\vector(0,1){9}}\put(300,250){\vector(1,0){9}}
\put(20,290){\vector(0,1){7}}\put(20,290){\vector(1,0){7}}
\put(60,290){\vector(0,1){9}}\put(60,290){\vector(1,0){9}}
\put(100,290){\vector(0,1){9}}\put(100,290){\vector(1,0){9}}
\put(140,290){\vector(0,1){9}}\put(140,290){\vector(1,0){9}}
\put(180,290){\vector(0,1){9}}\put(180,290){\vector(1,0){9}}
\put(220,290){\vector(0,1){9}}\put(220,290){\vector(1,0){9}}
\put(260,290){\vector(0,1){9}}\put(260,290){\vector(1,0){9}}
\put(300,290){\vector(0,1){7}}\put(300,290){\vector(1,0){7}}
\put(187,190){\makebox(0,0){${\bf B}$}}
\put(198,177){\makebox(0,0){${\bf E}$}}
\end{picture}
\vskip.2truecm

\caption{\small{\sl Example of electromagnetic field
smoothly reducing to zero at the boundary of the square.
The Poynting vector is  orthogonal to the page at each point.
The Maxwell equations are satisfied in the central part.
Instead, approaching the boundary, the divergence of the fields
turns out to be different from zero. }}
\end{figure}

\par\medskip
To prove what we claimed before, we show using very standard arguments that it is not possible to
construct solutions to  Maxwell equations, having finite energy and travelling unperturbed at constant
speed along a straight-line. We assume that the speed is $c$ and the straight-line is the $z$-axis.
Without loss of generality, such a signal-packet is supposed to be of the following type:
$${\bf E}~=~\Big(E_1(x,y), E_2(x,y), E_3(x,y)\Big)~ g(t-z/c)$$
\begin{equation}\label{eq:foten}
{\bf B}~=~\Big(B_1(x,y), B_2(x,y), B_3(x,y)\Big)~ g(t-z/c)
\end{equation}
where $g$ is a bounded function and all the components $E_1$, $E_2$, $E_3$,
$B_1$, $B_2$, $B_3$ are zero outside a bidimensional set $\Omega$.
It is not difficult to check that (\ref{eq:rotb}) and (\ref{eq:rote}) only hold
 when ${\bf E}$  and ${\bf B}$ are identically zero. Actually, it is
straightforward to discover that $E_3$ e $B_3$ must be constant (and the sole
constant allowed is zero). Then, one finds out that $E_1$, $E_2$, $B_1$, $B_2$ must
be harmonic functions in $\Omega$. Since they have to vanish at the boundary,
they must vanish everywhere.
\par\smallskip

Due to the above mentioned reasons, solitonic solutions are not described by the classical theory of
electromagnetism. Efforts have been made in the past to generalize the Maxwell model, in a nonlinear
way, in order to include solitons. Just to mention an example, the Born-Infeld theory (see \cite{borni})
predicts the existence of finite-energy soliton-like solutions (that have been successively called
BIons). These last equations have no relation with the ones we are going to develop in this paper.
However, they point out the necessity of looking for nonlinear versions of the model. We will come back
to the subject of solitary waves in section 5.
\par\smallskip

In many applications, a standard approach is to reconstruct the bidimensional profile of the fields
inside $\Omega$ with the help of a truncated Fourier series. This is accomplished by a complete
orthogonal set of plane waves, each one carrying a suitable eigenfunction in the variables $x$ and $y$.
We must pay attention, however, to the fact that these eigenfuntions are of the periodic type.
Therefore, they reproduce the same profile, not only inside $\Omega$, but in a lattice of infinite
contiguous domains. In this way, the represented solution turns out to have infinite energy. Considering
only one of these profiles, thus forcing to zero the solution outside $\Omega$, unavoidably brings us
again to a violation of the Maxwell equations near the boundary of $\Omega$. Some clarifying comments on
this issue can be found in \cite{goodman}, p.42.
\par\smallskip

We recognize that the techniques based on Fourier expansions provide excellent results in many practical
circumstances, as for example the study of diffraction. Nevertheless, in this last case and in the ones
treated before, it is necessary to adapt the solutions, introducing some approximation, if we want them
to correspond to the real phenomenon. Indeed, these adjustments are within the so-called limits of the
model. Hence, we could just stop our analysis here, with the trivial (well-known) conclusion that the
Maxwell model is not perfect. We believe instead that the discrepancies pointed out are not just
imperfections, but symptoms of a more profound pathology affecting
 the theory of  classical electromagnetism.
\par\smallskip

What we learn in these pages is that there are plenty of simple and interesting phenomenon, which  are
inadequately explained by the Maxwell model, because the equations impose too many restrictions.
Consequently, the idea we shall follow in the next section is of weakening the equations, with the aim
of widening the range of solutions.

\par\bigskip

\setcounter{equation}{0}
\section{Modified Maxwell equations}
\smallskip

The demolition process is finished, now it is time to rebuild.
To begin, we propose the following model:
\begin{equation}\label{eq:rotbm}
{\partial {\bf E}\over \partial t}~=~ c^2 ~{\rm curl} {\bf B}~
-~c ~({\rm div}{\bf E}){{\bf E} \times  {\bf B}\over \vert
{\bf E} \times  {\bf B}\vert}
\end{equation}
\begin{equation}\label{eq:rotem}
{\partial {\bf B}\over \partial t}~=~ -{\rm curl} {\bf E}
\end{equation}

\begin{equation}\label{eq:divbm}
{\rm div}{\bf B} ~=~0
\end{equation}
that will be further adjusted in the subsequent sections. The norm
$\vert \cdot\vert$ is the usual one in ${\bf R}^3$, i.e.:
$\vert (x,y,z)\vert=\sqrt{x^2+y^2+z^2}$.
We define  ${\bf J}= {\bf P}/\vert {\bf P}\vert
= ({\bf E} \times  {\bf B})/ \vert {\bf E} \times  {\bf B}\vert$.
Note that, when  ${\bf P}=0$, the direction of ${\bf J}$ is not determined (see also the
comments at the end of section 7). The vector ${\bf J}$ is supposed to be adimentional (or,
equivalently, ${\bf P}/\vert{\bf P}\vert$ is multiplied by a constant,
equal to 1, whose dimension is the inverse of the dimension of ${\bf P}$).
Consequently, $c{\bf J}$ is a velocity vector.
\par\smallskip

As the reader may notice, the ``awkward'' relation  ${\rm div}{\bf E}=0$ has been eliminated. It is also
evident that in all the points in which ${\rm div}{\bf E}=0$, we find again the classical Maxwell
system. This states that the solutions of
(\ref{eq:rotb})-(\ref{eq:dive})-(\ref{eq:rote})-(\ref{eq:divb}) are also solutions of
(\ref{eq:rotbm})-(\ref{eq:rotem})-(\ref{eq:divbm}). Therefore, the replacement of
(\ref{eq:rotb})-(\ref{eq:dive}) by (\ref{eq:rotbm}) brings us to the property we wanted, that is the
enlargement of the range of solutions.
\par\smallskip

Afterwards, we have to understand and justify what is happening from the point of view of Physics. Let
us  recall that we are in empty space, and that
 there are no electrical charges or masses anywhere. We are only examining
the behavior of waves. In spite of that, we pretend that there may be regions where ${\rm div}{\bf
E}\not =0$. The situation is not alarming, since we checked that the condition ${\rm div}{\bf E}\not =0$
is quite frequent in the study of waves. Anyway, such a hypothesis is acceptable, as long as it is
coherent with the basic laws of Physics. First of all, we observe that the equation (\ref{eq:rotbm}) has
been obtained by adding a nonlinear term to (\ref{eq:rotb}). The term has strong analogy with the
corresponding one of classical electromagnetism,  appearing on the right-hand side of (\ref{eq:rotb}) as
a consequence of the Amp\`ere law, and  due to the presence of moving charges. In fact, by setting
$\rho={\rm div}{\bf E}$, the vector $\rho c{\bf J}$ can be assimilated, up to dimensional constants,  to
an electric current density. Thus, even if in our case there are no real charged particles, we have to
deal  with a continuous time-varying medium, consisting  of infinitesimal electrical charges, living
with the electromagnetic wave during its evolution. Moreover, the added term does not compromise the
theoretical study of a functioning device like an antenna, since, at a certain distance, the quantity
${\rm div}{\bf E}$ is negligible.
\par\smallskip

\noindent By taking the divergence of (\ref{eq:rotbm}), we get a very important relation:
\begin{equation}\label{eq:continu}
{\partial \rho\over\partial t}~=~-c~{\rm div}(\rho {\bf J})
\end{equation}
which is, actually, the continuity equation for the density $\rho={\rm div}{\bf E}$. The equation
(\ref{eq:continu}) testifies to the presence of a transport, at the speed of light, along the direction
determined by ${\bf J}$. Hence, something  is flowing  together with the electromagnetic fields;
something that later, in sections 9 and 10, will be compared to a true mechanical fluid. On the other
hand,  this was  also the interpretation at the end of the 19th century, before the theory of fields was
rigorously developed. The fluid changes in density, but preserves its quantity, as stated by the
continuity equation. It is extremely significant to remark that this property comes directly from
(\ref{eq:rotbm}), so it is not an additional hypothesis. In section 12, based on the density $\rho$, we
will construct a mass tensor that, due to (\ref{eq:rotbm}), can be perfectly combined with the standard
electromagnetic energy tensor. The skilful reader has already understood that this will allow us to find
the link between electromagnetic and gravitational fields.
\par\smallskip

We are now going to collect other properties about  the new set of equations.
Considering that  ${\bf E}\times {\bf B}$ is orthogonal to both  ${\bf E}$
and ${\bf B}$, a classical result is obtainable:
\begin{equation}\label{eq:energ}
{1\over 2}{\partial \over \partial t} (\vert {\bf E}\vert^2
+c^2\vert {\bf B}\vert^2)~=~
~c^2\big( {\rm curl} {\bf B}\cdot {\bf E}-{\rm curl} {\bf E}\cdot {\bf B}\big)
~=~- c^2~{\rm div}{\bf P}
\end{equation}
where the quantity $\vert {\bf E}\vert^2 + c^2\vert {\bf B}\vert^2$, up to
a multiplicative dimensional constant, is related to the energy of the
electromagnetic field. Thus, the nonlinear term in (\ref{eq:rotbm}) is
not disturbing at this level, and the Poynting vector ${\bf P}$ preserves
its meaning.
\par\smallskip

 By noting that ${\bf J}\cdot {\bf J}=1$ and that
${\bf E}\cdot {\bf J}=0$, we get another interesting relation:
\begin{equation}\label{eq:interes}
{\bf E}\cdot{\partial {\bf J}\over \partial t}~=~
{\partial ({\bf E}\cdot {\bf J})\over \partial t} ~-~
{\partial {\bf E}\over\partial t}\cdot {\bf J}~=~
-c^2 ~({\rm curl}{\bf B})\cdot {\bf J} ~+~ c~{\rm div}{\bf E}
\end{equation}

\noindent Finally, one has:
$${\partial^2 {\bf B}\over \partial t^2}~=~-{\rm curl}{\partial {\bf E}
\over \partial t}~=~-c^2~{\rm curl}({\rm curl}{\bf B})~+~c~{\rm curl}
(\rho {\bf J})~$$
\begin{equation}\label{eq:preon}
=~-c^2~\nabla ({\rm div}{\bf B})~+~c^2 ~\Delta {\bf B}~+~
c~{\rm curl}(\rho {\bf J})
\end{equation}
from which we deduce the following second-order vector equation with a nonlinear forcing term:
\begin{equation}\label{eq:ondenon}
{\partial^2 {\bf B}\over \partial t^2}~=~c^2 ~\Delta {\bf B}~+~
c~{\rm curl}(\rho {\bf J})
\end{equation}
that generalizes the second equation in (\ref{eq:eonde}). We are sorry to announce that the ``wave''
equations for the fields ${\bf E}$ and ${\bf B}$ are no longer true. On the other hand, it has emerged
in section 2 that, in vector form, they are only a source of a lot of trouble.
\par\smallskip

In the classical Maxwell equations the role of the field ${\bf E}$ can be interchanged with that of
field $c{\bf B}$. This is not true for the new formulation. We will later see, in section 9, how to
solve this problem. For the moment, we keep working with
(\ref{eq:rotbm})-(\ref{eq:rotem})-(\ref{eq:divbm}), just because the theory will be more easy. In the
coming sections 4 and 5, we will see how elegantly it is possible to solve the problems raised in
section 2.

\par\bigskip

\setcounter{equation}{0}

\section{Perfect spherical waves}

In the case of a plane wave of infinite extension, for both  the Maxwell model and the new one, we are
able to enforce the condition ${\rm div}{\bf E}=0$ and realize the orthogonality of the Poynting vectors
with respect to the propagation fronts. Concerning a ``fragment'' of plane wave, the classical method
runs into problems. However, with the new approach
 the situation radically improves. Let us see why.
\par\smallskip

With the same assumptions of section 2,
let $\Omega$ be the square $[0,1]\times [0,1]$. We already noted that, on the sides
$\{0\}\times ]0,1[$ and $\{1\}\times ]0,1[$, the quantities
${\rm div}{\bf E}$ and ${\rm curl}{\bf B}$  become infinite. Nevertheless, when
 substituted into equation (\ref{eq:rotbm}), they come to a difference of the type $+\infty
-\infty$. The two singular terms  reciprocally cancel out, leaving a finite quantity, so that the
equation has a chance to be satisfied.  To show this, we can create a layer around the boundary of
$\Omega$.  Then, without developing singularities, we pass to the limit for the width of the layer
tending to zero. The trick now works, because, in contrast to the classical Maxwell case, equation
(\ref{eq:rotbm}) can be satisfied exactly in all the points, since it is compatible with the condition
${\rm div}{\bf E}\not =0$. In the limit process we can also guarantee that the Poynting vectors remain
parallel to the $z$-axis. Therefore, the fragment does not change its transversal shape. Explict
computations will be carried out in section 5,  in the case in which $\Omega$ is a circle.
\par\smallskip

For example, the situation represented in figure 2 is perfectly allowed for by our equations, except
near the lower and the upper sides. Actually, on the sides  $]0,1[\times \{0\}$ and  $]0,1[\times
\{1\}$, given that ${\rm div}{\bf B}$ and  ${\rm curl}{\bf E}$ are singular, we still have problems
(clearly equation (\ref{eq:rotem}) and (\ref{eq:divbm}) are
 not true). Similar problems
are encountered by modifying the polarization of the wave. These troubles will be solved in section 9,
by unifying  (\ref{eq:rotem}) and (\ref{eq:divbm}) in a single equation similar to (\ref{eq:rotbm}), in
such a way that the roles of
 ${\bf E}$ and $c{\bf B}$ are interchangeable.
\par\smallskip

The case of a spherical wave is very interesting. Let us consider the transformation of coordinates
given in (\ref{eq:coor}). Let us also suppose that the fields are given as in  (\ref{eq:campi}), with
 $u$, $v$, $w$ not depending on  $\theta$. We have:
\begin{equation}\label{eq:poyn}
{\bf E} \times  {\bf B}~=~(uw,~-uv,~0)
\end{equation}

\begin{equation}\label{eq:j}
{\bf J}~=~{{\bf E} \times  {\bf B}\over \vert {\bf E} \times  {\bf B}\vert}~=
~{s(u)\over \sqrt{v^2 +w^2}}\big(w,~-v,~0\big)
\end{equation}
where $s(u)=u/\vert u\vert$ is the sign of $u$.
\par\smallskip

\noindent The new equations in spherical coordinates become:

\begin{equation}\label{eq:utn}
u_t~=~-\Big(w_r+{w\over r}\Big)~+~{v_\phi\over r}
\end{equation}

\begin{equation}\label{eq:vtn}
v_t=~{c^2\over r}\Big(u {\cos\phi\over\sin\phi}~+~u_\phi \Big)
~-~c~s(u)~{\displaystyle{v_r+{2\over r}v + {1\over r}w_\phi +
{\cos\phi\over r\sin\phi}w}\over\sqrt{v^2+ w^2 }}~w
\end{equation}

\begin{equation}\label{eq:wtn}
w_t=~-c^2\Big({u\over r}~+~u_r\Big)
~+~c~s(u)~{\displaystyle{v_r+{2\over r}v + {1\over r}w_\phi + {\cos\phi\over
r\sin\phi}w}\over\sqrt{v^2+ w^2 }}~v
\end{equation}
These expressions may seem rather complicated, but there is nothing to be afraid of.
\par\smallskip

To avoid discontinuities, we also impose the boundary conditions
(\ref{eq:concon1}) and (\ref{eq:concon2}).
Now, by choosing $v=0$ and $w=cu$, one obtains:

\begin{equation}\label{eq:poydiv}
{{\bf E} \times  {\bf B}\over \vert
{\bf E} \times  {\bf B}\vert}{\rm div}{\bf E}~=~\Big( {cu\cos\phi
\over r\sin\phi}+{c\over r}u_\phi,~0, ~0\Big)
\end{equation}

\noindent Therefore, one gets:

\begin{equation}\label{eq:uts}
 u_t~=~-\Big(w_r+{w\over r}\Big)
\end{equation}

\begin{equation}\label{eq:vts}
v_t~=~0
\end{equation}

\begin{equation}\label{eq:wts}
 w_t~=~-c^2\Big(u_r+{u\over r}\Big)
\end{equation}

Once again, the first and the last equations are equivalent, providing the general solution
(\ref{eq:solgen}). Anyway, this time,  thanks to the nonlinear corrective term of (\ref{eq:rotbm}), the
second equation is compatible with  $v=0$. We are not obliged to choose $f(\phi )=1/\sin\phi$, in order
to enforce the condition ${\rm div}{\bf E}=0$, because this constraint is no longer required. The
conclusion is that perfect spherical waves are admissible with the new model. The functions $f$ e $g$
may be truly arbitrary (the only restriction is $f(0)=f(\pi)=0$). Continuing with our analysis, we will
construct later infinite other solutions which are unobtainable with the classical Maxwell model.
\par\smallskip

In the perfect spherical case, the Poynting vector ${\bf P}=(cu^2,~0, ~0)$ only has the radial component
different from zero. As expected, this component has a constant sign (even if $u$ and $w$ oscillate).
Since the set of equations is of a hyperbolic type, we can introduce the characteristic curves. In the
example of the spherical wave, such curves are semi straight-lines emanating from the point $r=0$, and
the vector ${\bf J}={\bf P}/\vert {\bf P}\vert =(1,~0, ~0)$ is aligned with them.
  The nonlinear term introduced in (\ref{eq:rotbm}) does not
adversely affect the behavior of the wave, because, with  $v=0$ and $w=cu$, the corresponding equations
(\ref{eq:uts}) and (\ref{eq:wts}) are linear. Therefore, the superposition principle is still valid. Any
piece of information, present at the boundary of the sphere of radius $r>0$, propagates radially at the
speed of light, without being disturbed (except by the natural decay in intensity). The nonlinear
effects of the model are latent. They show up when we try to force, with some external solicitations,
the Poynting vector not to follow the characteristic lines. This circumstance will be taken into account
in sections 7 and 8.
 \par\smallskip

In a very mild form, we can state that the divergence vanishes, by observing that, for any $T$:
\begin{equation}\label{eq:divnul}
\int_T^{T+2\pi /\omega} {\rm div}{\bf E}~dt~=~0
\end{equation}
that is ${\rm div}{\bf E}$ has zero average when integrated over a period of time. Nevertheless, in
section 13, we will get an astonishing result. We will see that an electromagnetic wave produces, during
its passage, a modification of space-time. In the new geometry, the 4-divergence of the electric field
is zero. This could make it difficult, or even impossible, to set up an experiment that emphasizes the
condition ${\rm div}{\bf E}\not=0$ at some point. The measure could be affected by the modified
space-time geometry in such a way the condition cannot be revealed.
\par\smallskip

\noindent Among the stationary solutions we find:
\begin{equation}\label{eq:solstaz}
u(t,r,\phi)={K_1\over r\sin\phi}~ ~~
 v(t,r,\phi)={K_2\over r^2}~ ~~
 w(t,r,\phi)={K_3\over r\sin\phi}
\end{equation}
as well as:
\begin{equation}\label{eq:solstaz2}
u(t,r,\phi)={K_1\over r\sin\phi}~ ~~
 v(t,r,\phi)={K_2\cos\phi}~ ~~
 w(t,r,\phi)={-K_2\sin\phi}
\end{equation}
with $K_1$, $K_2$, $K_3$ arbitrary constants. In particular, we recognize the
classical stationary electric field:
$${\bf E}~=~\big(K_2r^{-2},~0, ~0\big)$$
whose divergence is zero for any $r>0$.  Unfortunately, most of these solutions
show singularities.
\par\smallskip

Due to the nonlinearity of the equations, the study of the interference of waves looks quite
complicated. As long as the waves are such that ${\rm div}{\bf B}=0$  and ${\rm div}{\bf E}=0$ (as in
the plane case) there are no problems, since the nonlinear terms do not actually activate. For waves of
different shape, the situation may be truly intricate. In first approximation, the nonlinear effects
should attenuate faster than the amplitude of the waves. Thus, at a certain distance, these anomalies
may not normally be observed. Although we do not wish to discuss it here, the subject is of crucial
relevance and deserves to be studied in more detail.

\par\bigskip

\setcounter{equation}{0}
\section{Travelling signal-packets}

In this section, it is convenient for us to express our new set of equations in cylindrical coordinates.
After taking $(x,y,z)=(r\cos\theta, r\cos\theta, z)$, we assume that the fields are of the form ${\bf
B}=(0,0,u)$, ${\bf E}=(v,w,0)$, where the first component is referred to the variable $r$, the second
one to the variable $z$ and the third one to the variable $\theta$. Moreover, for simplicity, the
functions $u$, $v$ and $w$ will not depend on $\theta$. In cylindrical coordinates, the counterparts of
equations (\ref{eq:utn})-(\ref{eq:vtn})-(\ref{eq:wtn}) are:

\begin{equation}\label{eq:utcil}
u_t~=~-w_r~+~v_z
\end{equation}

\begin{equation}\label{eq:vtcil}
v_t=~c^2 u_z
~-~c~s(u)~{{\displaystyle{v_r+{v\over r} + w_z}}
\over\sqrt{v^2+ w^2 }}~w
\end{equation}

\begin{equation}\label{eq:wtcil}
w_t=~-c^2\Big({u\over r}~+~u_r\Big)
~+~c~s(u)~{{\displaystyle{v_r+{v\over r} + w_z}}
\over\sqrt{v^2+ w^2 }}~v
\end{equation}
By setting $v=0$, we can easily find solutions when $u$ and $w$ do not depend on $z$. In this case one
has ${\rm div}{\bf B}=0$ and ${\rm div}{\bf E}=0$. From (\ref{eq:utcil}) and (\ref{eq:wtcil}) it is easy
to get the equations:
\begin{equation}\label{eq:uew}
u_{tt}~=~c^2\left({u\over r}+u_r\right)_{\hskip-.1truecm r}~~~~~~~~~~~
w_{tt}~=~c^2\left({w\over r}+w_r\right)_{\hskip-.1truecm r}
\end{equation}
whose solutions are related to Bessel functions.
\par\smallskip

Anyhow, extremely interesting solutions in cylindrical coordinates turn out to be
the following ones:
$$u(t,r,z)=g(t\pm z/c)f(r)~ ~~~~~~v(t,r,z)=\pm c~g(t\pm z/c)f(r)$$
\begin{equation}\label{eq:solstc}
~~ w(t,r,z)=0
\end{equation}
Note that the divergence of ${\bf E}$ is equal to $\rho =v_r+r^{-1}v$, so that
it is different from zero, unless $f$ is proportional to $1/r$.
The functions $f$ and $g$ can be arbitrary (to guarantee the continuity of the vector
fields, we only impose $f(0)=0$). The relations in (\ref{eq:solstc}) give raise
 to electromagnetic waves  shifting at the speed of light along the $z$-axis.
If $f$ and $g$  vanish outside a finite measure interval, for any fixed time $t$
the wave is constrained inside a bounded cylinder. The packet travels unperturbed for an indefinite
amount of time. The corresponding  field ${\bf E}$ is perfecly radial and the vector ${\bf J}$ is
parallel to the $z$-axis.
\par\smallskip

Given $r_0>0$, suppose that $f$ is zero for $r>r_0$. Suppose also that $f$, in a neighborhood of $r_0$
has a sharp gradient. It is evident that the vector $c^2{\rm curl}{\bf B}-c({\rm div}{\bf E}) {\bf J}=
(c^2 u_z, 0 ,0)= (\pm c g^\prime f, 0,0)$ remains bounded even if we let the derivative of $f$ tend to
$\infty$ at $r_0$. Therefore, as we anticipated at the beginning of section 4, we can give a meaning to
equation (\ref{eq:rotbm}), even if $f$ is a discontinuous function in $r_0$.
\par\smallskip

We can get a transport equation for the unknown $\rho ={\rm div}{\bf E}$ by using
the  equation (\ref{eq:continu}), i.e.:
\begin{equation}\label{eq:cont}
{\partial \rho\over\partial t}~=~-c~{\rm div}(\rho {\bf J})
~=~-c\rho ~{\rm div}{\bf J}~-~c\nabla \rho \cdot {\bf J}
\end{equation}
which, thanks to the fact that ${\bf J}$ is a constant field, takes the simplified form of:
\begin{equation}\label{eq:trasp}
{\partial \rho\over\partial t}~=~\pm c~{\partial \rho\over\partial z}
\end{equation}
with the sign depending on the orientation of ${\bf J}$.
\par\smallskip

We recall that the Maxwell equations do not allow for the existence of solitary waves, as the ones we
have introduced right now. Therefore, here we obtained another important result.
\par\smallskip

The energy ${\cal E}$ of these solitary waves is obtained by integrating the energy density,
given by: ${1\over 2}\epsilon_0 (\vert {\bf E}\vert^2 +c^2 \vert {\bf B}\vert^2)$.
Thus, one gets:
\begin{equation}\label{eq:energia}
{\cal E}~=~2\pi\epsilon_0c^2 \int_0^{+\infty }\hskip-.2truecm f^2(r)rdr
\int_{-\infty}^{+\infty}g^2(\xi ) d\xi
\end{equation}

Suppose that, at an initial time $t_0$, the electromagnetic fields are assigned compatibly with
(\ref{eq:solstc}). The vector ${\bf J}$ turns out to be automatically determined, then the wave is
forced to move  in the direction  of ${\bf J}$ at speed $c$. There are no stationary solutions, unless
$g$ is constant. But, in this last case, due to (\ref{eq:energia}), the energy is not going to be
finite. The wave-packets take their energy far away, with no dissipation, until they react with other
waves or more complicated structures (like, for instance, particles).
\par\smallskip

Let us study more closely the expressions given in (\ref{eq:solstc}). Assume that, for $r\geq 0$, the
function  $f$ is non negative, and that  $f(0)=0$. If the function $g$ has a constant sign, we
distinguish between two cases, depending if the sign is positive or negative (see figure 3). The sign
determines the ``orientation''  of the vector ${\rm curl}{\bf B}$ (note however that ${\rm curl}{\bf B}$
is not exactly parallel to the $z$-axis, despite what is shown in figure 3). Then, we have  subcases,
depending whether ${\bf E}$ is directed toward the $z$-axis or not. In conclusion, two possible types of
solitary waves can occur, depending on the orientation of the electric field (external or internal).
These will be denoted by  $\gamma^+$ and  $\gamma^-$, respectively. In figure 3, ${\bf J}$ indicates the
direction of motion. Of course, $g$ could also have a non-constant sign. In this case, the corresponding
wave can be seen as a sequence of waves of type  $\gamma^+$ and $\gamma^-$, shifting one after the
other.
\par\smallskip

\begin{figure}[t]
\begin{picture}(400,290)
\put(0,200){\vector(1,0){150}}
\put(200,200){\vector(1,0){150}}
\put(140,195){\makebox(0,0){$z$}}
\put(340,195){\makebox(0,0){$z$}}

\put(75,230){\vector(0,-1){26}}
\put(65,222){\vector(0,-1){18}}
\put(85,222){\vector(0,-1){18}}
\put(55,216){\vector(0,-1){12}}
\put(95,216){\vector(0,-1){12}}
\put(45,212){\vector(0,-1){8}}
\put(105,212){\vector(0,-1){8}}
\put(35,209){\vector(0,-1){5}}
\put(115,209){\vector(0,-1){5}}

\put(75,170){\vector(0,1){26}}
\put(65,178){\vector(0,1){18}}
\put(85,178){\vector(0,1){18}}
\put(55,184){\vector(0,1){12}}
\put(95,184){\vector(0,1){12}}
\put(45,188){\vector(0,1){8}}
\put(105,188){\vector(0,1){8}}
\put(35,191){\vector(0,1){5}}
\put(115,191){\vector(0,1){5}}

\put(275,204){\vector(0,1){26}}
\put(265,204){\vector(0,1){18}}
\put(285,204){\vector(0,1){18}}
\put(255,204){\vector(0,1){12}}
\put(295,204){\vector(0,1){12}}
\put(245,204){\vector(0,1){8}}
\put(305,204){\vector(0,1){8}}
\put(235,204){\vector(0,1){5}}
\put(315,204){\vector(0,1){5}}

\put(275,196){\vector(0,-1){26}}
\put(265,196){\vector(0,-1){18}}
\put(285,196){\vector(0,-1){18}}
\put(255,196){\vector(0,-1){12}}
\put(295,196){\vector(0,-1){12}}
\put(245,196){\vector(0,-1){8}}
\put(305,196){\vector(0,-1){8}}
\put(235,196){\vector(0,-1){5}}
\put(315,196){\vector(0,-1){5}}

\put(0,60){\vector(1,0){150}}
\put(200,60){\vector(1,0){150}}
\put(140,55){\makebox(0,0){$z$}}
\put(340,55){\makebox(0,0){$z$}}

\put(75,90){\vector(0,-1){26}}
\put(65,82){\vector(0,-1){18}}
\put(85,82){\vector(0,-1){18}}
\put(55,76){\vector(0,-1){12}}
\put(95,76){\vector(0,-1){12}}
\put(45,72){\vector(0,-1){8}}
\put(105,72){\vector(0,-1){8}}
\put(35,69){\vector(0,-1){5}}
\put(115,69){\vector(0,-1){5}}

\put(75,30){\vector(0,1){26}}
\put(65,38){\vector(0,1){18}}
\put(85,38){\vector(0,1){18}}
\put(55,44){\vector(0,1){12}}
\put(95,44){\vector(0,1){12}}
\put(45,48){\vector(0,1){8}}
\put(105,48){\vector(0,1){8}}
\put(35,51){\vector(0,1){5}}
\put(115,51){\vector(0,1){5}}

\put(275,64){\vector(0,1){26}}
\put(265,64){\vector(0,1){18}}
\put(285,64){\vector(0,1){18}}
\put(255,64){\vector(0,1){12}}
\put(295,64){\vector(0,1){12}}
\put(245,64){\vector(0,1){8}}
\put(305,64){\vector(0,1){8}}
\put(235,64){\vector(0,1){5}}
\put(315,64){\vector(0,1){5}}

\put(275,56){\vector(0,-1){26}}
\put(265,56){\vector(0,-1){18}}
\put(285,56){\vector(0,-1){18}}
\put(255,56){\vector(0,-1){12}}
\put(295,56){\vector(0,-1){12}}
\put(245,56){\vector(0,-1){8}}
\put(305,56){\vector(0,-1){8}}
\put(235,56){\vector(0,-1){5}}
\put(315,56){\vector(0,-1){5}}
\put(15,235){\makebox(0,0){$g>0$}}
\put(15,95){\makebox(0,0){$g<0$}}
\put(215,235){\makebox(0,0){$g>0$}}
\put(215,95){\makebox(0,0){$g<0$}}

\put(125,220){\vector(1,0){18}}
\put(345,220){\vector(-1,0){18}}
\put(143,80){\vector(-1,0){18}}
\put(327,80){\vector(1,0){18}}
\put(132,214){\makebox(0,0){${\bf J}$}}
\put(336,214){\makebox(0,0){${\bf J}$}}
\put(132,74){\makebox(0,0){${\bf J}$}}
\put(336,74){\makebox(0,0){${\bf J}$}}
\put(5,180){\vector(1,0){18}}
\put(205,180){\vector(1,0){18}}
\put(23,40){\vector(-1,0){18}}
\put(223,40){\vector(-1,0){18}}
\put(15,170){\makebox(0,0){${\rm curl}{\bf B}$}}
\put(213,170){\makebox(0,0){${\rm curl}{\bf B}$}}
\put(15,30){\makebox(0,0){${\rm curl}{\bf B}$}}
\put(213,30){\makebox(0,0){${\rm curl}{\bf B}$}}

\end{picture}
\vskip.1truecm
\caption{\small{\sl Behavior of the electric field for the two possible wave-packets
(section for a fixed angle $\theta$):
 $\gamma^-$ shifting from left to right,
$\gamma^+$ shifting from right to left,
$\gamma^-$ shifting from  right to left,
$\gamma^+$ shifting from left to right.}}
\end{figure}
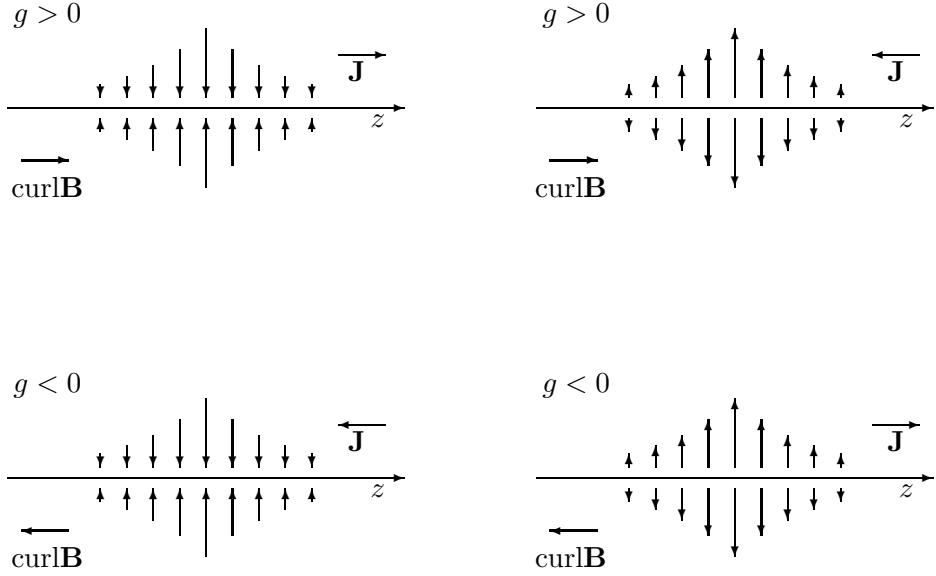

In vacuum, the electromagnetic fields at rest,  are assumed to be identically zero. During the passage
of a soliton, the calm is momentarily broken only at the points ``touched'' by the wave. The information
shifts, but does not irradiate. Thus, also if the wave-packet displays a negative or positive sign,
depending on the orientation of ${\bf E}$, this is not in relation to what is usually called electric
charge. Hence, as long as the cylinders containing two different solitons do not collide, they can cross
very near without influencing each other.  On the contrary, we expect some scattering phenomena, through
a mechanism that will be studied in section 8.
\par\smallskip

If we place a mirror parallel to the $z$-axis, at some distance from it, the reflected image of the
travelling wave-packet  will be the same wave-packet shifting in the opposite direction (because ${\rm
curl}{\bf B}$ changes sign, while ${\bf E}$ maintains its orientation). This disagrees with our common
sense. In other words, equation (\ref{eq:rotbm}) does not preserve plane symmetries. The same is true
for the Maxwell equations. In both cases we have no elements to decide the correct sign of the vector
product $\times$ (left-hand or right-hand). As a matter of fact, without modifying the equations, a
change of the sign of $\times$ can be compensated for by a change of the sign of the electric (or the
magnetic) field. To learn more about this problem we need to study the interactions between waves and
matter. Hence, for the moment, we have no sufficient information to understand from which part of the
mirror is our universe. An answer to these crucial questions will be given in section 8.
\par\smallskip

 In cylindrical coordinates, we can find many other solitonic solutions. Here is
another example:
\begin{equation}\label{eq:solscil}
{\bf E}~=~(\pm cu,~0,~\mp cv)~~~~~~{\bf B}~=~(v, ~0,~u)
\end{equation}
with $u=f_1(r,\theta)g(t\pm z/c)$ and $v=f_2(r,\theta)g(t\pm z/c)$. In order to fulfill the
condition  ${\rm div}{\bf B}=0$, it is necessary to impose:
\begin{equation}\label{eq:divbcil}
{\partial f_1\over\partial \theta}~+~{\partial (rf_2)\over\partial r}
~=~0
\end{equation}
Note that ${\bf E}\cdot{\bf B}=0$ and ${\bf J}=(0,\mp 1,0)$. Except for the condition
(\ref{eq:divbcil}), the functions $f_1$ and $f_2$ are arbitrary, so that the new solutions are very
general. Actually, they include the previous ones (take $f_1=f$ and $f_2=0$). In section 9, we will
remove the condition ${\rm div}{\bf B}=0$, allowing for the existence of even more solutions. We can
force $f_1$ and $f_2$  to vanish outside a bidimensional domain
 $\Omega$. Again, this plane front, modulated by the function  $g$, travels along the
$z$-axis at the speed of light.
\par\smallskip

We may reasonably expect that these solitary solutions are modified, or even destroyed, when they
encounter another external electromagnetic field. In fact, the equations being nonlinear, the
superposition principle does not hold, in particular if the motion is disturbed in a way that is in
contrast to the natural evolution along the characteristic curves. As far as we know, there are no
documented experiments evidencing these facts. In section 11, we instead examine the behavior of
solitary waves under the action of gravitational fields.
\par\smallskip

An electromagnetic radiation can be suitably considered as an envelope of solitary waves, travelling in
the same direction. If, transversally, these solitons are of infinitesimal size, they can be compared to
``light rays''. This observation clarifies how a wave can be viewed at the same time as a whole
electromagnetic phenomenon and as a bundle of infinite microscopic rays.
\par\smallskip

And then there are photons. They are also pure electromagnetic manifestations, but, unfortunately, they
are not modelled by the Maxwell equations. Modern atomic and subatomic physics would not exist without
photons, yet they find no place in the classical theory of electromagnetism. This is an unpleasant gap.
Although physicists are acquainted with this dualism, the general framework remains blurred. From our
new standpoint, we contend that the photons observed in nature have very good chances to be modelled by
the equations introduced here. As a matter of fact, we have enough freedom to be able to build solutions
(no matter how complicated) resembling real photons. We can assign
 a frequency to them, longitudinally or transversally. Then, we know that they must move at
the speed of light and can have finite energy, given by the energy of their electromagnetic vector
fields. They can be ``positive'' or ``negative'' without being electrical charges.
Even with no mass at rest, their motion can be distorted by gravitational fields (see section 11).
A concept of spin can be also introduced (see section 15). If what we are proposing here is a new functioning
 model for electromagnetism (and we will collect other evidence supporting this hypothesis),
 then it  explains why photons can be self-contained
elementary entities and electromagnetic emissions at the same time.
In this case, a first link between classical and quantum physics is set forth.

\par\bigskip

\setcounter{equation}{0}

\section{Lagrangian formulation}

In order to recover the equations (\ref{eq:rotbm})-(\ref{eq:rotem})-(\ref{eq:divbm}) from
the principle of least action, we follow the same path  bringing to the classical
Maxwell equations. Thus, we introduce the scalar potential $\Phi$ and the vector potential
${\bf A}=(A_1,A_2,A_3)$, such that:
\begin{equation}\label{eq:potenz}
{\bf B}~=~{1\over c}~{\rm curl}{\bf A}~~~~~~~~~{\bf E}~=~-{1\over c}
{\partial {\bf A}\over \partial t}~-~\nabla \Phi
\end{equation}
From the above definitions we easily get the equations: ${\rm div}{\bf B}=0$ and  ${\partial
\over\partial t}{\bf B}=-{\rm curl}{\bf E}$. The third equation (\ref{eq:rotbm}) is going
to be deduced from the minimization of a suitable action function.
\par\smallskip

Let us first note that, by taking the pontential $\Phi$ equal to zero and  setting
$\xi = t-(xJ_1+yJ_2+zJ_3)/c$, one obtains:
\begin{equation}\label{eq:campiort}
{\bf E}~=~-{1\over c}{\partial {\bf A}\over \partial \xi}~~~~~~~~~ c{\bf B}~=~\nabla\xi
\times{\partial {\bf A}\over \partial \xi}~=~-{\bf J}\times {1\over c}
{\partial {\bf A}\over \partial \xi}
\end{equation}
This allows us to infer that  ${\bf B}$ is orthogonal to ${\bf E}$ and that
$\vert {\bf E}\vert =c\vert {\bf B}\vert$ (see also \cite{landau}, p.126).
\par\smallskip

 Successively, for $i$ and $k$ between 0 and 3, we introduce the electromagnetic tensor:
\begin{equation}\label{eq:tensore}
F_{ik}~=~{\partial A_k\over\partial x_i}
~-~{\partial A_i\over\partial x_k}
\end{equation}
where $A_0=\Phi$ and $(x_0,x_1,x_2,x_3)=(ct,-x,-y,-z)$.
Explicitly, we have:
\begin{equation}\label{eq:tens1}
F_{ik}~=~\left(\matrix{ 0 & -E_1 & -E_2 & -E_3 \cr
E_1 & 0  & -cB_3 & cB_2 \cr E_2 & cB_3 & 0 & -cB_1 \cr
E_3 & -cB_2 & cB_1 & 0 \cr}\right)
\end{equation}
with ${\bf E}=(E_1,E_2,E_3)$ and ${\bf B}=(B_1, B_2,B_3)$.
 Replacing ${\bf E}$ by $-{\bf E}$, one gets instead the contravariant tensor $F^{ik}$:
\begin{equation}\label{eq:tens2}
F^{ik}~=~\left(\matrix{ 0 & E_1 & E_2 & E_3 \cr
-E_1 & 0 & -cB_3 & cB_2 \cr -E_2 & cB_3 & 0 & -cB_1 \cr
-E_3 & -cB_2 & cB_1 & 0 \cr}\right)
\end{equation}

\noindent Therefore, up to multiplicative constants, the action turns out to be
(see for instance \cite{jackson}, p.596):
\begin{equation}\label{eq:azione}
S~=~-\int F_{ik}F^{ik} dx_0 dx_1 dx_2 dx_3~=~ c\int F_{ik}F^{ik} dxdydzdt
\end{equation}
where, summing-up over repeated indices, the Lagrangian is $L=F_{ik}F^{ik}=2(c^2\vert {\bf B}\vert^2 -
\vert {\bf E}\vert^2)$. As customary, the variations are  functions
 $\delta A_i$, having compact support both in  space and time
(between two fixed instants).
\par\smallskip

\noindent With well-known results, one obtains:
\begin{equation}\label{eq:variaz}
\delta S~=~-4 \int {\partial F^{ik}\over\partial x_k} ~\delta A_i~
dx_0 dx_1 dx_2 dx_3
\end{equation}
Imposing $\delta S=0$, the corresponding Euler equations are exactly the standard
Maxwell equations. As a matter of fact, due to the arbitrariness of the variations
$\delta A_i$, one recovers:
$~-{\partial\over\partial x_k}F^{ik}  =0$ (for $i=0,1,2,3$), that is equivalent to write
${\rm div}{\bf E}=0$ and ${\partial\over\partial t}{\bf E}= c^2{\rm curl}{\bf B}$.
\par\smallskip

Let us now introduce a novelty. We require that the variations $\delta A_i$ are subjected
to a certain constraint, so that the conclusions are going to be different.
Actually, we impose the condition:
\begin{equation}\label{eq:vincolo}
\delta \Phi~-~ {\bf J}\cdot\delta {\bf A}~=~ \delta A_0 -
J_1 \delta A_1-J_2 \delta A_2-J_3 \delta A_3~=~0
\end{equation}
where ${\bf J}$ is the vector
$({\bf E}\times {\bf B})/\vert {\bf E}\times {\bf B}\vert $, already defined in section 3.
The relation (\ref{eq:vincolo}) says for instance that, if the vector variation
 $(\delta A_1$, $\delta A_2$, $\delta A_3)$  locally belongs to the tangent plane generated by
 ${\bf E}$ and ${\bf B}$, then the variation $\delta A_0$ is zero. Although for the moment
we can only  provide vague explanations, we assert that (\ref{eq:vincolo}) is the germ of the Huygens
principle. The picture will become more focused as we proceed with our analysis.
 \par\smallskip

Consequently, we discover that the 4-vector  $-{\partial\over\partial x_k}F^{ik}$
is not identically vanishing, but, due to (\ref{eq:vincolo}), it must have
a component along the 4-vector $(1, -{\bf J})$. This leads to:
\begin{equation}\label{eq:vinc1}
{\rm div}{\bf E}~=~\lambda
\end{equation}
\begin{equation}\label{eq:vinc2}
{1\over c}{\partial {\bf E}\over\partial t}~-~c~{\rm curl}{\bf B}~=~
-\lambda {\bf J}
\end{equation}
where the parameter $\lambda$ is a Lagrange multiplier.
By eliminating  $\lambda$ we easily arrive at equation (\ref{eq:rotbm}).
Thanks to (\ref{eq:vincolo}), the set of possible variations is smaller
than the set in which we impose no restrictions at all. Therefore, the
equation $\delta S=0$ is now less stringent. As we already know,
this shows that (\ref{eq:rotbm}) admits a space of solutions which is larger
than the one corresponding to (\ref{eq:rotb}) and (\ref{eq:dive}) together.
\par\smallskip

\noindent Using the electromagnetic tensor, the equation (\ref{eq:rotbm})
can be written as:
\begin{equation}\label{eq:forminv}
-c\left({\partial F^{ik}\over\partial x_k}~+~{\partial F^{0k}\over\partial x_k}J_i
\right)=0 ~~~~~~~~~~{\rm for}~i=1,2,3
\end{equation}
By defining $J_0=-1$, the above relation is also trivially satisfied for $i=0$. In section 11, within
the framework of general relativity, we will be able to write (\ref{eq:forminv}) in invariant tensor
form.
\par\smallskip

Let us try to  understand the reason for the constraint (\ref{eq:vincolo}). As far as the evolution of a
plane or a spherical wave is concerned (and, surely, in more general cases), it is easy to check that:
\begin{equation}\label{eq:vvinc}
{\bf A}~=~ \Phi {\bf J}
\end{equation}
A straightforward way to get (\ref{eq:vvinc}) is from explicit solutions.
For example, in spherical coordinates, we can use (\ref{eq:solgen}) in order to find:
$${\bf E}~=~\Big(0, ~{1\over r}f(\phi )g(t-r/c), ~0\Big)~~~~~~~~
{\bf B}~=~\Big(0, ~0, ~{1\over cr}f(\phi )g(t-r/c)\Big)$$
\begin{equation}\label{eq:soles}
{\bf J}=(1,0,0)~~~~{\bf A}=\Big(-F(\phi )g(t-r/c), ~0, ~0\Big)
~~~~\Phi = -F(\phi )g(t-r/c)
\end{equation}
where  $F$ is such that $F^\prime =f$. The relation (\ref{eq:vvinc}) is also true in the
case of solitons. In fact, in cylindrical coordinates, thanks to (\ref{eq:solstc}) one has:
$${\bf E}~=~\Big(\pm cf(r )g(t\pm z/c), ~0, ~0\Big)~~~~~~~~
{\bf B}~=~\Big(0, ~0, f(r )g(t\pm z/c)\Big)$$
\begin{equation}\label{eq:solesf}
{\bf J}=(0, \mp 1 ,0)~~~~{\bf A}=\Big(0,~ cF(r )g(t\pm z/c),
 ~0\Big) ~~~~\Phi = \mp cF(r )g(t\pm z/c)
\end{equation}
where $F$ is such that $F^\prime =f$ and $f(0)=0$. Note that, in general,  $\Phi$
and ${\bf A}$ are not uniquely determined. However, there exists at least one choice of
$\Phi$ and ${\bf A}$ such that (\ref{eq:vvinc}) is satisfied.
\par\smallskip

Now, the equation ${\bf A}=\Phi {\bf J}$ implies $\vert \Phi\vert =\vert {\bf A}\vert$, or equivalently:
$\Phi^2 -\vert {\bf A}\vert^2=0$. Taking the variation of the last relation  brings us to the
constraint:
\begin{equation}\label{eq:eqvinc}
2\Big( \Phi ~\delta\Phi~-~{\bf A}\cdot\delta {\bf A}\Big)~=~
2\Phi ~\Big(\delta \Phi ~-~ {\bf J}\cdot \delta {\bf A}\Big)~=~0
\end{equation}
which is the same as in (\ref{eq:vincolo}). Another way to get
(\ref{eq:vincolo}) is  by directly evaluating the variation of (\ref{eq:vvinc}):
\begin{equation}\label{eq:eqvinc2}
\delta {\bf A}=\delta (\Phi {\bf J})=\delta \Phi ~{\bf J}+\Phi~
\delta {\bf J}~~\Rightarrow~~ {\bf J}\cdot \delta {\bf A}=\delta \Phi
\end{equation}
where we used that $\vert {\bf J}\vert =1$ and $\delta {\bf J}\cdot
{\bf J}=0$ (a normalized vector is orthogonal to its variation).
\par\smallskip

\noindent Obviously, the vector relation ${\bf A}=\Phi {\bf J}$ implies the scalar relation:
\begin{equation}\label{eq:vincsca}
\Phi~=~{\bf J}\cdot {\bf A}
\end{equation}
obtainable after scalar multiplication  by ${\bf J}$ and by observing that $\vert {\bf J}\vert =1$.
Another way of expressing (\ref{eq:vincsca}) is to require that the scalar product between the 4-vectors
$(\Phi , {\bf A})$ and $(1, -{\bf J})$ is zero. This makes (\ref{eq:vincsca}) an invariant in the
context of general relativity.
\par\smallskip

In conclusion, the equation (\ref{eq:rotbm}) can be recovered from the constrained minimization of the
action function associated with the classical Lagrangian. The constraint originates from
(\ref{eq:vvinc}) which says, in particular, that ${\bf A}$ is lined up with ${\bf J}$. As will be better
explained in section 10, such a condition characterizes the propagation of waves, whose evolution is
ruled by the Huygens principle. From now on, these will be called ``free waves''. In sections 7 and 8,
we will see that not all waves are of this type.
\par\smallskip

An interesting relation, that is  directly obtained by checking (\ref{eq:soles}) or
(\ref{eq:solesf}), is the following one:
\begin{equation}\label{eq:lorentz}
{\bf E}~+~c~{\bf J}\times {\bf B}~=~0
\end{equation}
The above equation is extremely important, since it represents another characterization of free waves,
perhaps simpler than (\ref{eq:vvinc}). Indeed, it is the analogous of the Lorentz law for moving
electric charges (recall that $c{\bf J}$ is a velocity), even if here there are no particles around. As
it will be explained in the coming sections, equation (\ref{eq:lorentz}) tells us that the  mechanical
forces acting on a free wave are zero. Therefore, the wave actually moves without external disturbances
and in agreement with the Huygens principle. We will
 prove all these statements in sections 9 and 10.
\par\smallskip

\noindent The covariant version of (\ref{eq:lorentz}) is:
\begin{equation}\label{eq:lore2}
-~c{\bf B}~+~{\bf J}\times {\bf E}~=~0
\end{equation}
obtained by vector multiplication of (\ref{eq:lorentz}) by ${\bf J}$.
Both (\ref{eq:lorentz}) and (\ref{eq:lore2}) can be trivially deduced from the orthogonality of
 ${\bf E}$ with respect to  ${\bf B}$, and by  the equality $\vert {\bf E}\vert = c \vert
{\bf B}\vert$ (see also the beginning of this section). Therefore, they have quite a general
validity.
\par\smallskip

Before ending this section, we will collect a few more properties. We begin by considering the following
writing:
$$\vert {\bf E}\vert^2 -c^2\vert {\bf B}\vert^2~=~
 {\bf E}\cdot\left( -{1\over c}{\partial {\bf A}\over
\partial t}~-~\nabla\Phi\right)~-~c~{\bf B}\cdot{\rm curl}{\bf A}$$
$$=~-{1\over c}~{\partial \over \partial t}({\bf E}\cdot {\bf A})~+
~{1\over c}~{\partial {\bf E}\over \partial t}\cdot {\bf A}
~-~ {\bf E}\cdot \nabla\Phi$$
$$-~c~{\bf A}\cdot {\rm curl}{\bf B}~-~c~{\rm div}
({\bf A}\times {\bf B})$$
$$=~-{1\over c} ~{\partial\over \partial t}( {\bf E}\cdot {\bf A})~+~
{1\over c}\left( {\partial {\bf E}\over \partial t} ~-~c^2
{\rm curl}{\bf B}\right)\cdot {\bf A}$$
\begin{equation}\label{eq:catena}
-~ {\rm div}(\Phi {\bf E})~+~\Phi~ {\rm div}{\bf E}~
-~c~ {\rm div}({\bf A}\times {\bf B})
\end{equation}
where we used the definitions in (\ref{eq:potenz}) and known formulas of
vector calculus. Introducing the constraint
 ${\bf A}=\Phi {\bf J}$ (hence also  $\Phi ={\bf J}\cdot {\bf A}$)
we have that ${\bf A}$ is orthogonal to  ${\bf E}$, because so it is ${\bf J}$.
Thus, (\ref{eq:catena}) can be simplified and becomes:
$$\vert {\bf E}\vert^2 -c^2\vert {\bf B}\vert^2
~=~- {\rm div}\Big( \Phi ({\bf E}~+~c~{\bf J}\times {\bf B})
\Big)$$
\begin{equation}\label{eq:catena2}
~+~{1\over c}\left( {\partial {\bf E}\over\partial t}~-
~c^2 {\rm curl}{\bf B}~+~c ({\rm div}{\bf E}){\bf J}\right)
\cdot {\bf A}
\end{equation}
Then, it is interesting to point out that, when both (\ref{eq:lorentz}) and (\ref{eq:rotbm}) are
satisfied, the Lagrangian vanishes. On the contrary, when ${\bf E}$ is orthogonal to ${\bf B}$ and
$\vert{\bf E}\vert = c\vert {\bf B}\vert$ (so that (\ref{eq:lorentz}) holds), the relation
(\ref{eq:catena2}) reveals that imposing equation (\ref{eq:rotbm}) is a natural requisite.
\par\smallskip

\noindent Finally, due to  (\ref{eq:potenz}), equation (\ref{eq:rotbm}) entails:
$$-{1\over c}~{\partial^2 {\bf A}\over \partial t^2}~-~\nabla {\partial\Phi\over
\partial t}~=~c~{\rm curl}({\rm curl}{\bf A})~+~c\left({1\over c}~{\rm div}
{\partial {\bf A}\over \partial t}~+~\Delta \Phi\right){\bf J}
$$
\begin{equation}\label{eq:ondeap}
=-c~\Delta {\bf A}~+~c\nabla ({\rm div}{\bf A})~+~\left(
{\partial ({\rm div}{\bf A})\over \partial t}~+~c~\Delta \Phi\right){\bf J}
\end{equation}
We also assume the following Lorentz condition:
\begin{equation}\label{eq:color}
{\rm div}{\bf A}~+~{1\over c}{\partial \Phi\over \partial t}~=~0
\end{equation}
which is known to be an invariant expression in general relativity. With the
help of (\ref{eq:color}), from (\ref{eq:ondeap}) we get:
\begin{equation}\label{eq:ondea}
{\partial^2 {\bf A}\over \partial t^2}~-~c^2\Delta {\bf A}~=~
\left({\partial^2\Phi\over \partial t^2}~-~c^2\Delta \Phi\right){\bf J}
\end{equation}
which is in perfect agreement with the continuity equation (\ref{eq:continu}),
once we define $\rho =c^{-2}{\partial^2\over \partial t^2}\Phi - \Delta \Phi$.
Note however that (\ref{eq:continu}) is true independently of
 (\ref{eq:color}) and (\ref{eq:ondea}). In the classical Maxwell case, the right
and the left terms of (\ref{eq:ondea})  both vanish, providing, together with  (\ref{eq:eonde}),
additional ``wave'' equations. This is not necessarily true in our case.

\par\bigskip

\setcounter{equation}{0}

\section{Encounter of a wave with an obstacle}

In this section and in the following one, we study what happens to a free wave when it meets an obstacle
that we define of ``mechanical type''. The phenomenon can be extremely complicated, therefore the
analysis will be carried out on very simple cases. For the moment, here we just discuss some basic
facts, trying to catch the underlying ideas. In section 9, we formalize the problem better, by writing
down the final equations.
\par\smallskip

We first take into account an example concerning the reflection of an electromagnetic radiation.  We
assume to have a monocromatic plane wave, linearly polarized, which is totally reflected by a
perfectly-conducting metallic wall. Hence, we suppose that there is no refraction at all. In Cartesian
coordinates, the wall corresponds to the plane $y=0$. Initially, the wave-front propagates forming
 an angle $\zeta \not=0$ with the $y$-axis. Referring to figure 4,
the incident wave is described by the fields:
$${\bf E}^{(i)}~=~\Big(0,~-c\sin\zeta ,~c\cos\zeta\Big)~ \sin\omega(t-(y\cos\zeta +
z\sin\zeta)/c)$$
\begin{equation}\label{eq:polariz}
{\bf B}^{(i)}~=~\Big(1,~0,~0\Big)~ \sin\omega(t-(y\cos\zeta +
z\sin\zeta)/c)
\end{equation}

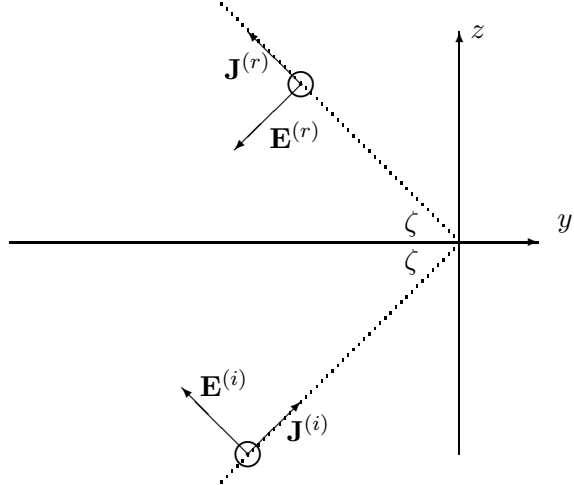
\begin{figure}[t]
\begin{picture}(400,200)
\put(240,10){\vector(0,1){160}}
\put(70,90){\vector(1,0){200}}
\multiput(150,0)(2,2){45}{\line(0,-1){1}}
\multiput(150,180)(2,-2){45}{\line(0,1){1}}
\put(247,170){\makebox(0,0){$z$}}
\put(280,97){\makebox(0,0){$y$}}
\put(160,10){\vector(1,1){20}}
\put(160,10){\makebox(0,0){$\bigodot$}}
\put(180,150){\vector(-1,1){20}}
\put(160,10){\vector(-1,1){25}}
\put(180,150){\makebox(0,0){$\bigodot$}}
\put(180,150){\vector(-1,-1){25}}
\put(222,97){\makebox(0,0){$\zeta$}}
\put(222,83){\makebox(0,0){$\zeta$}}
\put(178,131){\makebox(0,0){${\bf E}^{(r)}$}}
\put(160,157){\makebox(0,0){${\bf J}^{(r)}$}}
\put(151,37){\makebox(0,0){${\bf E}^{(i)}$}}
\put(183,20){\makebox(0,0){${\bf J}^{(i)}$}}
\end{picture}
\vskip.5truecm

\caption{\small \sl Reflection of a plane wave when the magnetic field
is normal to the plane of incidence $x=0$. The vectors ${\bf B}^{(i)}$
and ${\bf B}^{(r)}$ are therefore orthogonal to the page.}

\end{figure}
\begin{figure}[t]
\begin{picture}(400,200)
\put(240,10){\vector(0,1){160}}
\put(70,90){\vector(1,0){200}}
\multiput(150,0)(2,2){45}{\line(0,-1){1}}
\multiput(150,180)(2,-2){45}{\line(0,1){1}}
\put(247,170){\makebox(0,0){$z$}}
\put(280,97){\makebox(0,0){$y$}}
\put(160,10){\vector(1,1){20}}
\put(160,10){\makebox(0,0){$\bigoplus$}}
\put(180,150){\vector(-1,1){20}}
\put(180,150){\makebox(0,0){$\bigodot$}}
\put(160,10){\vector(-1,1){25}}
\put(180,150){\vector(1,1){25}}
\put(222,97){\makebox(0,0){$\zeta$}}
\put(222,83){\makebox(0,0){$\zeta$}}
\put(204,161){\makebox(0,0){${\bf B}^{(r)}$}}
\put(160,157){\makebox(0,0){${\bf J}^{(r)}$}}
\put(151,35){\makebox(0,0){${\bf B}^{(i)}$}}
\put(183,20){\makebox(0,0){${\bf J}^{(i)}$}}
\end{picture}
\vskip.5truecm

\caption{\small\sl Reflection of a plane wave when the electric field
is normal to the plane of incidence $x=0$. The vectors ${\bf E}^{(i)}$
and ${\bf E}^{(r)}$ are therefore orthogonal to the page.}
\end{figure}
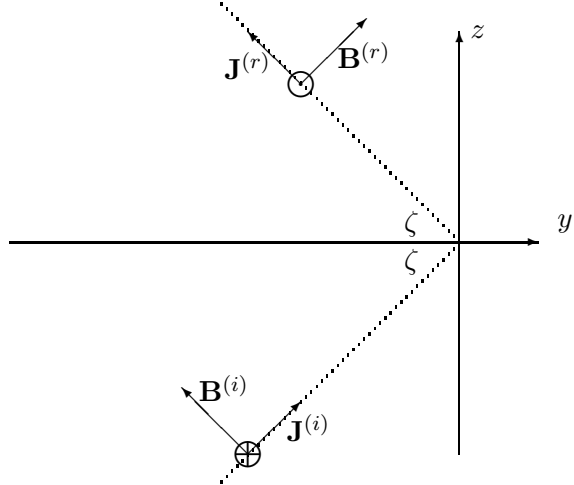

The reflected electric field  ${\bf E}^{(r)}$ is such that, for $y=0$, the component parallel to the
obstacle, of the vector ${\bf E}^{(r)}+{\bf E}^{(i)}$, vanishes (see for instance \cite{bleaney},
p.270). Concerning the magnetic field at $y=0$, we have: ${\bf B}^{(r)}={\bf B}^{(i)}$. Then, one easily
gets:
$${\bf E}^{(r)}~=~\Big(0,~-c\sin\zeta ,~-c\cos\zeta\Big)~ \sin\omega(t+(y\cos\zeta -
z\sin\zeta)/c)$$
\begin{equation}\label{eq:polarizb}
{\bf B}^{(r)}~=~\Big(1,~0,~0\Big)~ \sin\omega(t+(y\cos\zeta -
z\sin\zeta)/c)
\end{equation}
To justify the imposition of the  boundary conditions on $y=0$, it is customary
 to assume the existence of instantaneous electric currents on the conducting surface,
 which force the tangential electric field to zero (see \cite{born}, p.558,
and \cite{jackson}, p.335).
\par\smallskip

After reflection, the wave is very similar to the incident one, with the exception that ${\bf
J}^{(i)}=(0, \cos\zeta ,\sin\zeta )$ has changed in ${\bf J}^{(r)}=(0, -\cos\zeta ,\sin\zeta )$. Note
that the relation (\ref{eq:lorentz}) is valid both for the incident and the reflected waves.
Nevertheless, during the impact, in which an instantaneous flip of the signs occurs,
 the wave-front at $y=0$ does not show a behavior consistent with
the one corresponding to a free wave. Each wave-front, when reaching the wall, evolves for a single
moment without respecting the eikonal equation. Of course, what we are considering here is just a
mathematical idealization. More realistically, the wall is made of matter, so it would be more correct
to check what happens to the wave when it interacts with the atoms of the wall. Anyway, we do not think
it is useful to study this more complicated situation, since it only modifies the form but not the
substance of  facts. We believe that the main idea has already been outlined: when encountering an
obstacle a free wave may lose its likeness and become, for a small amount of time,  a ``constrained
wave''. The reaction of the obstacle can be so strong that, as in the case of the reflecting wall, the
wave-fronts are forced to modify abruptly the direction of their movement. What we would like to do in
the following pages, is to understand what happens at those instants.
\par\smallskip

We recall that, for $y=0$, the vector ${\bf E}^{(r)}+{\bf E}^{(i)}$ does not have the same length of
${\bf E}^{(i)}$ before the impact (or ${\bf E}^{(r)}$ after the impact). Therefore, some electromagnetic
energy turns out to be missing. We conjecture that it has been spent, with the help of the mechanical
constraint, to allow the variation of the direction of the wave-front of an angle $\pi-2\zeta$. For a
moment, this energy has gone elsewhere, compensated by something which is not of an electromagnetic
kind. We would like to find out what is. To this end, let us introduce a new vector field:
\begin{equation}\label{eq:gifi}
{\bf G}~ =~ \cases{ 0 & if $y\not =0$ \cr\cr {\displaystyle{c^2\lim_{y\rightarrow 0}
{ {\bf J}^{(r)}-{\bf J}^{(i)} \over y}}} ~~~~& if $y=0$}
\end{equation}
Note that, dimensionally, the vector ${\bf G}$ corresponds to an acceleration.
In the case under study, at $y=0$, ${\bf G}$ is oriented as the vector $(0,-1,0)$,
with an infinite magnitude. If we imagine the wave as a bundle of rays (we saw
in section 5 that the two things are strictly related), then
${\bf G}$ provides a measure of their curvature. When we are dealing with a free wave,
 the rays proceed along straight-lines. Corresponding to this situation,
we have ${\bf G}=0$. When the rays hit the wall, then ${\bf G}$ becomes different from zero. In the
particular case we are examining, ${\bf G}$ is a singularly distributed field, but if we allow the rays
to change their trajectories smoothly, then ${\bf G}$ is going to be finite. This would give the idea of
a centripetal time-varying acceleration, responsable for the rotation of the rays and the corresponding
wave-fronts.
\par\smallskip

In the coming sections, we will make evident that a nonvanishing vector ${\bf G}$ appears each time a
wave evolves without following the Huygens principle, as a consequence of external perturbations. Some
suitable way of estimating the magnitude ${\bf G}$ should allow us to compensate the missing
electromagnetic energy. By the way, the theory is not going to be easy. In section 13, we will discover
that, in order to change the trajectories of the rays, it is necessary to pass through a modification of
the space-time structure. Thus,  we cannot be more precise, until we are ready to carry out our analysis
in the framework of general relativity. Before that, we have to work a little more on the definition
(\ref{eq:gifi}). This will be done in section 9. For the moment, let us put together other basic facts.
\par\smallskip

We now vary the polarization of the incident wave of 90 degrees (see
 figure 5). For $y=0$, the component, orthogonal to the reflecting wall,
of the magnetic field must be zero. At the same time, the whole electric field vanishes (since, for $y$
approaching to zero, the field ${\bf E}^{(r)}$ is opposite to ${\bf E}^{(i)}$). Therefore, more
electromagnetic energy is missing when the wave hits the wall. Actually, in this case, the effects of
the obstacle are stronger: together with the deviation of the wave-front, we also observe a torsion that
modifies the polarization by 180 degrees. Such a torsion process is instantaneuos, but qualitatively
similar to that corresponding to a circularly-polarized plane wave, like the one for example expressed
by the following fields:
$${\bf E}~=~\big(c\cos \omega (t-y/c),~0,~c\sin \omega (t-y/c)\big)$$
\begin{equation}\label{eq:polcir}
{\bf B}~=~\big(\sin \omega (t-y/c),~0,~-\cos \omega (t-y/c)\big)
\end{equation}
where the polarization constantly changes at finite speed. We note that this wave is also a solution to
Maxwell equations and is more ``energetic'' than the corresponding one with constant polarization.
\par\smallskip

If, together with  reflection, we also have refraction within a medium of different density, the study
is more involved. When reaching the plane of reflection the rays bifurcate. In our opinion, this is due
to the arbitrariness of the vector ${\bf J}$ at the time of the impact (for example because ${\bf P}$ is
zero). The incident wave splits into two solutions (the reflected and the refracted waves), both
compatible with the same boundary conditions on the plane $y=0$, brought by the incoming solution. We do
not necessarily have bifurcation each time that ${\bf P}$ is zero (this actually happens very
frequently), but only when, in addition to this, suitable uncommon conditions are verified.
\par\smallskip

Based on the above observations, we conclude this section with a little philosophical dissertation. The
equations (\ref{eq:rotbm})-(\ref{eq:rotem})-(\ref{eq:divbm}) are of local and deterministic type. Hence,
for given initial data, the solution is uniquely determined. Nevertheless, there may be circumstances in
which, following the evolution of a certain solution, several other branches of solutions of the
equation (\ref{eq:rotbm}) may be admissible. As we noticed, this could be true because the evolution of
the vector ${\bf J}$  turns out to be compatible with different options. Thus, the original solution
splits, and this event is also deterministic. As a matter of fact, when the right conditions are
fulfilled, an incident wave bifurcates, giving raise to a reflected and a refracted wave. There is no
uncertainty: both the solutions are systematically observed. However, this situation
 becomes extremely unstable when reversing time. We are indeed  allowed to
think that, marching backward in time, the inverse of a bifurcation phenomenon could occur. In this
case,  as a film runs in reverse, two waves would meet in perfect  phase and melt, producing a single
wave. Nevertheless, this event has zero probability of happening. A little perturbation is sufficient to
modify completely the evolution of the two waves, with no chances of seeing their fusion.
\par\smallskip

In conclusion, our equations are of hyperbolic type, deterministic and energy preserving. Nevertheless
the particular nature of the nonlinear term can be a source of instabilities when reversing the sign of
time. The consequence is that some original information can get lost, and there is no practical way to
recover it by following the reverse path. We ask ourselves if this could be  an explanation (at least in
part) for the second law of thermodynamics.

\par\bigskip

\setcounter{equation}{0}

\section{Diffraction phenomena}

We continue our qualitative analysis on the interaction of waves with matter.
The second  example that we take into account is the
developing of diffraction, where an electromagnetic wave encounters
an aperture. Once again, for simplicity, the device used to generate
the phenomenon is a  pure mathematical abstraction. So, let us
suppose that a plane monocromatic wave propagates in the direction
of the $y$-axis (with $y$ increasing) and hits a perfect-conducting
wall at $y=0$. This time, however,  there is a passage through the
strip  $0\leq z \leq a$, for some positive width $a$.
\par\smallskip

For $y<0$, the incident wave is polarized as in (\ref{eq:polariz}) with
 $\zeta =0$. As a consequence, we have:
$$ {\bf E}~=~(0,~0,~c\sin\omega (t-y/c))~~~~~{\bf B}~=~(\sin\omega (t-y/c),~0,~0)$$
$${\bf E}\times{\bf B}~=~{\bf P}~=~\big(0,~c[\sin\omega(t-y/c)]^2,~0\big)$$
\begin{equation}\label{eq:relaz}
{\bf J}~=~{{\bf P}\over \vert{\bf P}\vert}~=~(0,~1,~0)~~~~~~~
{\rm div}{\bf E}~=~0~~~~~~{\rm div}{\bf B}~=~0
\end{equation}
For $z < 0$ and $z > a$ the wave is reflected. Actually, it is not a perfect
reflection, since it is affected by some perturbations originating at
the boundaries of the aperture. Here, we neglet this aspect and
focus our attention on the study of the sources of these disturbances.
\par\smallskip

We assume that along the two straight-lines
$y=0$, $z=0$ and  $y=0$, $z=a$, instantaneous electric currents push the electric field
to zero, so that the  condition ${\bf E}=0$ is enforced (see \cite{born},
p. 559). At the instant in which the wave reaches the obstacle, the electric field
 develops discontinuities. Therefore, its divergence is a  concentrated Dirac
 distribution along the above mentioned straight-lines.
Thus, for  $y=0$, it is easy to realize that:
\begin{equation}\label{eq:divdi}
{\rm div}{\bf E}~=~c~[\delta_0(z)-\delta_a(z)]\sin\omega t
\end{equation}
On the other hand, for $0<z<a$, we can expect that:
\begin{equation}\label{eq:lim1}
\lim_{y\rightarrow 0^+}{\bf P}~=~(0,~c(\sin\omega t)^2,~0)
\end{equation}
because these are points in which the wave does not hit the obstacle.
\par\smallskip
Next, let us examine more in detail what happens on the straight-lines
 $y=0, z=0$ and  $y=0, z=a$.
Using equation (\ref{eq:rotbm}) and neglecting the term ${\rm curl}{\bf B}$, the remaining nonlinear
term brings an instantaneous rotation of the electric field. As a matter of fact, let us define in
Cartesian coordinates
 ${\bf E}=(0,v,w)$ and ${\bf B}=(u,0,0)$, where $u$, $v$ and $w$ do not depend on $x$.
When $y<0$, we trivially have $v=0$. In terms of the new unknowns, the equation
(\ref{eq:rotbm}) takes the form:
\begin{equation}\label{eq:compce}
v_t~=~u_z~-~\Xi w~~~~~~~~~~~~~w_t~=~-u_y~+~\Xi v
\end{equation}
where $\Xi =s(u)c({\rm div}{\bf E})/\sqrt{v^2+w^2}=s(u)c(v_y+w_z)/\sqrt{v^2+w^2}$
and $s(u)$ is the sign of  $u$.
\par\smallskip

We have $u_z=0$. Neglecting $u_y$ (which is bounded), the system (\ref{eq:compce}) is
equivalent to a rotation with angular speed $\Xi$ (note that, for $z=0$ and $z=a$,
 $\Xi$  is infinite
because so it is $w_z$). The rotation is anti-clockwise at the points $(x,0,0)$. It is clockwise at the
points $(x,0,a)$. This is true whatever the sign of the electric field (note that ${\bf E}$ and ${\bf
B}$ change sign together and  ${\bf J}$ always maintains the same orientation). Even if the rotation is
at infinite speed, the rotation angle is finite and depends on the magnitude of the fields. In practice,
by equation (\ref{eq:rotbm}), a sudden change of the quantity ${\rm div}{\bf E}$ is balanced by a
modification of ${\partial\over\partial t}{\bf E}$, forcing the field to vary in the direction  $\pm
{\bf J}$.
\par\smallskip

\begin{figure}[t]
\begin{picture}(400,220)
\put(30,0){\vector(0,1){200}}
\put(160,10){\line(0,1){70}}
\put(260,10){\line(0,1){70}}
\put(160,120){\line(0,1){70}}
\put(260,120){\line(0,1){70}}
\put(20,200){\makebox(0,0){$z$}}
\put(30,80){\circle*{4}}
\put(30,120){\circle*{4}}
\put(25,80){\makebox(0,0){$0$}}
\put(25,120){\makebox(0,0){$a$}}
\put(30,80){\vector(1,0){40}}
\put(65,73){\makebox(0,0){$y$}}
\put(120,60){\vector(1,0){20}}
\put(120,80){\vector(1,0){20}}
\put(120,100){\vector(1,0){20}}
\put(120,120){\vector(1,0){20}}
\put(120,140){\vector(1,0){20}}
\put(120,60){\vector(0,-1){10}}
\put(120,80){\vector(0,-1){10}}
\put(120,100){\vector(0,-1){10}}
\put(120,120){\vector(0,-1){10}}
\put(120,140){\vector(0,-1){10}}
\put(126,147){\makebox(0,0){$\bf J$}}
\put(114,135){\makebox(0,0){$\bf E$}}
\put(260,60){\vector(-1,0){20}}
\put(260,60){\vector(0,1){10}}
\put(260,100){\vector(1,0){20}}
\put(260,140){\vector(0,1){10}}
\put(260,140){\vector(-1,0){20}}
\put(260,80){\vector(-1,-1){10}}
\put(260,100){\vector(0,-1){10}}
\put(260,120){\vector(1,-1){10}}
\put(260,80){\vector(1,-1){14}}
\put(260,120){\vector(1,1){14}}
\put(260,60){\circle*{4}}
\put(260,140){\circle*{4}}
\put(250,55){\makebox(0,0){$\bf J$}}
\put(250,146){\makebox(0,0){$\bf J$}}
\put(270,95){\makebox(0,0){$\bf J$}}
\put(273,79){\makebox(0,0){$\bf J$}}
\put(274,125){\makebox(0,0){$\bf J$}}

\end{picture}
\vskip.5truecm

\caption{\small\sl Plane wave encountering a reflecting wall with
an aperture. Qualitative behavior of the fields ${\bf E}$ and ${\bf J}$
just before the impact and during  the impact. The vector  ${\bf B}$ is
orthogonal to the page.}
\end{figure}
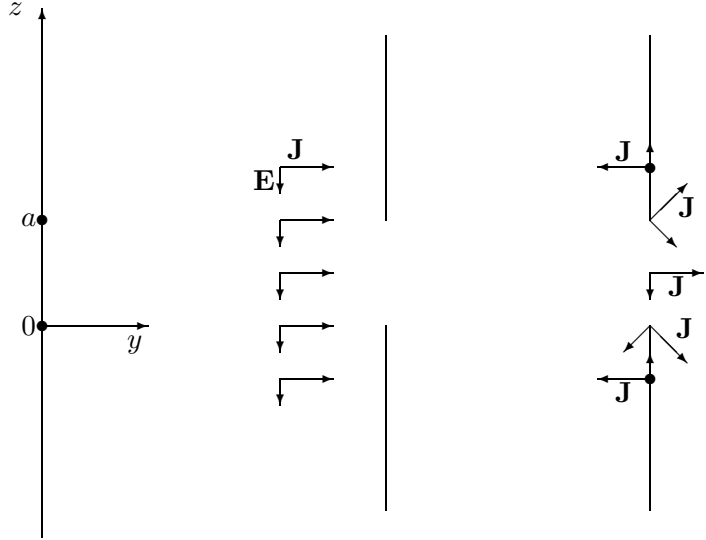

This behavior is not at all in agreement with what is usually observed. The diffraction is a diffusive
phenomenon. Therefore, referring to figure 6, we should expect a clockwise rotation at $(x,0,0)$ and an
anti-clockwise rotation at $(x,0,a)$. We can correct this inconsistency by changing the sign to the
vector product $\times$, which amounts, in other terms, to inverting the sign of the electric field. The
conclusion we can draw is quite astonishing. According to our equations, the standard right-handed
product $\times$ is not suitable for describing a natural event like diffraction. Answering the question
raised in section 5, the correct side of the mirror is the one where $\times$ is left-handed. This means
that, if we do not wish to modify the model equations, we need to switch the polarity of the electric
fields, in such a way an electron turns out to be positive and a proton negative. We will return to this
subject in section 15. The asymmetry of our universe and, consequently, the determination of its parity,
is a problem that emerged about 50 years ago. The effects of this dichotomy were predicted by Lee and
Yang (see for instance \cite{wehr}, p.534), but the reasons for preferring left or right have still to
be found. If the above arguments are free from errors, then here may lie the solution to the problem.
\par\smallskip

In a more realistic situation, the change of direction of the Poynting vector field is not
instantaneous, but develops smoothly. It is very important to point out that the diffraction comes as a
consequence of the nonlinearity of the model equation  (\ref{eq:rotbm}), and not from the imposition of
some artificial boundary conditions at the points $(x,0,0)$ and $(x,0,a)$, as supposed by other theories
(see for instance \cite{born}, chapter XI). The successive evolution of the wave after the obstacle
follows the Huygens principle. In fact, for $y>0$, the wave is free. It displays a slight diffusion due
to the rearrangement  of the electromagnetic fields, described before. It is well-known that the
behavior depends on the ratio between the width $a$ and the frequency $\omega /2\pi$. We do not
investigate this aspect, since it has been intensively studied in the past. Here, we were only concerned
with detecting the mechanism that leads to the deflection of the rays, when they meet the border of the
aperture. The same mechanism takes place in the scattering of two solitons (see section 5), when their
electromagnetic fields collide, with reciprocal influence.
\par\smallskip

Some quantitative information can be recovered by examining equation (\ref{eq:energ}). We know that the
Poynting vector  ${\bf P}$ presents a natural pulsating variation along the direction of propagation of
the wave. But, during the encounter with the obstacle, we have to add another variation, due to the
instantaneous transversal change to the flux of energy.  As we can see from figure 6, during the impact,
the vector field ${\bf P}$ shows some dispersion and its divergence suddenly grows. Thus, for just a
moment, the quantity  $-c^2{\rm div}{\bf P}$ registers a negative peek. To restore the normal energy
flux in (\ref{eq:energ}), some corrective term, taking care of the ``reaction'' of the obstacle, should
be added. In fact, as in the previous section, the change of curvature of the rays is accompanied by the
creation of a new vector field ${\bf G}$, concentrated on the obstacle. We better formalize this idea in
section 9.
\par\smallskip

Anyhow, in spite of the good achievements, we still have some problems to fix. Suppose that the incident
wave is polarized in a different way, for instance by applying a rotation of 90 degrees:
\begin{equation}\label{eq:polariz2}
{\bf E}~=~\big(c\sin\omega(t-y/c),~0,~0\big)
~~~~~~{\bf B}~=~\big(0,~0,~-\sin\omega(t-y/c)\big)
\end{equation}
On the contact with the straight-lines $y=0, z=0$ and $y=0, z=a$, we should now observe a prompt change
of the magnetic field  ${\bf B}$, along the direction $z$. This event is not modelled by our equations,
since we need to suppose that ${\rm div}{\bf B}$ can be different from zero. In order to proceed, it is
necessary to correct the model in such a way that the fields ${\bf E}$ and $c{\bf B}$ have the same
role, as in the classical Maxwell equations. We also discuss this in the coming section.

\par\bigskip

\setcounter{equation}{0}

\section{Adding the mechanical terms}

Based on some problems emerged in the previous sections, we make a first adaptation of the set of
equations (\ref{eq:rotbm})-(\ref{eq:rotem})-(\ref{eq:divbm}), with the aim of bringing to the same level
the two fields  ${\bf E}$ and $c{\bf B}$. Thus, the new formulation is:
\begin{equation}\label{eq:sfep}
{\partial {\bf E}\over \partial t}~=~ c^2 {\rm curl} {\bf B}~
-~c ~{\rm div}{\bf E}~{{\bf E}\times {\bf B} \over \vert {\bf E}
\times {\bf B}\vert}
\end{equation}

\begin{equation}\label{eq:sfbp}
{\partial {\bf B}\over \partial t}~=~ -{\rm curl} {\bf E}~
-~c ~ {\rm div}{\bf B}~{{\bf E}\times {\bf B} \over \vert {\bf E}
\times {\bf B}\vert}
\end{equation}
Now,  (\ref{eq:sfep})-(\ref{eq:sfbp}) do not change if we replace  ${\bf E}$ by $c{\bf B}$ and $c{\bf
B}$ by $-{\bf E}$, as in the Maxwell equations. This makes the model invariant under change of
polarization. Clearly, if the divergence of  ${\bf B}$  vanishes, we come back to equations
(\ref{eq:rotem})-(\ref{eq:divbm}). The possibility for
 ${\rm div}{\bf B}$  to be different from zero, does not imply
the existence of magnetic monopoles, as  the condition
 ${\rm div}{\bf E}\not =0$ does not imply the existence of electrical charges.
However, the issue is delicate, and will be further discussed at the end of
section 14 and in section 15.
\par\smallskip

The modified version (\ref{eq:sfep})-(\ref{eq:sfbp}) allows for an even greater
space of solutions. The spherical waves analyzed in section
4 can be now constructed with the electric field following the parallels,
and the magnetic field following the meridians. In this case, we
have ${\rm div}{\bf E}=0$. Different other intermediate polarizations,
which may also vary in time, can be considered as well.
\par\smallskip

Finally, we can also get solutions as the one shown in figure 2.
It is sufficient to set:
$${\bf E}=\big(cf(x,y)g(t-z/c),~0,~0\big)~~~~~~~$$
\begin{equation}\label{eq:cf2}
{\bf B}=\big(0,~f(x,y)g(t-z/c),~0\big)~~~~~~~~{\bf J}=\big(0,~0,~1\big)
\end{equation}
where $f$ is an arbitrary function, decaying to zero near the boundary of a bidimensional domain
$\Omega$. The function $g$ can also be arbitrary. If $g$ vanishes outside a finite interval, then the
solution in (\ref{eq:cf2}) is a travelling soliton, as the ones considered in section 5. The  difference
with respect to section 5 is that those solutions satisfied the condition ${\rm div}{\bf B}=0$. Hence,
in cylindrical coordinates, the solutions  given in (\ref{eq:solscil}) satisfy
(\ref{eq:sfep})-(\ref{eq:sfbp}),
 without the need to enforce  relation (\ref{eq:divbcil}).
\par\smallskip

However, we are not completely satisfied yet. There is too much symmetry now, while we know that, in
most natural phenomena, the role of  fields  ${\bf E}$ and $c{\bf B}$ is very well differentiated.
Actually, the difference is detectable when a wave interacts with matter. We can take for example the
case of the  wire-grid polarizers, where an incident wave hits a grate of parallel metallic wires. If
the wave is polarized with the electric field orthogonal to the wires, then it
 passes the obstacle almost undisturbed (if its wave-length is much
smaller than the distance between two wires of the grate). If the
electric field has a component along the direction of the wire, then
the wave changes the polarization  of a certain angle.
\par\smallskip

Insisting on a similar example, we can go back to the end of section 8. We are now able to study the
diffraction of the wave given in (\ref{eq:polariz2}), where the electric field is parallel to the
$x$-axis. Using the equations (\ref{eq:sfep})-(\ref{eq:sfbp}), we come to the same conclusions obtained
for the wave given in (\ref{eq:relaz}), polarized in another way. But this result is incorrect, because
in the case of the wave (\ref{eq:polariz2}), together with the diffraction of the rays, there should be
a change of the  polarization after passing the obstacle, which is not modelled by the equations
(\ref{eq:sfep})-(\ref{eq:sfbp}), and which is not present in the case of the wave polarized as in
(\ref{eq:relaz}).
\par\smallskip

In addition to the above observations, we also note in the reflection-refraction phenomenon, that the
way the incident wave is polarized affects the final result. Thus, it is necessary to further improve
the model. For this purpose it will be useful the material collected in sections 6, 7 and 8.
\par\smallskip

We need to introduce  new vector fields (not of electromagnetic type), which are activated each time a
free wave becomes a constrained wave. Let us begin by defining  a velocity vector field ${\bf V}$. We
will ask  all the vectors to be of constant norm, in particular:
 $\vert{\bf V}\vert =c$, where $c$ is the speed of light. Therefore,
what really matters is the  orientation of the vectors. The idea is that
${\bf V}$ is the tangent vector field to a bundle of light rays.
An example is given by the vector field $c{\bf J}$, introduced in section 3,
representing the direction of propagation of a wave-front.
\par\smallskip

Afterwords, we propose the following system of time-dependent partial differential
equations, with three unknown vector fields:

\begin{equation}\label{eq:sfe}
{\partial {\bf E}\over \partial t}~=~ c^2 {\rm curl} {\bf B}~
-~({\rm div}{\bf E}) {\bf V}
\end{equation}

\begin{equation}\label{eq:sfb}
{\partial {\bf B}\over \partial t}~=~ -{\rm curl} {\bf E}~
-~({\rm div}{\bf B}){\bf V}
\end{equation}

\begin{equation}\label{eq:sfg}
{\partial {\bf V}\over \partial t}~=~-({\bf V}\cdot \nabla){\bf V}
~+~ \mu\big({\bf E}~+~{\bf V}\times {\bf B}\big)
\end{equation}
\par\smallskip

The constant  $\mu$ is dimensionally equivalent to an electric
charge divided by a mass. Finally, we add the condition previously anticipated:
\begin{equation}\label{eq:normal}
\vert {\bf V}\vert ~=~c
\end{equation}
Concerning the choice of the norm in (\ref{eq:normal}), the discussion is
postponed to  section 15.
\par\smallskip

It is customary, in fluid mechanics, to introduce the material (or substantial) derivative:
\begin{equation}\label{eq:material}
{\bf G}~=~{D{\bf V}\over Dt}~=~{\partial {\bf V}\over \partial t}~+~
({\bf V}\cdot \nabla){\bf V}
\end{equation}
where ${\bf G}$ is an acceleration field.
 Hence, the equation  (\ref{eq:sfg}) is equivalently written as:
\begin{equation}\label{eq:sfgr}
{D{\bf V}\over Dt}~=~\mu\big({\bf E}~+~{\bf V}\times{\bf B}\big)
\end{equation}
Geometrically, the vector ${D\over Dt}{\bf V}$ provides a measure of the curvature of the stream-lines,
which in this case are identified with the rays (recall (\ref{eq:gifi})). As will become clearer, the
knowledge of the vector field ${\bf V}$ is secondary with respect to the determination of its variation
${\bf G}$.
\par\smallskip

Again, we assume to be in vacuum, with no particles of any kind around. In spite of that, the  equation
(\ref{eq:sfgr}) has strong similarity with  the Lorentz law for a density of charge moving at the speed
of light. Actually, all the ingredients are there. Multiplying by a mass, the left-hand side in
(\ref{eq:sfgr}) is a force: its component along the direction  of motion  turns out to be proportional
to  the electric force field, while the orthogonal component is proportional to the  magnetic force
field. As we can see,  the symmetry is broken, so that the electric and the magnetic fields cannot be
interchanged anymore. But, this only happens in the case of constrained waves (${\bf G}\not = 0$). For
free waves, we recall that the relation (\ref{eq:lorentz}), corresponding to ${\bf V} = c{\bf J}$ and
${\bf G} = 0$, is true. Moreover, replacing  ${\bf E}$ by $c{\bf B}$ and  $c{\bf B}$ by $-{\bf E}$, we
obtain the relation (\ref{eq:lore2}), which is also true. Therefore, all the free waves, no matter what
kind of polarization they have, are included in the new model. The interesting part is to study the
behavior of constrained waves. We will discuss some general properties in section 10.
\par\smallskip

Before going ahead, we feel that some clarification is necessary concerning the meaning of the word
``mass'', used, perhaps improperly, several times in the paper. In our discussion, there are no masses
in classical sense, since there are no elementary particles of any sort. Nevertheless, we needed to make
distinction, in terms of dimensionality, between electromagnetic and mechanical (later they will be
called gravitational) fields. This responsability has been given to the constant $\mu$, which provides
the dimensional link between the two ``flavors''. Although other names could have been appropriate to
this purpose, the choice of the term ``mass'' is not incidental, since, as we proceed with our
arguments, it will come out to be consistent with the standard setting.
\par\bigskip

\setcounter{equation}{0}

\section{Properties of the new set of equations}

The new system of equations (\ref{eq:sfe})-(\ref{eq:sfb})-(\ref{eq:sfg})
is able to describe electromagnetic phenomena where the wave-front, locally
evolving in the direction determined by ${\bf V}$, could be subjected to
transversal perturbations modifying the trajectories of the rays.
The propagation of the wave is governed by the first two equations.
Through a feed-back process, the third equation,
 from the current knowledge of the local electromagnetic fields,
allows for the determination of ${D\over Dt}{\bf V}$, setting up the
new direction of motion. This coupling is possible because we have
been able to include the vector ${\bf V}$ in the description of
the electromagnetic part, in the same way the term $c{\bf J}$ was
added to (\ref{eq:rotbm}). Thus, we got a remarkable result: a
link  between electromagnetic and mechanical forces. Using the standard
Maxwell equations such a connection could never be established.

\par\smallskip

Let us continue with our analysis. From known formulas of vector calculus,
we first deduce that:
\begin{equation}\label{eq:variort}
{D{\bf V}\over Dt}\cdot{\bf V}~=~{\partial {\bf V}\over \partial t}
\cdot{\bf V}~+~\Big({1\over 2} \nabla\vert {\bf V}\vert^2~-~{\bf V}
\times {\rm curl}{\bf V}\Big)\cdot {\bf V}~=~0
\end{equation}
where we used that $\nabla \vert {\bf V}\vert^2=0$ and that
${\partial\over\partial t}{\bf V}$ is orthogonal to ${\bf V}$, since ${\bf V}$ has
constant norm.
\par\smallskip

We recall that, by definition, ${\bf E}\cdot{\bf J}=0$.
Similarly, by (\ref{eq:variort}) and by scalar multiplication of  (\ref{eq:sfgr}) by
${\bf V}$, one easily gets:
\begin{equation}\label{eq:ortho}
{\bf E}\cdot{\bf V}~=~0
\end{equation}
Although one has ${\bf B}\cdot{\bf J}=0$, nothing can be deduced however for the scalar product ${\bf
B}\cdot{\bf V}$.
\par\smallskip

\noindent By vector multiplication of (\ref{eq:sfgr}) by ${\bf V}$, we get:
\begin{equation}\label{eq:torsion}
{\bf V}\times {D{\bf V}\over Dt}~=~\mu\Big({\bf V}\times {\bf E}~-~c^2 {\bf B}
~+~({\bf V}\cdot {\bf B}){\bf V}\Big)
\end{equation}
that generalizes  (\ref{eq:lore2}). Finally,
by scalar multiplication of  (\ref{eq:sfe}) by ${\bf E}$ and (\ref{eq:sfb})
by ${\bf B}$, one obtains:
\begin{equation}\label{eq:energ2}
{1\over 2}{\partial \over \partial t} (\vert {\bf E}\vert^2
~+~c^2\vert {\bf B}\vert^2)~=~- c^2~{\rm div}({\bf E}\times {\bf B})~-~
c^2({\bf V}\cdot {\bf B})~{\rm div}{\bf B}
\end{equation}
which is the counterpart of (\ref{eq:energ}).
\par\smallskip

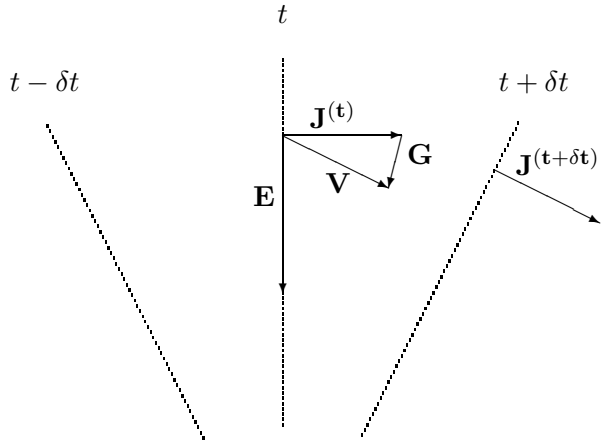
\begin{figure}[t]
\begin{picture}(400,240)
\put(160,160){\vector(0,-1){60}}
\put(160,160){\vector(1,0){45}}
\put(160,160){\vector(2,-1){40}}
\put(205,160){\vector(-1,-4){5}}
\multiput(160,50)(0,2){70}{\line(0,1){1}}
\multiput(190,46)(1,2){60}{\line(0,1){1}}
\multiput(130,46)(-1,2){60}{\line(0,-1){1}}
\put(153,137){\makebox(0,0){${\bf E}$}}
\put(179,168){\makebox(0,0){${\bf J^{(t)}}$}}
\put(181,142){\makebox(0,0){${\bf V}$}}
\put(212,153){\makebox(0,0){${\bf G}$}}
\put(240,147){\vector(2,-1){40}}
\put(264,150){\makebox(0,0){${\bf J^{(t+\delta t)}}$}}
\put(160,206){\makebox(0,0){$t$}}
\put(255,180){\makebox(0,0){$t+\delta t$}}
\put(70,180){\makebox(0,0){$t-\delta t$}}
\end{picture}
\vskip-.2truecm

\caption{\small\sl Typical behavior of different fields, when a wave-front
is forced to turn.}
\end{figure}

Referring to figure 7, let ${\bf J}^{(t)}$ be the normalized Poynting vector
at  time $t$ and ${\bf J}^{(t+\delta t)}$ the one at time $t+\delta t$.
Let ${\bf V}$ be the vector at time $t$, obtained by backward parallel transport along
the stream-lines of the vector ${\bf J}^{(t+\delta t)}$.
Then, we have:
\begin{equation}\label{eq:limite}
{D{\bf V}\over Dt}~=~\lim_{\delta t\rightarrow 0}
{{\bf V} - c{\bf J}^{(t)}\over\delta t}
\end{equation}
We recall that the same was done in section 7 in order to define the vector
${\bf G}$ (see (\ref{eq:gifi})).
Therefore, for small time variations $\delta t$, we are allowed to write:
\begin{equation}\label{eq:sviluppo}
{\bf V}~\approx~ c{\bf J}~+~{\bf G}\delta t~=~c{\bf J}~+~
\mu ({\bf E}~+~c{\bf J}\times {\bf B})\delta t
\end{equation}
with  ${\bf J}=({\bf E}\times {\bf B})/\vert {\bf E}\times {\bf B}\vert$.
If ${\bf E}\cdot {\bf B}=0$, the relation  (\ref{eq:sviluppo}) can be
rewritten as:
\begin{equation}\label{eq:svilupp2}
{\bf V}~\approx~ c{\bf J}~+~\mu ~{{\bf E}\over \vert {\bf E}\vert}~
\Big (\vert{\bf E}\vert~-~c\vert {\bf B}\vert \Big)\delta t
\end{equation}
after noting that: $({\bf E}\times {\bf B})\times {\bf B}=
({\bf E}\cdot {\bf B}){\bf B}- \vert{\bf B}\vert^2 {\bf E}=
-\vert {\bf B}\vert^2 {\bf E}$. This shows that it is sufficient to have
$\vert{\bf E}\vert \not = c\vert {\bf B}\vert$, in order to activate the
transversal field ${\bf G}$.
\par\smallskip

We can compare the evolution of an electromagnetic phenomenon to that of an inviscid fluid, whose mass
density,  up to dimensional constants, is given by  $\rho={\rm div}{\bf E}$. Note, however, that a real
``mass'' does not exist. Note also that $\rho$ can also attain negative values. We do not define the
density $\rho={\rm div}{\bf B}$ for reasons that will be detailed at the end of section 14. The
following continuity equation holds (see also (\ref{eq:continu})):
\begin{equation}\label{eq:conti2}
{\partial \rho\over\partial t}~=~-{\rm div}(\rho{\bf V})
\end{equation}
obtainable by taking the divergence of (\ref{eq:sfe}). The equation
(\ref{eq:conti2}) can be also written as:
\begin{equation}\label{eq:conti3}
{D\rho\over Dt}~=~-\rho~ {\rm div}{\bf V}
\end{equation}
For a plane wave (or soliton) having $\rho\not =0$, we obtain ${\rm div}{\bf V}=0$ as well as ${\bf
G}=0$. Then,  (\ref{eq:conti3}) tells us that the fluid shifts, without modifications, along the
direction determined by ${\bf V}$. The fluid travels at constant speed $c$, showing rarefactions and
compressions,. More properly, it evolves like an incompressible fluid, but with density not equally
distributed in space. Regarding a spherical wave, one has ${\rm div}{\bf V}>0$ and ${\bf G}=0$. As
expected, this implies that the mass density diminishes (in absolute value), while time passes, since it
spreads on spheres of growing area. In both examples (the plane and the spherical) we have
 ${\rm curl}{\bf V}=0$. In other words, the fluid is irrotational.
\par\smallskip

Let us suppose that ${\bf V}$ is a gradient, i.e.: ${\bf V}=\nabla \Psi$,
where $\Psi$ is a scalar potential not depending on time. Then, the corresponding fluid is
irrotational. Thanks to (\ref{eq:normal}), we trivially have:
\begin{equation}\label{eq:eiko2}
\vert\nabla\Psi\vert~=~c
\end{equation}
which is the eikonal equation.
Then, we observe that (\ref{eq:eiko2}) and the relation ${\partial \over\partial t}
\Psi =0$, imply:
\begin{equation}\label{eq:equa2}
{D{\bf V}\over Dt}~=~\nabla \Big({\partial\Psi\over\partial t}~+~
{1\over 2}\vert\nabla \Psi\vert^2\Big)~=~0
\end{equation}
This confirms a remarkable result: the eikonal equation (hence, the evolution of the wave-fronts based
on the Huygens principle) is perfectly compatible with the condition  ${\bf G}=0$. This analytic
property, obtained without approximations, goes beyond the famous limits of  geometrical optics. So,
here, with a very elementary proof, we obtained another important result.
\par\smallskip

In the equation (\ref{eq:sfgr}),  the term  ${\bf V}\times {\bf B}= -{\bf B}\times {\bf V}={\cal T}({\bf
V})$ can be viewed as a suitable stress tensor ${\cal T}$ applied to the vector normal to the front of
propagation (see for instance \cite{batchelor}, p.10). As we already know, forced variations of the
electric field produce changes in the motion of the fronts. If these are combined with forced variations
of the magnetic field, a  torsion is also introduced, which modifies the polarization of the wave. We
guess, that, when the external perturbations end, the electromagnetic fields return to their natural
equilibrium in which $\vert {\bf E}\vert =c\vert {\bf B}\vert$ and (\ref{eq:lorentz}) is satisfied, so
that the fluid takes again an irrotational motion. From the examples discussed in sections 7 and 8, this
behavior corresponds to what is commonly observed in nature, and certainly comes from the minimization
of some Lagrangian. At the moment, we do not however have theoretical explanation for this conjecture.
\par\smallskip

Furthermore, we note that stationary electric fields, for example with ${\bf B}=0$, are not longer
solutions. We can check this by examining relations (\ref{eq:sfgr}) and (\ref{eq:ortho}). They force the
velocity field ${\bf V}$ to turn itself around (${\bf V}$ deviates in the direction of ${\bf E}$, but
${\bf E}$ remains orthogonal to it). More generally, equation (\ref{eq:sfg}) requires the solutions to
be in continuous evolution. We contend that the new system of equations does not admit stationary
solutions having finite energy. We made the same consideration in section 5, in the particular case of
solitons. Nevertheless, there could be nonstationary solutions localized in space. We can imagine for
instance the case of two (or more) solitons, in such a way that they are constrained, by influencing
their electromagnetic fields each other, to revolve around a common center. We still do not have all the
elements to study these phenomena, which, as we will see in the following pages, need the environment of
general relativity to be stated properly. Some remarks about the case of two rotating solitons will be
given in section 15.
\par\smallskip

We are unable at the moment to obtain the equations (\ref{eq:sfe})-(\ref{eq:sfb})-(\ref{eq:sfg}) from
the minimization of a suitable action function as we did in section 6 (concerning (\ref{eq:sfg}) alone,
something in this direction will be obtained in the next section). One may consider the usual Lagrangian
$L=2(c^2\vert {\bf B}\vert^2 - \vert {\bf E}\vert^2)$ and  the generalization of the relation
(\ref{eq:vvinc}), i.e.:
\begin{equation}\label{eq:nvinc}
{\bf A}~=~{1\over c}~\Phi {\bf V}
\end{equation}
Then, we could minimize the same action function given in
(\ref{eq:azione}) using the constraint (\ref{eq:nvinc}). Nevertheless, we
would not obtain the desired result, since (\ref{eq:nvinc})
is too restrictive.  In this way, we only get a set of equations describing
free waves. As a matter of fact, we can prove that, if ${\bf V}$ has the same direction
as ${\bf A}$, then one automatically has  ${\bf G}=0$. Assuming ${\rm div}{\bf B}=0$, this
check can be done as follows. Considering  (\ref{eq:potenz})
and (\ref{eq:nvinc}), we have:
$${D\over Dt}\left( {c\over \Phi}{\bf A}~+~{\mu\over c}{\bf A}\right)~=~
{D{\bf V}\over Dt}~+~{\mu\over c}{\partial{\bf A}\over \partial t}~+~
{\mu\over c}({\bf V}\cdot \nabla ){\bf A}$$
$$=~{D{\bf V}\over Dt}~-~\mu {\bf E}~-~\mu \nabla \Phi~+~{\mu\over\Phi}
({\bf A}\cdot\nabla ){\bf A}$$
$$=~{D{\bf V}\over Dt}~-~\mu {\bf E}~+~{\mu\over\Phi}
\left( -{1\over 2}\nabla \Phi^2 ~+~({\bf A}\cdot\nabla ){\bf A}\right)$$
$$=~{D{\bf V}\over Dt}~-~\mu {\bf E}~+~{\mu\over\Phi}
\left( -{1\over 2}\nabla \vert {\bf A}\vert^2~+~({\bf A}\cdot\nabla ){\bf A}\right)$$
\begin{equation}\label{eq:dera}
=~{D{\bf V}\over Dt}~-~\mu \left({\bf E}~+~{1\over\Phi}{\bf A}\times {\rm curl}
{\bf A}\right)~ =~{D{\bf V}\over Dt}~-~\mu ({\bf E}~+~{\bf V}\times {\bf B})~=~0
\end{equation}
where we used that $\Phi^2 =\vert {\bf A}\vert^2$. The last equality is true thanks
to (\ref{eq:sfgr}). Then, noting that $(c/\Phi +\mu /c) {\bf A} =(1+\mu\Phi /c^2){\bf V}$,
the relation (\ref{eq:dera}) leads to:
\begin{equation}\label{eq:dera2}
{D\over Dt}\left[ \left( 1+{\mu\Phi\over c^2}\right){\bf V}\right]~=~
{\mu\over c^2}{D \Phi \over Dt} {\bf V}~+~
 \left (1+{\mu\Phi\over c^2}\right){D{\bf V}\over Dt}~=~0
\end{equation}
By scalar multiplication by ${\bf V}$, due to (\ref{eq:variort}), we recover:
\begin{equation}\label{eq:dera3}
{\mu\over c^2}{D \Phi \over Dt} \vert {\bf V}\vert^2~+~
 \left (1+{\mu\Phi\over c^2}\right){D{\bf V}\over Dt}\cdot {\bf V}~=~
\mu {D\Phi \over Dt}~=~0
\end{equation}
Thus, $\Phi$ turns out to be constant along the stream-lines. For this reason, from  (\ref{eq:dera}),
also  ${\bf A}$ is constant along the stream-lines. Therefore, one has  ${D\over Dt}{\bf V}={\bf G}=0$.
We also conclude that, when the rays bend (${\bf G}\not =0$), then the vector  ${\bf A}$  cannot be
aligned in the direction of motion.
\par\smallskip

We mentioned in the previous sections that the mechanical effects are
implicitly included in the term $c^2{\rm div}{\bf P}$, where it is necessary
to distinguish between the contribution due to the variation of the Poynting
vector along the actual direction of propagation of the front, and the transversal
contribution (which is zero when ${\bf G}=0$). Differentiating with respect
to time the expression ${\bf J}={\bf P}/\vert{\bf P}\vert$, we get:
\begin{equation}\label{eq:gp}
{\bf G}~=~{1\over \vert{\bf P}\vert}\left({\partial {\bf P}\over
\partial t}~-~{{\bf P}\cdot {\partial\over\partial t}{\bf P}\over
\vert{\bf P}\vert^2}{\bf P}\right)
\end{equation}
In particular, by scalar multiplication of ${\bf G}$ by ${\bf P}$, (\ref{eq:gp}) shows the orthogonality
relation  ${\bf G}\cdot {\bf P}=0$. Furthermore, from (\ref{eq:energ}), the energy can be described as a
work by integrating $-2c^2{\rm div}{\bf P}$ with respect to time. This yields:
\begin{equation}\label{eq:lavoro}
-2c^2\int {\rm div}{\bf P}~dt~=~-2\int_\Gamma {\rm div}{\bf P}~~{\bf V}
\cdot d{\bf s}
\end{equation}
where we set $d{\bf s}={\bf V}dt$, which implies ${\bf V}\cdot d{\bf s}=
\vert {\bf V}\vert^2 dt= c^2 dt$. The last integration is made along the curve
 $\Gamma$ representing the path of the light ray.
\par\smallskip

We end this section by illustrating another interesting relation. Let us define as usual $\rho = {\rm
div}{\bf E}$. Afterwords, let us assume that $\rho \not = 0$ and define  $\bar\omega = {\bf F}/\rho$,
where ${\bf F}={\rm curl}{\bf V}+\mu {\bf B}$. Then, along the stream-lines we have:
\begin{equation}\label{eq:vortici}
{D\bar\omega\over Dt}~=~(\bar\omega\cdot \nabla ){\bf V}
\end{equation}
Note that  $\bar\omega$ is dimensionally equivalent to a time multiplied by a charge and divided by a
mass. The equation (\ref{eq:vortici}) recalls the analogous one for isentropic flows, which is
introduced in fluid dynamics by defining $\bar\omega$ as the curl of the velocity field divided by the
mass density (see \cite{chorin}, p.24). Using (\ref{eq:potenz}), the field  $\bar\omega$ also takes the
following form:
\begin{equation}\label{eq:omegabar}
\bar\omega~=~{{\rm curl}({\bf V}+\mu {\bf A}/c)\over
\displaystyle{-{1\over c}{\partial\over \partial t}{\rm div}{\bf A}
-\Delta\Phi}}
\end{equation}

\noindent The equation (\ref{eq:vortici}) can be proven as follows:
$$\rho\left[{D\bar\omega\over Dt}~-~(\bar\omega\cdot \nabla ){\bf V}\right]~=~
\rho\left[{1\over \rho}{D{\bf F}\over Dt}~-~{1\over \rho^2}{D\rho \over Dt}{\bf F}\right]~-~
({\bf F}\cdot \nabla ){\bf V}$$
$$=~{\rm curl}\left({\partial {\bf V}\over \partial t}\right)~+~\mu{\partial {\bf B}
\over \partial t}~+~({\bf V}\cdot \nabla ){\bf F} ~+~ {\bf F}~{\rm div}
{\bf V}~-~ ({\bf F}\cdot \nabla ){\bf V}$$
$$=~{\rm curl}\Big(-({\bf V}\cdot \nabla ){\bf V}~+~\mu ({\bf E}~+~
{\bf V}\times {\bf B})\Big)
~+~\mu (-~{\rm curl}{\bf E}~-~{\bf V}~{\rm div}{\bf B})$$
$$~~~~~~~~~~~~~~+~({\bf V}\cdot \nabla ){\bf F} ~+~
{\bf F}~{\rm div}{\bf V}~-~ ({\bf F}\cdot \nabla ){\bf V}$$
$$=~\Big[{\rm curl}({\bf V}\times {\bf F})~-~{\bf V}~{\rm div}{\bf F}
+~({\bf V}\cdot \nabla ){\bf F} ~+~
{\bf F}~{\rm div}{\bf V}~-~ ({\bf F}\cdot \nabla ){\bf V}\Big]$$
$$~-~\Big[{\rm curl}[({\bf V}\cdot \nabla ){\bf V}]~+~{\rm curl}
({\bf V}\times {\rm curl}{\bf V})\Big]$$
\begin{equation}\label{eq:civo}
~=~-{\rm curl}\Big(({\bf V}\cdot \nabla ){\bf V}~+~
({\bf V}\times {\rm curl}{\bf V})\Big)~=~-{1\over 2}~{\rm curl}
\big(\nabla\vert {\bf V}\vert^2\big)~=~0
\end{equation}
where, in the order,  we used (\ref{eq:conti3}),  (\ref{eq:sfg}), (\ref{eq:sfb}),
some well-known calculus properties and the fact that $\nabla\vert {\bf V}\vert^2=0$.
\par\smallskip

In the case of plane solitary waves, we have ${\rm curl}{\bf V}=0$, hence $\bar\omega =\mu{\bf B}/\rho$
(when $\rho \not =0$). Therefore, $\bar\omega$ remains orthogonal to ${\bf V}$, so that the relation
(\ref{eq:vortici}) becomes ${D\over Dt}\bar\omega =0$. Then, the quantity $\bar\omega $  shifts,
remaining constant along the stream-lines determined by the velocity field ${\bf V}$ (which are
straight-lines in this case).

\par\bigskip

\setcounter{equation}{0}

\section{Towards general relativity}
\smallskip

Our first step, in this section, is to recover the equation (\ref{eq:sfgr}) through the minimization of
a suitable Lagrangian. To this end we  work in space-time using 4-vectors. Let us start by defining
$(x_0, x_1, x_2, x_3) =(ct, -x, -y, -z)$ and $(e_0, e_1, e_2, e_3)=(1, -1, -1, -1)$. Then, for the
vector $(V_0, V_1, V_2, V_3)=(V_0, {\bf V})$, one has:
\begin{equation}\label{eq:scalar}
\sum_{i=0}^3 e_i V_i^2~=~V_0^2~-~\vert {\bf V}\vert^2
\end{equation}
As in section 6, we assume that ${\rm div}{\bf B}=0$ and introduce the potentials
$\Phi$ and ${\bf A}$ by (\ref{eq:potenz}). Let also be $(A_0, A_1, A_2, A_3)=(\Phi, {\bf
A})$. Up to multiplicative constants, we can define a Lagrangian in the following way
(see also \cite{landau}, p.50):
\begin{equation}\label{eq:lagrv}
L~=~c~\sqrt{V_0^2-\vert {\bf V}\vert^2}~+~\mu \left(\Phi ~-~{1\over
c}~{\bf A}\cdot {\bf V}\right)
\end{equation}
The quantities $V_i, ~i=0,1,2,3$, are the independent variables, while the potentials
depend on $x_i, ~i=0,1,2,3$. By setting $V_0=c$, the term in parenthesis of (\ref{eq:lagrv})
can be written as: $c^{-1}\sum_{i=0}^3 e_iA_iV_i$. For the moment, we do not impose
the condition (\ref{eq:normal}), implying that the sum in (\ref{eq:scalar}) is zero.
\par\smallskip

Suppose that we are moving along a stream-line (or curved ray), between two
instants  of time $t_1$ and $t_2$, then the action function takes the form:
\begin{equation}\label{eq:aziov}
S~=~-\int_{t_1}^{t_2} L~dt
\end{equation}
Its minimization brings to the Euler-Lagrange equation (see \cite{jackson}, p.577):
\begin{equation}\label{eq:eula}
{d\over dt}\left({\partial L\over \partial V_0}\right)={\partial L
\over \partial t}=c~{\partial L\over \partial x_0}~~~~~~~~~~
{d\over dt}\left({\partial L\over \partial V_i}\right)=-{\partial L
\over \partial x_i}~~~~~ i=1,2,3
\end{equation}
where we observed that $({1\over c}{\partial\over\partial t}, -\nabla )=
({\partial\over \partial x_0}, {\partial\over \partial x_1},
{\partial\over \partial x_2}, {\partial\over \partial x_3})$.
In particular, for $i=1,2,3$, we have:
$$
{d\over dt}\left({\partial L\over \partial V_i}\right)~=~{d\over dt}
\left( {-c V_i\over \sqrt{V_0^2 -\vert {\bf V}\vert^2}}~-~{\mu\over c}~A_i
\right)
$$
\begin{equation}\label{eq:eula2}
=~-{DV_i\over Dt} ~-~{\mu\over c}{\partial A_i\over\partial t}~-~{\mu\over c}
\sum_{k=1}^3{\partial A_i\over \partial x_k}~{d x_k\over d t}
\end{equation}
\begin{equation}\label{eq:eula3}
{\rm and}~~~~~~~~{\partial L\over \partial x_i}~=~\mu {\partial \Phi\over\partial x_i}~-~{\mu
\over c}~{\bf V}\cdot {\partial {\bf A}\over \partial x_i}
\end{equation}
where, in (\ref{eq:eula2}), the substantial derivative ${D\over Dt}V_i$ gives the variation,
along the stream-lines, of the coordinates of the velocity  field, parametrized with respect
to the arc-length: $s= \vert c\vert ^{-1}\int_{t_1}^t\sqrt{V_0^2 -\vert {\bf V}\vert^2} d\xi$.
\par\smallskip

If we now define  ${d\over dt}x_k=V_k$, for $k=1,2,3$, thanks to (\ref{eq:tensore}) and
(\ref{eq:tens2}), we get:
$$
{DV_i\over Dt}~=~\mu\left({\partial \Phi\over\partial x_i}~-~{1\over c}{\partial A_i
\over\partial t}\right)
$$
\begin{equation}\label{eq:eula4}
-~{\mu\over c}~{\bf V}\cdot\left(\nabla A_i~+~{\partial {\bf A}\over\partial x_i}
\right)~=~-{\mu\over c}~F^{ik}V_k
\end{equation}
When $V_0=c$, the last term in (\ref{eq:eula4}) is equal to the $i$-th component of the vector
$\mu({\bf E}+{\bf V}\times {\bf B})$. This implies the equation (\ref{eq:sfgr}).
\par\smallskip

 Concerning $k=0$, if we fix $V_0$ to be constantly equal to $c$, one obtains
 ${D\over Dt}V_0 =0$. Therefore, we have:
\begin{equation}\label{eq:eula5}
{d\over dt}\left({\partial L\over\partial V_0}\right)~=~{DV_0\over Dt}~+~
~\mu ~{d \Phi\over dt}~=~\mu
~{\partial\Phi\over \partial t}~+~\mu ~{\bf V}\cdot \nabla\Phi
\end{equation}
\begin{equation}\label{eq:eula6}
{\rm and}~~~~~~~~{\partial L\over\partial t}~=~\mu ~{\partial \Phi\over\partial t}~
-~{\mu\over c}~{\bf V}\cdot {\partial {\bf A}\over\partial t}
\end{equation}
Due to (\ref{eq:eula}), by equating these two last expressions, we recover:
\begin{equation}\label{eq:eula7}
0~=~-\mu\left(\nabla\Phi~+~{1\over c}{\partial {\bf A}\over\partial t}\right)\cdot {\bf V}
~=~\mu~F^{0k}V_k~=~\mu~{\bf E}\cdot{\bf V}
\end{equation}
which corresponds to (\ref{eq:ortho}). Considering (\ref{eq:eula7}), by scalar
multiplication of  (\ref{eq:sfgr}) by
${\bf V}$, we deduce that the field minimizing the action satisfies
 ${\bf V}\cdot{D\over Dt}{\bf V}=0$.  Hence, the norm $\vert {\bf V}\vert$ is constant.
If such a constant is $c$, we finally obtain the relation (\ref{eq:normal}), that says that the
solutions  evolve on the light-cone.
\par\smallskip

At this point, it should be noted that, by choosing $\vert {\bf V}\vert^2 =c^2$, the first part of the
Lagrangian in  (\ref{eq:lagrv}) vanishes. This does not mean that it vanishes identically, but only in
correspondence to the minimum. Instead, the second part of the Lagrangian is zero when  ${\bf A}\cdot
{\bf V}=c\Phi$, which is very similar to the condition (\ref{eq:vincsca}), obtained from the constraint
(\ref{eq:vvinc}) (see also (\ref{eq:nvinc})). This coincidence is quite significant. Perhaps, in the
future, it will suggest a way to build a Lagrangian for the entire set of equations
(\ref{eq:sfe})-(\ref{eq:sfb})-(\ref{eq:sfg}).
\par\smallskip

By multiplying the equation (\ref{eq:eula4}) by $V_i$, $i=1,2,3$, and the equation
(\ref{eq:eula7}) by  $V_0$, we get:
\begin{equation}\label{eq:2forma}
F^{ik}V_kV_i~=~0
\end{equation}
where the sum is for $i$ and  $k$ going from 0 to 3. This also trivially follows from the
anti-symmetry of the tensor  $F^{ik}$. The equation (\ref{eq:forminv}) is also written as:
\begin{equation}\label{eq:forminv2}
\left({\partial F^{ik}\over\partial x_k}\right)V_0~-\left({\partial F^{0k}\over\partial x_k}
\right)e_iV_i~=~0 ~~~~~~{\rm for}~i=0,1,2,3
\end{equation}
Otherwise, the equations (\ref{eq:rote}) and (\ref{eq:divb}), can be recovered from the expression (see
for instance \cite{fock}, p.150):
\begin{equation}\label{eq:forminv3}
F_{ikj}~=~{\partial F_{ik}\over\partial x_j}~+~{\partial F_{kj}\over\partial x_i}~+
~{\partial F_{ji}\over\partial x_k}~=~0
\end{equation}
where there is no sum on repeated indices. The rank-three tensor $F_{ikj}$ is
anti-symmetric and called the cyclic derivative of $F_{ik}$. On the other hand,
the equation (\ref{eq:sfb}) follows from the expression:
\begin{equation}\label{eq:forginv3}
V_0\left({\partial F_{ik}\over\partial x_j}+{\partial F_{kj}\over\partial x_i}+
{\partial F_{ji}\over\partial x_k}\right)=
\pm ~e_mV_m\left({\partial F_{23}\over\partial x_1}+{\partial F_{31}\over\partial x_2}+
{\partial F_{12}\over\partial x_3}\right)
\end{equation}
where the indices $m,j,i,k$ (taken in this order) are all different. The sign $\pm$
depends on the permutation (even or odd) of the indices  (the sign is plus if $m=0$, $j=1$,
$i=2$, $k=3$). In (\ref{eq:forginv3}) the term in parenthesis on the right-hand side is
equal to $c{\rm div}{\bf B}$.
In a more contracted form, the last equation reads as follows:
\begin{equation}\label{eq:forginv5}
V_0~F_{jik}~=~\pm e_m V_m~F_{123}
\end{equation}

\par\smallskip

In the results obtained above, we basically considered ${\bf V}$ as the velocity field of an
infinitesimal particle moving at the speed of light. On the other hand, in a wave there are infinite
contiguous trajectories. As a matter of fact, the evolution of a wave is a global phenomenon, that
should be taken as a whole, and not studied independently along each stream-line. For such a more in
depth analysis, we need to  work in the context of general relativity. We are going to show that the
passage of a wave modifies the space-time structure. For a free wave this does not affect the evolution
of the wave itself (see section 13), but for constrained waves the change of the geometry influences their
entire behavior. The analysis will allow us to find the coupling between the fields describing the wave
and  space-time geometry, hence the link between electromagnetic and gravitational  phenomena.
\par\smallskip

We first need  to introduce some classical definitions. Mainly, we adopt the notation
used in \cite{fock}. The space-time geometry is locally determined by a symmetric
bilinear form (the metric tensor), whose coefficients are denoted by $g_{ij}$. Then,
the Christoffel  symbols are defined in the following way:
\begin{equation}\label{eq:chris}
\Gamma^i_{kj}~=~{g^{im}\over 2}\left( {\partial g_{mk}\over\partial x_j}~+~
{\partial g_{mj}\over\partial x_k}~-~{\partial g_{kj}\over\partial x_m}\right)
\end{equation}
where we sum over the index $m$. The coefficients  $g^{ij}$ are in such a way that:
\begin{equation}\label{eq:cambinv}
g_{im}~g^{mj}~=~\delta_{ij}
\end{equation}
The coefficients $g_{ij}$ are adimensional, while the  Christoffel symbols are the inverse of a
distance. If the space is ``flat'' (Euclidean or Minkowski space), all the  Christoffel symbols vanish.
In this case, one has $g^{ik}=g_{ik}=e_i\delta_{ik}$. As usual, we denote by $g$ the determinant (which
is negative) of the tensor $g_{ik}$. A lemma due by Ricci (see \cite{fock}, p.129) claims that the
4-divergence of the metric tensor is zero. In detail, one has:
\begin{equation}\label{eq:divtenm}
\nabla_k g^{ik}~=~{1\over\sqrt{-g}}{\partial (\sqrt{-g}~g^{ik})\over\partial
x_k}~+~\Gamma^i_{jm}g^{jm}~=~0
\end{equation}
where $\nabla_k$ is the covariant differentiation operator.
The same is true for the coefficients $g_{ik}$.  Moreover, the coefficients $g^{ik}$
are said to be harmonic when:
\begin{equation}\label{eq:armon}
{1\over \sqrt{-g}}{\partial (\sqrt{-g} ~g^{ik})\over \partial x_k}~=~0
\end{equation}

\par\smallskip
Next, we define $V^i=g^{im}V_m$. The values $V_m$ are the entries of a velocity vector
expressed in the coordinates system $x_0, x_1, x_2, x_3$.  Let us set $V^0=c$. Then, the condition
 (\ref{eq:normal}) is generalized in the following way:
\begin{equation}\label{eq:normal2}
V^iV_i~=~g^{ik}V_kV_i~=~g_{im}V^iV^m~=~0
\end{equation}

In this more general framework, the equations  (\ref{eq:eula4}) and (\ref{eq:eula7})
are rewritten as:
\begin{equation}\label{eq:geofor}
{DV^i\over Dt }~+~\Gamma^i_{jk}V^jV^k~=~-{\mu\over c}F^{im}V_m~~~~~{\rm for}~
i=0,1,2,3
\end{equation}
For $i=0,1,2,3$, we also define (see \cite{fock}, p.217):
\begin{equation}\label{eq:defg}
G^i~=~{DV^i\over Dt }+\Gamma^i_{jk}V^jV^k~=~V^m\nabla_mV^i
\end{equation}
From  (\ref{eq:geofor}) we easily recover the orthogonality relations:
\begin{equation}\label{eq:ortog}
G_iV^i~=~G^iV_i ~=~0
\end{equation}
where $G_i=g_{im}G^m$. Finally, let us define ${\bf G}=(G_1,G_2,G_3)$, which is dimensionally
equivalent to an acceleration.
\par\smallskip

In general relativity, the gravitational field is somehow identified with the tensor $g_{ij}$. Of
course, the vector ${\bf G}$ may vanish without having that the space is flat. Although ${\bf G}$ does
not fully characterize the properties of space-time, for us it will be the ``real'' gravitational field,
that is the one we can measure in our everyday life. In the following, ${\bf G}$ will be called the
vector gravitational field.
\par\smallskip

The equation (\ref{eq:geofor}) enables us to understand how the trajectory of a ``thin'' solitary wave
can be distorted when immerged in a given gravitational field (having ${\bf  G}\not = 0$). Being the
soliton a free wave, the right-hand side of (\ref{eq:geofor}) vanishes (see (\ref{eq:lorentz})). Thus,
its path follows a suitable geodesic in space-time, the shape of which is determined by the external
gravitational field. This should correspond to some transversal bending in the direction locally
individuated by the vector  ${\bf G}$. With this reasoning, we have to neglect a couple of facts, both
due to the change of direction: the modification of the electromagnetic fields  and the ``gravitational
reaction'' of the soliton (a curving wave produces new gravitational field). As we argued in section 10,
these should be minor effects, since the wave, for some principle of least action, tries to compensate
the electromagnetic fields in order to enforce (\ref{eq:lorentz}). From the point of view of the
soliton, the path followed is straight, even if it actually travels on a curved geodesic. To get more
reliable quantitative results, we must solve a quite complicated system of equations. Comparing the
computed results with the experimental evidence, one could probably evaluate the constant $\mu$. This is
however an exercise that we would prefer to avoid here. Also if some theoretical passages may be
formally similar, the important clue is that there is no need to suppose that a soliton has an
infinitesimal mass to justify that  is attracted by a gravitational field.
\par\smallskip

After recalling that $F^{ik}$ is an anti-symmetric tensor, in general coordinates,
the equation (\ref{eq:forminv2}) becomes:
\begin{equation}\label{eq:forminv4}
{1\over \sqrt{-g}}\left({\partial (\sqrt{-g}~F^{ik})\over\partial x_k}V^0~-
~{\partial (\sqrt{-g}~F^{0k})\over\partial x_k}V^i\right)
=0 ~~~~~i=0,1,2,3
\end{equation}
or, in more contracted form:
\begin{equation}\label{eq:forminvc}
(\nabla_k F^{ik})V^0~-~(\nabla_k F^{0k})V^i
~=~0 ~~~~~~~~i=0,1,2,3
\end{equation}

The equation (\ref{eq:forminv3}) remains unchanged. However, it can be also written
in the following way (see \cite{fock}, p.133):
\begin{equation}\label{eq:ciclina}
F_{ikj}~=~\nabla_j F_{ik}~+~\nabla_i F_{kj}~+~\nabla_k F_{ji}~=~0
\end{equation}
Besides, equation (\ref{eq:forginv5}) becomes:
\begin{equation}\label{eq:forgex}
V^0~F_{jik}~=~\pm V^m~F_{123}
\end{equation}
By taking the 4-divergence of the contravariant vector in (\ref{eq:forminv4})
and considering once again that the tensor $F^{ik}$
is anti-symmetric, we arrive at the continuity equation:
\begin{equation}\label{eq:congen}
{1\over \sqrt{-g}}{\partial (\sqrt{-g}~\rho_{\bf E} V^i)\over\partial x_i}=0 ~~~~~
{\rm with}~~\rho_{\bf E}={1\over \sqrt{-g}}{\partial (\sqrt{-g}~F^{0k})\over\partial x_k}
\end{equation}
We got an analogous result in section 3, by taking the standard divergence of the vector equation
(\ref{eq:rotbm}). The time derivative came from the term ${\rm
div}\hskip-1.truemm\left({\partial\over\partial t}{\bf E}\right)$ and the term ${\rm div}({\rm curl}{\bf
B})$ was zero. All the pieces here combine in a completely different manner. Nevertheless, the final
result is extraordinarily similar.
\par\bigskip

\setcounter{equation}{0}

\section{The energy tensor}
\smallskip

Let us first work in Cartesian coordinates. We will define the symmetric electromagnetic stress tensor
in the classical way (see \cite{fock}, p.96), i.e.:
\begin{equation}\label{eq:tensen}
U_{ik}~=~-\left(\sum_{j=0}^3 e_jF_{ij}F_{kj}~-~{1\over 2}\Big( c^2\vert {\bf B}\vert^2
-\vert {\bf E}\vert^2\Big)e_i\delta_{ik}\right)
\end{equation}
We have $U_{00}={1\over 2}( \vert {\bf E}\vert^2
+c^2\vert {\bf B}\vert^2)$ and $\sum_{i=0}^3 e_iU_{ii}=0$.

\noindent Its contravariant version is given by $U^{ik}=e_ie_kU_{ik}$ and of course we have
$\sum_{i=0}^3 e_iU^{ii}=0$. The explicit expression of the contravariant tensor is the following:

$$\left({\small\matrix{{1\over 2}( \vert {\bf E}\vert^2 +c^2\vert {\bf B}\vert^2) & cB_2E_3-cE_2B_3
& cE_1B_3-cB_1E_3 & cB_1E_2-cE_1B_2 \cr\cr
cB_2E_3-cE_2B_3 & {{-E_1^2+c^2B^2_2+c^2B^2_3}\atop{-{1\over 2}( c^2\vert {\bf B}\vert^2
-\vert {\bf E}\vert^2)}} & -E_1E_2-c^2B_1B_2 & -E_1E_3-c^2B_1B_3 \cr\cr
cE_1B_3-cB_1E_3 & -E_1E_2-c^2B_1B_2 & {{-E_2^2+c^2B^2_1+c^2B^2_3}\atop{-{1\over 2}( c^2\vert
{\bf B}\vert^2 -\vert {\bf E}\vert^2)}}& -E_2E_3-c^2B_2B_3\cr\cr
cB_1E_2-cE_1B_2 & -E_1E_3-c^2B_1B_3  & -E_2E_3-c^2B_2B_3 &
{{-E_3^2+c^2B^2_1+c^2B^2_2}\atop{-{1\over 2}( c^2\vert {\bf B}\vert^2 -\vert {\bf E}\vert^2)}}\cr
}}\right)
$$
\par\smallskip

If (\ref{eq:sfe}) and (\ref{eq:sfb}) are satisfied, then an important property of this last tensor is
that, in the case of free waves (hence in absence of mechanical terms), its 4-divergence vanishes.
Indeed, we have for $i=0,1,2,3$:
\begin{equation}\label{eq:divnul}
{\partial U^{ik}\over \partial x_k}~=~0
\end{equation}
provided (\ref{eq:lorentz}) (or (\ref{eq:lore2})) is satisfied. This implies that ${\bf E}\cdot
{\bf B}=0$ and $\vert {\bf E}\vert = c\vert {\bf B}\vert$. These hypotheses also imply that
 $\vert {\bf E}\times {\bf B}\vert =\vert{\bf E}\vert\vert{\bf B}\vert$
and  ${\bf V}\cdot {\bf B}=0$. Let us prove (\ref{eq:divnul}) starting from $i=0$.
Thanks to (\ref{eq:energ}),  one has:
\begin{equation}\label{eq:divnul0}
{\partial U^{0k}\over \partial x_k}~=~{1\over 2c}{\partial\over\partial t}
\big( \vert {\bf E}\vert^2
+c^2\vert {\bf B}\vert^2\big)~+~c~{\rm div}({\bf E}\times {\bf B})~=~0
\end{equation}
As far as the other values of $i$ are concerned, let us begin to define:
$${\bf N}~=~(N_1, N_2,N_3)~=~{\partial {\bf E}\over \partial t}~-~ c^2 {\rm curl} {\bf B}~
+~({\rm div}{\bf E}) {\bf V}
$$
\begin{equation}\label{eq:sfaz}
{\bf M}~=~(M_1, M_2,M_3)~=~{\partial {\bf B}\over \partial t}~+~{\rm curl} {\bf E}~
+~({\rm div}{\bf B}){\bf V}
\end{equation}
Thus, if the equations  (\ref{eq:sfe}) and (\ref{eq:sfb}) are true, then we get
${\bf N}=0$ and ${\bf M}=0$.
We are ready to check (\ref{eq:divnul}) for $i=1$ (the other cases are treated in a very similar way).
We have:
$$
{\partial U^{1k}\over \partial x_k}~=~{\partial\over\partial t}(B_2E_3-E_2B_3)-
{\partial\over \partial x}(-E_1^2+c^2B^2_2+c^2B^2_3)
$$
$$
+{1\over 2}~{\partial\over\partial x}\Big( c^2\vert {\bf B}\vert^2
-\vert {\bf E}\vert^2\Big)
~-~{\partial\over\partial y}(-E_1E_2-c^2B_1B_2)-
{\partial\over \partial z}(-E_1E_3-c^2B_1B_3)
$$
$$
=~(M_2E_3-M_3E_2+N_3B_2-N_2B_3) ~+~(E_1+V_2B_3-V_3B_2)~{\rm div}{\bf E}
$$
\begin{equation}\label{eq:divnul1}
~~~~~~~+~(c^2B_1+V_3E_2-V_2E_3)~{\rm div}{\bf B}~=~0
\end{equation}
The last three terms in (\ref{eq:divnul1}) are actually zero for the following reasons. In the first one
we recognize the second and the third components of ${\bf N}$ and ${\bf M}$, which  vanish, if we assume
that the equations (\ref{eq:sfe}) and (\ref{eq:sfb}) are satisfied. The second one contains the first
component of the vector ${\bf E}+{\bf V}\times {\bf B}$, which vanishes in the case of a free wave.
Concerning the last term, the part in parenthesis is the first component of the vector $c^2{\bf B}-{\bf
V}\times {\bf E}$, which  also vanishes (see (\ref{eq:lore2})).
\par\smallskip

The property (\ref{eq:divnul}) is reported in many texts (see for instance \cite{fock}, p.97). But, it
is extremely important to observe that, in the case of Maxwell equations, the last two terms are zero
because ${\rm div}{\bf E}=0$ and ${\rm div}{\bf B}=0$. This is the reason why we decided to double check
equation (\ref{eq:divnul}), which turns out to be fulfilled even when the divergence of the fields ${\bf
E}$ and ${\bf B}$ is not zero (the assumption we supported throughout this paper). Therefore
(\ref{eq:divnul}) holds under weaker hypotheses.
\par\smallskip

As expected, a converse statement also holds:
assuming that  (\ref{eq:divnul}) is true, then we can recover the equations
 (\ref{eq:sfe}) and (\ref{eq:sfb}). This amounts to differentiate the equation
of energy conservation, in order to obtain the corresponding Euler equations.
Arguing as we did to get (\ref{eq:divnul1}), we arrive at:
$$
\left({\partial U^{1k}\over \partial x_k},~ {\partial U^{2k}\over \partial x_k},
~{\partial U^{3k}\over \partial x_k}\right)~=~({\bf M}\times {\bf E}~-~{\bf N}
\times {\bf B})
$$
\begin{equation}\label{eq:divvet}
+~({\bf E}~+~{\bf V}\times {\bf B})~{\rm div}{\bf E}
~+~(c^2{\bf B}~-~{\bf V}\times {\bf E})~{\rm div}{\bf B}
\end{equation}

Assuming, as previously done, that we are dealing with a free wave, after eliminating
 in (\ref{eq:divvet}) the vanishing terms, we are left with
 $({\bf M}\times {\bf E}-{\bf N}\times {\bf B})$. Since, by hypothesis, the equation
(\ref{eq:divnul}) is true, if the vector ${\bf N}$ is zero, then ${\bf M}$ must be also zero (likewise,
if ${\bf M}$ is zero, then  ${\bf N}$ is zero). Therefore,  (\ref{eq:sfe}) is satisfied if and only if
(\ref{eq:sfb}) is satisfied. This is the same situation  encountered in the classical Maxwell equations,
where ${\partial \over\partial t}{\bf B} + {\rm curl}{\bf E}$ and ${\rm div}{\bf B}$ both vanish if and
only if ${\partial \over\partial t}{\bf E}-c^2 {\rm curl}{\bf B}$ and
 ${\rm div}{\bf E}$ are both zero. In the standard approach, the first pair of
equations are satisfied by choosing the pontentials ${\bf A}$ and $\Phi$ as in (\ref{eq:potenz}). The
second pair is obtained by means of variational type arguments.
\par\smallskip

Of course, we can find ``intermediate'' situations, by suitably redefining the two potentials. Let us
take for example:
$$
{\cal B}~=~{\rm curl}{\bf A}~~~~~~~~~ ~~{\cal E}~=~
-{1\over c}{\partial {\bf A}\over \partial t}-\nabla \Phi
$$
\begin{equation}\label{eq:potnuo}
{\rm with} ~~~~ {\cal E}~=~{\lambda {\bf E}+(1-\lambda )c{\bf B}\over \sqrt{\lambda^2 +(1-\lambda)^2}}
~~~{\rm and}~~~ {\cal B}~=~{\lambda c{\bf B}-(1-\lambda ){\bf E}\over \sqrt{\lambda^2 +(1-\lambda)^2}}
\end{equation}
where $\lambda$ is a real parameter. From the relations  (\ref{eq:potnuo}) we can explicitly
compute the fields  ${\bf E}$ and ${\bf B}$ in terms of ${\cal E}$ and ${\cal B}$. These also
imply:
\begin{equation}\label{eq:ecalb}
{\rm div}{\cal B}~=~0~~~~~~~~~~~~
~~{\partial{\cal B}\over\partial t}~=~-c~{\rm curl}{\cal E}
\end{equation}
Then, it is a matter of minimizing the usual Lagrangian.  At this point, introducing
 the constraint ${\bf A}\cdot{\bf V}=\Phi$, one gets the equation:
\begin{equation}\label{eq:ecale}
{\partial {\cal E}\over \partial t}~=~c~{\rm curl}{\cal B}~-~
{\bf V}~{\rm div}{\cal E}
\end{equation}
that, for $\lambda =1$, is equivalent to equation (\ref{eq:sfe}). The equations in (\ref{eq:ecalb}) are
equivalent to require $(\lambda -1){\bf N}+\lambda {\bf M}=0$, while the one in (\ref{eq:ecale}) brings
to $\lambda {\bf N}+(1-\lambda ) {\bf M}=0$.
\par\smallskip

It is to be noted that ${\bf V}$ has the same direction of  ${\bf E}\times {\bf B}$, which is also like
that of ${\cal E}\times {\cal B}$. So, from the energy tensor it is not possible to figure out what the
parameter $\lambda$ is, as well as the polarization of the wave. This information has to be provided
with the initial conditions. For instance, the wave in  (\ref{eq:polcir}), circolarly polarized,
produces the same tensor  $U^{ik}$ of a linearly polarized wave, moving in the same direction with twice
the  intensity. As a further consequence, we finally observe that $U^{ik}$ does not change if ${\bf E}$
takes the place of  $-c{\bf B}$ and $c{\bf B}$ takes the place of ${\bf E}$. Such a permutation
corresponds to the choice $\lambda =0$.
\par\smallskip

We can now argue in a general framework. For a given metric tensor $g_{ik}$, the electromagnetic stress
tensors must be modified in the following way (see \cite{fock}, p.151):
$$
U_{ik}~=~-\left( g^{mj}F_{im}F_{kj}~-~{\textstyle{1\over 4}}g_{ik}F_{mj}F^{mj}\right)
$$
\begin{equation}\label{eq:tennuo}
U^{ik}~=~-\left( g_{mj}F^{im}F^{kj}~-~{\textstyle{1\over 4}}g^{ik}F_{mj}F^{mj}\right)
\end{equation}
where $F_{ik}$ is given in (\ref{eq:tens1}), while $F^{ik}$ comes from the relation:
\begin{equation}\label{eq:tinvf}
F^{ik}~=~g^{im}g^{kl}F_{ml}
\end{equation}
Assuming to be as in the case of a free wave, the equation (\ref{eq:divnul}) has to be replaced by the
following one:
\begin{equation}\label{eq:divurg}
\nabla_k U^{ik}~=~{1\over \sqrt{-g}}{\partial (\sqrt{-g }~U^{ik})\over \partial x_k}
~+~\Gamma^i_{mj}U^{mj}~=~0
\end{equation}
The proof of  (\ref{eq:divurg}) is given for instance in \cite{jackson}, p.606,
for the classical Maxwell equations. This is also true in the case of our new set of  equations
(hence under weaker hypotheses). At the end of this section,
we will evaluate the 4-divergence of the tensor $U^{ik}$ in general coordinates.
 Such generalizations are unavoidable since, even the simple case of a plane wave provokes a
modification of the space-time geometry, requiring to work with tensors of the form (\ref{eq:tennuo}).
These aspects will be better studied in the next section.
\par\smallskip

When the electromagnetic phenomenon is not a free wave, we cannot expect that (\ref{eq:divnul}) and
its generalization (\ref{eq:divurg}) are verified. This means that the system constituted by the sole
electromagnetic part is not energy preserving. We know  that, in this case, the energy balance
has to take care of the mechanical effects. Thus, we study how to introduce them.
We start by assuming  that ${\rm div}{\bf B}=0$, leaving the discussion of the more general case
 to section 14. Then, let us define a mass tensor as follows:
\begin{equation}\label{eq:tenmas}
M_{ik}~=~V_iV_k~{\rm div}{\bf E}
\end{equation}
 The contravariant version is given by $M^{ik}=e_ie_kM_{ik}$, which is explicitly written as:
\begin{equation}\label{eq:tenmas2}
M^{ik}~=~\rho_{\bf E}\left(\matrix{V_0^2 & -V_0V_1 & -V_0V_2 & -V_0V_3 \cr\cr
-V_0V_1 & V_1^2 & V_1V_2 & V_1V_3 \cr\cr
-V_0V_2 & V_1V_2 & V_2^2 & V_2V_3 \cr\cr
-V_0V_3 & V_1V_3 & V_2V_3 & V_3^2 \cr}\right)
\end{equation}
where $V_0=c$ and $\rho_{\bf E} ={\rm div}{\bf E}$ is a kind of mass density (dimensionally this is not
correct, but this aspect will be altered later). We recall that  $\rho_{\bf E}$ can also be negative.
Let us check what happens to ${\partial \over \partial x_k}M^{ik}$. For $i=0$ we have:
$$
{\partial M^{0k}\over \partial x_k}~=~c\left( c~{\partial \rho_{\bf E}\over\partial x_0}
~-~{\partial (\rho_{\bf E} V_1)\over\partial x_1}~-
~{\partial (\rho_{\bf E} V_2)\over\partial x_2}
~-~{\partial (\rho_{\bf E} V_3)\over\partial x_3}\right)
$$
\begin{equation}\label{eq:divmas0}
=~c\left({\partial \rho_{\bf E}\over\partial t}~+~{\rm div}(\rho_{\bf E} {\bf V})\right)~=~0
\end{equation}
This is true because of the continuity equation (\ref{eq:conti2}) with
$\rho=\rho_{\bf E}$. For the other indices $i=1,2,3$, we have:
$$
{\partial M^{ik}\over \partial x_k}~=~-\left( c~{\partial (\rho_{\bf E} V_i)\over\partial x_0}
~-~{\partial (\rho_{\bf E} V_1V_i)\over
\partial x_1}~-~{\partial (\rho_{\bf E} V_2V_i)\over\partial x_2}
~-~{\partial (\rho_{\bf E} V_3V_i)\over\partial x_3}\right)
$$
$$
=~-\left({\partial (\rho_{\bf E} V_i)\over\partial t}~+
~{\rm div}(\rho_{\bf E} V_i {\bf V})\right)~=~
-V_i\left({\partial \rho_{\bf E}\over\partial t}~+~{\rm div}(\rho_{\bf E} {\bf V})\right)
$$
\begin{equation}\label{eq:divmasi}
-\rho_{\bf E}\left({\partial  V_i\over\partial t}~+~({\bf V}\cdot \nabla)V_i\right)
~=~ - {DV_i\over Dt}~{\rm div}{\bf E}
\end{equation}
where we again used the continuity equation. We conclude for instance that, if the light rays are
straight-lines (that is: ${\bf G}={D\over Dt}{\bf V}=0$), then one gets
  ${\partial \over \partial x_k}M^{ik}=0$, for $i=0,1,2,3$.
\par\smallskip

In non Euclidean geometry, it is necessary to generalize the mass tensors in the following way:
\begin{equation}\label{eq:tenmasg}
M_{ik}=\rho_{\bf E}V_iV_k~~~~~~M^{ik}=\rho_{\bf E}V^iV^k~~~~{\rm with}~~~
\rho_{\bf E}={1\over\sqrt{-g}}{\partial (\sqrt{-g}~F^{0k}) \over\partial x_k}
\end{equation}
With the help of the continuity equation (\ref{eq:congen}) and the definition
(\ref{eq:defg}), it is easy to get, for $i=0,1,2,3$:
$$
\nabla_k M^{ik}~=~ {1\over\sqrt{-g}}{\partial (\sqrt{-g}~M^{ik}) \over\partial x_k}~+~
\Gamma^i_{mj}M^{mj}
$$
\begin{equation}\label{eq:ledmot}
=~\rho_{\bf E}V^k {\partial V^i\over \partial x_k}~+~ \rho_{\bf E}\Gamma^i_{mj}V^mV^j
~=~\rho_{\bf E}G^i~=~-{\mu\over c}\rho_{\bf E} F^{ik}V_k
\end{equation}
where the last equality is a consequence of (\ref{eq:geofor}). Hence, in a flat space
($G^i=0$), the 4-divergence of the mass tensor vanishes. Moreover, we observe that the mass
tensor does not contain the pressure term (on the other hand, an equation of state
is not defined).
\par\smallskip

In order to combine electromagnetic and mechanical effects, we sum up the corresponding
tensors, by defining:
\begin{equation}\label{eq:tensor}
T_{ik}~=~{\mu\over c^4}\Big( \mu U_{ik}~+~M_{ik}\Big)
\end{equation}
where the constant $\mu$ is dimensionally equivalent to a charge divided by a mass. It follows that
$T_{ik}$ has the same dimension of a curvature, that is the inverse of the square of a distance. Now, in
a flat space-time, even if we are not dealing with a free wave, we may write:
\begin{equation}\label{eq:divnult}
{\partial T^{ik}\over \partial x_k}~=~0
\end{equation}
As a matter of fact, due to (\ref{eq:sfgr}), if the term $\mu({\bf E}+{\bf V}\times {\bf B}) {\rm
div}{\bf E}$ of the electromagnetic part does not vanish (see (\ref{eq:divvet})), it is anyway
compensated by the corresponding term $-{\rm div}{\bf E}{D\over Dt}{\bf V}$ of the mechanical part (see
(\ref{eq:divmasi})).
\par\smallskip

\noindent In the general case, the relation (\ref{eq:divnult}) is substituted by:
\begin{equation}\label{eq:divnulrg}
\nabla_k T^{ik}~=~{1\over \sqrt{-g}}{\partial (\sqrt{-g }~T^{ik})\over \partial x_k}
~+~\Gamma^i_{mj}T^{mj}~=~0
\end{equation}
Before ending this section, we would like to verify that (\ref{eq:divnulrg}) actually corresponds to the
Euler equations. As a matter of fact, (\ref{eq:divnulrg}) is satisfied when (\ref{eq:forminvc}),
(\ref{eq:forgex}) and (\ref{eq:geofor}) are true. We recall that these three last equations are the
generalizations of (\ref{eq:sfe}), (\ref{eq:sfb}), (\ref{eq:sfg}), respectively. For the moment, we will
only treat the case in which ${\rm div}{\bf B}=0$, leaving the general discussion to section 14.
 Let us start by computing the 4-divergence of the
tensor $U^{ik}$. First of all, we have:
$$
\nabla_k U^{ik}~=~-\nabla_k(g_{mj} F^{im}F^{kj})~+~{\textstyle{1\over 4}}
\nabla_k(g^{ik}F_{mj}F^{mj})
$$
\begin{equation}\label{eq:divu1}
=~ g^{im}\nabla_k( F_{mj}F^{jk})~+~{\textstyle{1\over 4}}g^{ik}
\nabla_k(F_{mj}F^{mj})
\end{equation}
where we notice that $g_{mj}F^{im}=g^{im}F_{mj}$ (thanks to (\ref{eq:tinvf})), that $F^{kj}=-F^{jk}$ and
that, due to (\ref{eq:divtenm}), it is allowed to exchange the metric tensor with the covariant
derivative (see also \cite{fock}, p130). Going ahead, one gets:
$$
\nabla_k U^{ik}~=~g^{im}(\nabla_kF^{jk})F_{mj}~+~g^{im}(\nabla_kF_{mj})F^{jk}
~+~{\textstyle{1\over 2}}g^{ik}(\nabla_kF_{mj})F^{mj}
$$
$$
=~ c^{-1}(\nabla_k F^{0k})F_{mj}g^{im}V^j~+~{\textstyle{1\over 2}}g^{im}
(\nabla_kF_{mj})F^{jk}~
$$
\begin{equation}\label{eq:divu2}
+~ {\textstyle{1\over 2}}g^{im}(\nabla_kF_{mj}+\nabla_m F_{jk})F^{jk}
\end{equation}
where we used (\ref{eq:forminvc}) (with $V^0=c$). The other passages have been obtained
by a suitable renaming of the indices. Recalling the definition of  $F_{ikj}$ given in (\ref{eq:ciclina}),
we have:
$$
\nabla_k U^{ik}~=~ { \rho_{\bf E}\over c} F_{mj}g^{im}V^j~+~{\textstyle{1\over 2}}g^{im}
(\nabla_kF_{mj})F^{jk}~
$$
$$
+~ {\textstyle{1\over 2}}g^{im} F_{mjk}F^{jk}~-~{\textstyle{1\over 2}}g^{im}
(\nabla_jF_{km})F^{jk}
$$
\begin{equation}\label{eq:divu3}
=~  {\rho_{\bf E}\over c} F^{im}V_m~+~{\textstyle{1\over 2}}g^{im} F_{mjk}F^{jk}
\end{equation}
In the last passage two terms have been deleted, since, after renaming the indices, they resulted in
being equal and with opposite signs. The last term in (\ref{eq:divu3}) is zero because of
(\ref{eq:ciclina}) (remember that we are studying the case ${\rm div}{\bf B}=0$, thus $F_{123}=0$). Of
course, the final result is zero when, for instance, $\rho_{\bf E}=0$, as in the classical Maxwell case.
But it is also zero when $F^{im}V_m=0$, which corresponds to the case of a free electromagnetic wave. On
the contrary, we need to consider the contribution of the mass tensor. If $F_{mjk}=0$, taking into
account the relations (\ref{eq:tensor}) and (\ref{eq:ledmot}), one finally obtains:
$$
\nabla_kT^{ik}~=~{\mu\over c^4}~\Big(\mu \nabla_k U^{ik}~+~\nabla_k M^{ik}\Big)
$$
$$
~~~~~~~~=~{\mu\over c^4}~\left({\mu\over c}\rho_{\bf E}F^{im}V_m~-~{\mu\over c}\rho_{\bf E}
F^{ik}V_k\right)~=~0
$$

\par\bigskip

 \setcounter{equation}{0}

\section{Unified fields equations}
\smallskip

In the previous section, we build the symmetric tensor $T_{ik}$ that includes both the energy
contribution of an electromagnetic wave and that of mechanical type, taking into account possible
deviations from the natural propagation path of the wave. The properties of  $T_{ik}$ insure the
preservation of energy and momentum.
 Hence, we can put $T_{ik}$ on the right-hand side of the Einstein equation:
\begin{equation}\label{eq:eins}
R_{ik}~-~{\textstyle{1\over 2}}g_{ik}R~=~\chi ~T_{ik}
\end{equation}
in which we recognize the Ricci tensor:
\begin{equation}\label{eq:ricci}
R_{ik}~=~{\partial \Gamma^m_{ik}\over \partial x_m}~-~{\partial \Gamma^m_{im}
\over \partial x_k}~+~\Gamma^j_{ik}\Gamma^m_{jm}~-~\Gamma^j_{im}\Gamma^m_{kj}
\end{equation}
the scalar curvature:
\begin{equation}\label{eq:cusca}
R~=~g^{ik}R_{ik}
\end{equation}
and an adimensional constant  $\chi$. We recall that the Christoffel symbols are defined
in (\ref{eq:chris}). Let us note that, by (\ref{eq:tennuo}) and (\ref{eq:tenmasg}),
the metric tensor, which is now our unknown, also
appears  on the right-hand side of (\ref{eq:eins}).
\par\smallskip

We soon examine the response of equation (\ref{eq:eins})  to the passage of the most elementary plane
wave. We take for instance the expression given in (\ref{eq:campip}), where we have $E_1=cB_2=c\sin
\omega(t-z/c)$, ${\rm div}{\bf E}=0$ and ${\rm div}{\bf B}=0$. We will verify that, even in this simple
case, the space-time geometry, that comes from the solution of  (\ref{eq:eins}), is not Euclidean. In
fact, we look for a metric tensor $g_{ik}$ of the following type:
\begin{equation}\label{eq:metrica1}
g_{ik}=\left(\matrix{1 & 0 & 0 & 0 \cr
0 & -p^2 & 0 & 0 \cr 0 & 0 & -1 & 0 \cr
0 & 0 & 0 & -1 \cr}\right)~~~~~
g^{ik}=\left(\matrix{1 & 0 & 0 & 0 \cr
0 & -1/p^2 & 0 & 0 \cr 0 & 0 & -1 & 0 \cr
0 & 0 & 0 & -1 \cr}\right)
\end{equation}
where $p$ is a function, to be determined, of the variable  $\xi =t-z/c$.
Somehow, we are expressing a preference for the direction of the
 $x$-axis, which is orientated with the electric field. The determinant $g$ of
$g_{ik}$ is equal to $-p^2$. The corresponding  Christoffel symbols are:
\begin{equation}\label{eq:chronda}
\Gamma^0_{11}={pp^\prime\over c}~~~~~~\Gamma^3_{11}={-pp^\prime\over c}~~~~~~
\Gamma^1_{01}=\Gamma^1_{10}=\Gamma^1_{13}=\Gamma^1_{31}={p^\prime\over c p}
\end{equation}
where the prime denotes the derivative with respect to $\xi$. All the other symbols  vanish. The
non-zero coefficients of the Ricci tensor are instead:
\begin{equation}\label{eq:ricconda}
R_{00}~=~R_{03}~=~R_{30}~=~R_{33}~=~-{p^{\prime\prime}\over c^2p}
\end{equation}
The scalar curvature $R$ is zero.
\par\smallskip

Being zero the divergence of ${\bf E}$, the mass tensors $M_{ik}$ e $M^{ik}$ vanish. Actually, one
should check that $\rho_{\bf E}=0$.  This is also true, as the comments at the end of this section
illustrate.  The tensors $U_{ik}$ and $U^{ik}$ have to be computed through (\ref{eq:tennuo}). First of
all, one has:
\begin{equation}\label{eq:campif0}
F_{ik}=c \left(\matrix{0 & -u & 0 & 0\cr
u & 0 & 0 & u \cr 0 & 0 & 0 & 0 \cr
0 & -u & 0 & 0 \cr}\right)
\end{equation}
\begin{equation}\label{eq:campif}
F^{ik}=c \left(\matrix{0 & u/p^2  & 0 & 0 \cr
-u/p^2 & 0 & 0 & u/p^2 \cr 0 & 0 & 0 & 0 \cr
0 &  -u/p^2 & 0 & 0\cr}\right)
\end{equation}
where $u=B_2=E_1/c$. Note that $(V_0, {\bf V})=(c,0,0,c)$  and $(V^0,V^1,V^2,V^3)= (c,0,0,-c)$. Hence,
one gets $F^{ik}V_k=0$, from which we deduce that the wave is free, as  is already known. Afterwards, we
have:
\begin{equation}\label{eq:campip0}
T_{ik}={\mu^2\over c^2}\left(\matrix{(u/p)^2 & 0 & 0 & (u/p)^2 \cr
0 & 0 & 0 & 0 \cr 0 & 0 & 0 & 0 \cr
(u/p)^2 & 0 & 0 & (u/p)^2 \cr}\right)
\end{equation}
\begin{equation}\label{eq:campip8}
T^{ik}={\mu^2\over c^2}\left(\matrix{(u/p)^2 & 0 & 0 & -(u/p)^2 \cr
0 & 0 & 0 & 0 \cr 0 & 0 & 0 & 0 \cr
-(u/p)^2 & 0 & 0 & (u/p)^2 \cr}\right)
\end{equation}
Thus, (\ref{eq:eins}) and (\ref{eq:campip8}) bring us to the equation:
\begin{equation}\label{eq:ecconda}
-p^{\prime\prime}p~=~\mu^2\chi u^2
\end{equation}
For $u=\sin \omega (t-z/c)$, we finally obtain $p=\big(\mu\sqrt{\chi}/\omega
 \big)\sin \omega (t-z/c)$, which is the solution we were looking for.
There are surely other geometries compatible with the same plane wave. Note that the one presented
here satisfies the relation (\ref{eq:armon}). We also observe that there are points in
which the metric becomes singular, that is, the determinant $g$ is zero.
Another equivalent possibility is to exchange $g_{11}$ and $g_{22}$ in
(\ref{eq:metrica1}), and make $p$ oscillate with the magnetic field. Comments
about this option will be given in the next section.
\par\smallskip

The solution just obtained can be assimilated to a transversal (perfectly plane) gravitational wave,
travelling in phase with the electromagnetic one. It must also be noted that, even if the space is
officially non Euclidean, the geodesics involved in the motion of the wave are straight-lines. The field
${\bf G}$, defined by (\ref{eq:defg}), is identically zero. This is in agreement with our viewpoint: the
geometry may be deformed, but there is no creation of a real gravitational vector field.
\par\smallskip

Pure gravitational solutions resembling plane waves, were formerly detected in \cite{bondi}. We have
been able to get the above explicit (and very simple) solution because we were resolute enough to assume
the dependence from the metric tensor of the right-hand side of the Einstein equation. As far as we
could deduce from the current literature, in contrast to our general approach to the problem, it is
customary to construct the electromagnetic energy tensor in vacuum (thus, in  Minkowski space-time),
also because such an assumption is supposed (erroneously) to simplify the computations.  Then, one comes
to a set of solutions, but, as we proved, this is not the correct setting. Note also that, commonly,
gravitational waves are searched among the solutions of the linearized homogeneous Einstein equation,
obtained after perturbation of the flat space-time.
\par\smallskip

We can recover the laws of motion by evaluating the 4-divergence of the tensor
 $T^{ik}$ in (\ref{eq:campip8}). The geometry is non Euclidean, therefore, the relation
(\ref{eq:divnult}) has to be replaced by (\ref{eq:divnulrg}), where $\sqrt{-g}=\vert p\vert$. For $i=0$,
one has:
$$
{1\over \sqrt{-g}}{\partial (\sqrt{-g }~T^{0k})\over \partial x_k}~+~\Gamma^0_{mj}T^{mj}
~=~{\mu^2\over c^2p}\left({1\over c}{\partial (u^2/p)\over\partial t}~+~{\partial (u^2/p)
\over\partial z}\right)
$$
\begin{equation}\label{eq:divgr1}
=~{\mu^2\over c^2p}\left[{1\over p}\left({1\over c}{\partial u^2\over\partial t}~+~{\partial u^2
\over\partial z}\right)~-~{u^2\over p^2}\left({1\over c}{\partial p\over\partial t}~+~{\partial p
\over\partial z}\right)\right]
\end{equation}
The situation is exactly the same for $i=3$.  The last term in (\ref{eq:divgr1}) is zero, when for
instance:
\begin{equation}\label{eq:emoto}
{1\over c}{\partial u\over\partial t}~+~{\partial u\over\partial z}~=~0~~~~~~~~~~
{1\over c}{\partial p\over\partial t}~+~{\partial p \over\partial z}~=~0
\end{equation}
Our plane electromagnetic-gravitational wave is actually the solution to both the above equations, once
the proper initial conditions have been assigned.
\par\smallskip

\noindent Let us now discuss the example of a circularly-polarized plane wave:
$$
{\bf E}~=~(c\cos \omega(t-z/c), ~c\sin \omega(t-z/c),~0)
$$
\begin{equation}\label{eq:ondapc}
{\bf B}~=~(-\sin \omega(t-z/c),~ \cos \omega(t-z/c),~0)
\end{equation}
The classical divergence of the electric field vanishes, as well as that of the magnetic field. Then,
let us take the following metric tensor:
\begin{equation}\label{eq:metrica3}
g_{ik}={\mu^2\chi\over \omega^2}\left(\matrix{1 & 0 & 0 & 0 \cr
0 & -[\cos \omega(t-z/c)]^2 & 0 & 0 \cr 0 & 0 & -[\sin \omega(t-z/c)]^2 & 0 \cr
0 & 0 & 0 & -1 \cr}\right)
\end{equation}
in such a way that the coordinates $x$ and $y$ are syncronized with the electric field (the reasons for
this choice will be explained at the end of section 14). In this case, the nonvanishing coefficients of
the Ricci tensor are: $R_{00}=R_{03}=R_{30}=R_{33}=2\omega^2/c^2$. They coincide with the respective
coefficients of the stress tensor:  $T_{00}=T_{03}=T_{30}=T_{33}=2\omega^2/c^2$. Therefore, once again,
the Einstein equation is verified. The wave is free and we have  $R=0$ and ${\bf G}=0$.
\par\smallskip

Slightly more complicated is the case of a plane wave where ${\rm div}{\bf E}$ is non-zero. This happens
for instance when  $u=B_2=E_1/c= f(x)\sin \omega (t-z/c)$. As we know, the solution satisfies the
equations (\ref{eq:rotbm}), (\ref{eq:rotem}), (\ref{eq:divbm}), but not the classical Maxwell equations.
We suggest looking for a metric tensor of the form:
\begin{equation}\label{eq:metrica2}
g_{ik}=\left(\matrix{1 & 0 & 0 & 0 \cr
0 & -p^2f^2 & 0 & 0 \cr 0 & 0 & -1 & 0 \cr
0 & 0 & 0 & -1 \cr}\right)
\end{equation}
where $p$ is function of the variable $\xi =t-z/c$. One has: $\sqrt{-g}=\vert fp \vert$. The tensor
$F_{ik}$ is the same as in (\ref{eq:campif0}). Regarding the other tensors, we get:
\begin{equation}\label{eq:camp17}
F^{ik}=c \left(\matrix{0 & ~~u/(pf)^2~~  & ~~~0~~~ & 0 \cr
-u/(pf)^2  & 0 & 0 & u/(pf)^2  \cr 0 & 0 & 0 & 0 \cr
0 &  -u/(pf)^2  & 0 & 0\cr}\right)
\end{equation}
\par\smallskip
\begin{equation}\label{eq:campi18}
T_{ik}={\mu^2\over c^2}\left(\matrix{(u/pf)^2 & ~~0~~ & ~~0~~ & (u/pf)^2 \cr
0 & 0 & 0 & 0 \cr 0 & 0 & 0 & 0 \cr
(u/pf)^2 & 0 & 0 & (u/pf)^2 \cr}\right)
\end{equation}
We must point out an extraordinary fact: in the new geometry, the 4-divergence of the electric field
turns out to be zero. As a matter of fact, by noting that
 $u/f$ and $p$ do not depend on $x$, one has:
$$
\rho_{\bf E}~=~{1\over\sqrt{-g}}{\partial (\sqrt{-g}~F^{0k})\over \partial x_k}
$$
\begin{equation}\label{eq:divnog}
~=~{-1\over \vert fp\vert}~{\partial ( \vert fp\vert ~F^{01})\over \partial x}~=~
{-c\over f p^2}~{\partial (u/f)\over\partial x}~=~0
\end{equation}
 Thus, the mass tensor is still vanishing. The Christoffel
symbols are a little  different from the ones in (\ref{eq:chronda}) (in particular $\Gamma^1_{11}$ is
not zero), but the coefficients of the Ricci tensor are exactly equal to those given in
(\ref{eq:ricconda}). Then, the equation (\ref{eq:ecconda}) must be modified as follows:
\begin{equation}\label{eq:ecconda2}
-p^{\prime\prime}p~=~\mu^2\chi \left({u\over f}\right)^2
\end{equation}
thereby admitting the same solution $p$ obtained in the case of the plane wave at uniform density. The
laws of motion are the same as in (\ref{eq:emoto}). They tell us that $u$ and $p$ shift at the speed of
light along the  $z$-axis. They do not specify however the function  $f$, which must be assigned through
the initial conditions.
\par\smallskip

We are ready to illustrate the case of a spherical wave. With the same notation
of sections 2 and 4, we set the coordinates in order to have: $(x_0,x_1,x_2,x_3)=(
ct,-r,-\phi ,-\theta )$. Let us assume that ${\bf E}=(0, cu, 0)$ and ${\bf B}=(0, 0, u)$,
 where $u={1\over r}f(\phi )\sin \omega (t-r/c)$. We recall that, to avoid
singularities at the poles, the function  $f$ is not allowed to be constant. So,
this is similar to the case of a  variable-density plane wave. We also have that
 $(V^0,V^1,V^2,V^3)=(c,-c,0,0)$. So, let us begin by giving the metric tensor:
\begin{equation}\label{eq:metrica4}
g_{ik}=\left(\matrix{1 & 0 & 0 & 0 \cr
0 & -1 & 0 & 0 \cr 0 & 0 & -p^2f^2 & 0 \cr
0 & 0 & 0 & -1 \cr}\right)
\end{equation}
where $p$ is a function of the variable $\xi =t-r/c$. Note that here the case $g_{22}=-1$
corresponds to the standard spherical system of coordinates. For the electromagnetic tensors
we get:
$$
F_{ik}=c \left(\matrix{0 & 0 & -ru & 0\cr
0 & 0 & -ru & 0 \cr ru & ru & 0 & 0 \cr
0 & 0 & 0 & 0 \cr}\right)
$$
$$
F^{ik}=c \left(\matrix{0 & 0  & ru/(pf)^2 & 0 \cr
0 & 0 & -ru/(pf)^2 & 0 \cr -ru/(pf)^2 & ru/(pf)^2 & 0 & 0 \cr
0 &  0 & 0 & 0\cr}\right)
$$
 In order to evaluate $F_{ik}$, we started from
(\ref{eq:tensore}), recalling that, by (\ref{eq:soles}), one has $(\Phi ,
{\bf A})=(-F(\phi)\sin \omega (t-r/c), -F(\phi)\sin \omega (t-r/c), 0,0)$, where
 $F$ is a primitive of $f$. The metric in (\ref{eq:metrica4}) is the same as
the one we would have obtained if we had worked with a plane electromagnetic wave. The fact that we are
in spherical coordinates is actually contained in the electromagnetic tensors (in which we find $ru$ in
place of $u$). Note that $(V_0,V_1,V_2,V_3)=(c,c,0,0)$, from which one obtains the relation
$F^{ik}V_k=0$, confirming that the wave is free. As far as energy is concerned, we get:
$$
T_{ik}={\mu^2\over c^2}\left(\matrix{(ru/pf)^2  & (ru/pf)^2 & ~~0~~ & ~~0~~\cr
(ru/pf)^2  & (ru/pf)^2 & ~~0~~ & ~~0~~ \cr 0 & 0 & 0 & 0 \cr
0 & 0 & 0 & 0 \cr}\right)
$$
Therefore, from  the Einstein equation, we come to:
\begin{equation}\label{eq:ecconda3}
-p^{\prime\prime}p~=~\mu^2\chi \left({ru\over f}\right)^2
\end{equation}
where we observe that the right-hand side only depends on the variable $\xi =t-r/c$. The equation
(\ref{eq:ecconda3})  once again gives the solution  $p=\big(\mu\sqrt{\chi}/\omega
 \big)\sin \omega (t-z/c)$.
\par\smallskip

Finally, by differentiating $T^{ik}$ (for $i=0$ and $i=1$), we find the Euler equation
in spherical coordinates:
\begin{equation}\label{eq:eules}
{1\over c}{\partial u \over \partial t}~+~{1\over r}{\partial (ru)\over \partial r}~=~0
\end{equation}
The function $u$ is the  solution to (\ref{eq:eules}), after assuming the appropriate initial
conditions.
\par\smallskip

The results of this section, although only restricted to the analysis of free waves, bring to  attention
some important issues. Up to now, we have claimed that a good theory of  electromagnetism was meaningful
only by allowing ${\rm div}{\bf E}$ to be different from zero. Here instead we find that $\rho_{\bf
E}=0$. In our opinion, what is happening can be explained as follows. The space-time ``reacts'' to the
passage of a wave, by varying itself in syncronism, in order to make
 the 4-divergence of the electric field  vanish. The perturbation of the geometry
is however weak enough to maintain   field curvature ${\bf G}$ equal to zero. The classical divergence
${\rm div}{\bf E}$ may instead attain arbitrary values. We ask ourselves if it is possible to set up an
experiment showing that, at some point and at a certain time, one has ${\rm div}{\bf E}\not =0$.
Perhaps, this is not possible since, due to the modification of the metric,
 the instruments are unavoidably affected by the deformation of
 time and distances (with respect to the Euclidean reference frame).
Therefore, in place of ${\rm div}{\bf E}$, we could end up  measuring $\rho_{\bf E}$. But the last
quantity is always zero (a least for free waves). As a consequence,
 we conclude that some divergence  vanishes, although is not the classical,
 but the relativistic one. According to this new interpretation of the facts,
in some sense the Maxwell theory was correct.
\par\smallskip

There is another point that needs to be clarified. The problem is why the geometry changes  depending on
 the electric field, and not  the magnetic field, expecially after we said that for free waves the
two fields have the same role. Firstly, we note that, in all  examples studied in this section, the
condition ${\rm div }{\bf B}=0$ was always fulfilled. In addition, if similarly to $\rho_{\bf E}$, we
define  $\rho_{\bf B}$, we discover that this new quantity is also zero (see section 14). If we imagine
the wave like a fluid in motion, then this condition says that there is no flow of some ``magnetic
density of matter''. In truth, it is reasonable to assume that a sole electromagnetic fluid exists (not
two, a separate electrical one and a magnetic one).  As  will become clear in the next section, where we
analyze the case ${\rm div }{\bf B}\not=0$, such a fluid turns out to  pulsate along a specific
tangential direction (in principle, not necessarily corresponding to that of the electric field).
Exchanging  cause with  effect, in section 14 we will support the following statement: from the behavior
of the natural events, we are inclined to name the direction of the electric field as being that
identified by the transversal oscillations of the fluid in motion.
\par\smallskip

Far more complicated phenomena show up, when we suppose that the waves are no longer free (thus, ${\bf
G}\not =0$). In this context, the real gravitational fields come into life. We do not have any specific
examples to discuss, due to the difficulty of the problems involved. Some hints will be given in section
15.

\par\bigskip

\setcounter{equation}{0}

\section{The divergence of the magnetic field}
\smallskip

In the previous sections, some situations were discussed under the hypothesis
 ${\rm div}{\bf B}=0$. Although our equations now have  a general validity, the assumption
is necessary, for instance when introducing the potentials ${\bf A}$ and $\Phi$. Regarding this
condition, we would like to add further comments in this section. It is standard to introduce a
transformation that exchanges the role of the electric and  magnetic fields. This can be done through
the pseudo-tensor:
\begin{equation}\label{eq:epsilon}
\epsilon_{mjik}=\cases{~0 & when at least two indices are equal\cr
~1 & if the indices form an even permutation
\cr -1 & if the indices form an odd permutation }
\end{equation}
The parity of the permutations is counted starting from the set: $\{ 0,1,2,3\}$. Then, we define:
\begin{equation}\label{eq:epsi2}
\epsilon^{mjik}~=~e_me_je_ie_k\epsilon_{mjik}~=~-\epsilon_{mjik}
\end{equation}
We may now introduce the duals of the tensors (\ref{eq:tens1}) and (\ref{eq:tens2}) in the following
way:
\begin{equation}\label{eq:duali}
\hat F_{mj}={\textstyle{1\over 2}}\epsilon_{mjik}F^{ik}~~~~~~
\hat F^{mj}=e_me_j\hat F_{mj}=-{\textstyle{1\over 2}}\epsilon^{mjik}F_{ik}
\end{equation}
Therefore, we obtain for example:
$$\hat F_{01}=F^{23}~~~\hat F_{02}=F^{31}~~~\hat F_{03}=F^{12}~~~
\hat F_{23}=F^{01}~~~\hat F_{31}=F^{02}~~~\hat F_{12}=F^{03}$$
The original tensors and their duals have the same structure, with the difference
that  ${\bf E}$ replaces $-c{\bf B}$ and $c{\bf B}$ replaces ${\bf E}$.
\par\smallskip
In a similar way, the dual of the anti-symmetric rank-three tensor $F_{jik}$ (defined
in (\ref{eq:forminv3})) is given by:
\begin{equation}\label{eq:duali2}
\hat F^m~=~-{\textstyle{1\over 6}}\epsilon^{mjik}F_{jik}
\end{equation}
Hence, up to even permutations of the lower indices, one has:
$$
\hat F^0=F_{123}~~~~~~ \hat F^1=-F_{023}~~~~~~ \hat F^2=F_{013}~~~~~~ \hat F^3=-F_{012}
$$
In general coordinates, it is customary to define:
$$
\in^{mjik}~=~\sqrt{-g}~g^{mm^\prime}g^{jj^\prime}g^{ii^\prime}g^{kk^\prime}
\epsilon_{m^\prime j^\prime i^\prime k^\prime}~=~-{1\over \sqrt{-g}}~\epsilon_{mjik}
$$
\begin{equation}\label{eq:epsig}
\in_{mjik}~=~\sqrt{-g}~\epsilon_{mjik}
\end{equation}
So that the duals in (\ref{eq:duali}) and in (\ref{eq:duali2}) are
generalized as follows:
\begin{equation}\label{eq:duali3}
\hat F_{mj}={\textstyle{1\over 2}}\in_{mjik}F^{ik}~~~~~~
\hat F^{mj}=-{\textstyle{1\over 2}}\in^{mjik}F_{ik}~~~~~~
\hat F^{m}=-{\textstyle{1\over 6}}\in^{mjik}F_{jik}
\end{equation}
where $F_{ik}$ is provided in (\ref{eq:tens1}) and $F^{ik}$ can be found in (\ref{eq:tinvf}).
\par\smallskip

\noindent Then, the following relation is known (see \cite{fock}, p.134):
\begin{equation}\label{eq:divderc}
{1\over \sqrt{-g}}~{\partial (\sqrt{-g}~\hat F^{mj})\over \partial x_j}~=~\hat F^m
\end{equation}
where we supposed that $\hat F^m$ is the dual of the cyclic derivative  $F_{jik}$
of the tensor $F_{ik}$ (of which $\hat F^{mj}$ is the dual).
\par\smallskip

Passing to the duals, the equation (\ref{eq:forgex}) becomes: $V^0\hat F^m=V^m\hat F^0$.
Therefore, by  (\ref{eq:divderc}) we get:
\begin{equation}\label{eq:forminv6}
{1\over \sqrt{-g}}\left({\partial (\sqrt{-g}~\hat F^{mj})\over\partial x_j}V^0~-
~{\partial (\sqrt{-g}~\hat F^{0j})\over\partial x_j}V^m\right)
=0 ~~~~~m=0,1,2,3
\end{equation}
which is the exact counterpart of (\ref{eq:forminv4}). The equation (\ref{eq:forminv6})
represents, in a general coordinates system, the equation (\ref{eq:sfb}), that is equivalent
to (\ref{eq:sfe}), after taking  ${\bf E}$ in place of $-c{\bf B}$ and $c{\bf B}$
in place of ${\bf E}$. From (\ref{eq:forminv6}), we can recover the continuity equation:
\begin{equation}\label{eq:congen2}
{1\over \sqrt{-g}}{\partial (\sqrt{-g}~\rho_{\bf B} V^i)\over\partial x_i}=0 ~~~~~
{\rm with}~~\rho_{\bf B}={1\over \sqrt{-g}}{\partial (\sqrt{-g}~\hat F^{0k})\over\partial x_k}
\end{equation}
\par\smallskip
It is worth noting that $\rho_{\bf B}$ has the same dimensions of $\rho_{\bf E}$. For example, according
to (\ref{eq:duali3}), the dual of  $F_{ik}$ in (\ref{eq:campif0}) is:
\begin{equation}\label{eq:camp19}
\hat F^{mj}={c\over \sqrt{-g}} \left(\matrix{~0~ & ~0~ & -u & 0 \cr
0  & 0 & 0 & 0  \cr u & 0 & 0 & -u \cr
0 &  0  & u & 0\cr}\right)
\end{equation}
Since we supposed that  $u$ does not depend on $y$,  we obtain $\rho_{\bf B}=0$. Based on the metric
given by  (\ref{eq:metrica2}) (where $\sqrt{-g}=\vert fp\vert$), we just checked  that, together with
the condition ${\rm div}{\bf B}=0$, the 4-divergence of the magnetic field also vanishes.
\par\smallskip

Going back to the equation (\ref{eq:divu3}), this time we cannot assume that $F_{mjk}=0$. On the other
hand, we can use (\ref{eq:divderc}) and (\ref{eq:forminv6}) with $V^0=c$, to get:
$$
{\textstyle{1\over 2}}g^{im} F_{mjk}F^{jk}~=~g^{im}\hat F^l\hat F_{lm}~=~
g^{im}{1\over\sqrt{-g}}~{\partial (\sqrt{-g}~
\hat F^{lj})\over\partial x_j}~\hat F_{lm}
$$
\begin{equation}\label{eq:divu4}
=~{\rho_{\bf B}\over c} g^{im}\hat F_{lm}V^l~=~ {\rho_{\bf B}\over c} \hat F^{im}g_{lm}V^l
~=~ {\rho_{\bf B}\over c} \hat F^{im}V_m
\end{equation}
The first passage follows on from a direct counting of the permutations of the indices, thanks to the
definitions provided  in (\ref{eq:duali3}). Substituting in  (\ref{eq:divu3}), one finally gets:
\begin{equation}\label{eq:divu5}
\nabla_k U^{ik}~=~{1\over c}\Big(\rho_{\bf E} F^{im}V_m
~+~ \rho_{\bf B} \hat F^{im}V_m\Big)
\end{equation}
Let us observe that, for $i=0$, we have  $F^{0m}V_m=0$ (due to (\ref{eq:eula7})),
while $\rho_{\bf B}\hat F^{0m}V_m$ recalls the product $-c^2({\bf B}\cdot
{\bf V}){\rm div}{\bf B}$. Thus, the first line of (\ref{eq:divu5}) turns out
to be equivalent to (\ref{eq:energ2}).
\par\smallskip

The equation (\ref{eq:divu5}) is the generalization of (\ref{eq:divvet}) with ${\bf N}=0$ and ${\bf
M}=0$. In spite of its elegance, it is not very convincing, since it involves two mass densities (see
also the comments at the end of section 13). Let us try to explain what is happening. Without going into
technical detail, we may make some remarks. We first note that $\hat{\hat {F}}_{ik}=-F_{ik}$, which
means that, after applying the dual twice, one gets the opposite of the original tensor. Then, for any
real $\lambda$, we consider the two tensors:
$$
{\cal F}_{ik}={1\over \sqrt{\lambda^2 +(1-\lambda)^2}}\Big[\lambda F_{ik}+(1-\lambda )
\hat F^{ik}\Big]
$$
\begin{equation}\label{eq:tenscomb}
\hat{\cal F}^{ik}={1\over \sqrt{\lambda^2 +(1-\lambda)^2}}\Big[\lambda \hat F^{ik}-
(1-\lambda )F_{ik}\Big]
\end{equation}
where the second one is the dual of the first one. As in (\ref{eq:tinvf}) we have: ${\cal F}^{ik}~
=~g^{im}g^{kl}{\cal F}_{ml}$. Moreover, we can check that the tensor $U_{ik}$ in
(\ref{eq:tennuo}) does not change if  in place of $F_{ik}$ and $F^{ik}$ we take ${\cal F}_{ik}$
and ${\cal F}^{ik}$, respectively. Therefore, the electromagnetic stress tensor does not
depend on $\lambda$, even if this parameter varies in space and time. Actually, we
already observed in section 12 that  the energy tensor does not recognize the polarization
of the electromagnetic field.
\par\smallskip

At this point, we can introduce the two new densities (see also (\ref{eq:potnuo})):
$\rho_{\cal E}=\nabla_k {\cal F}^{0k}$ and
$\rho_{\cal B}=\nabla_k \hat{\cal F}^{0k}$, where ${\cal E}=\big(\lambda {\bf E}+(1-\lambda )
c{\bf B}\big)/\sqrt{\lambda^2 +(1-\lambda )^2}$ and ${\cal B}=\big(\lambda c{\bf B}-(1-\lambda )
{\bf E}\big)/\sqrt{\lambda^2 +(1-\lambda )^2}$.
\par\smallskip

\noindent So, another equivalent way to write equation (\ref{eq:divu5}) is:
\begin{equation}\label{eq:divu7}
\nabla_k U^{ik}~=~{1\over c}\Big(\rho_{\cal E} {\cal F}^{im}V_m
~+~ \rho_{\cal B} \hat {\cal F}^{im}V_m\Big)
\end{equation}
For $\lambda =1$ the two versions are actually the same. Now, by letting  $\lambda$ to vary, suppose
that it is possible to modify the polarization of the fields ${\cal E}$ and ${\cal B}$  at each point,
in order to get $ \rho_{\cal B}=0$. In this way, we are left with a single density $\rho_{\cal E}$,
which is the one to be used in constructing the mass tensor $M_{ik}=\rho_{\cal E}V_iV_k$.
\par\smallskip

Let us restate the situation in brief.  Every non-trivial electromagnetic wave presents regions where
the classical divergence of any of the two fields is non-zero. The eletromagnetic energy tensor does not
distinguish between the two types of fields (electric or magnetic). In the end, what matters is the
intensity of the wave and the modality of propagation of its fronts, without paying attention to the way
each front has been parametrized. We can associate a fluid in motion at the speed of light with the
wave. Independently of the actual orientation of the fields ${\bf E}$ and ${\bf B}$, we can locally
build two other fields ${\cal E}$ and ${\cal B}$, so that the first one oscillates together with the
fluid and the second one satisfies ${\rm div}{\cal B}=0$. This fictitious change of polarization has no
influence on the electromagnetic energy tensor. The 4-divergence of ${\cal E}$, when different from
zero, represents the mass density of the fluid and it is used to construct the mass tensor. This last
tensor is added to the electromagnetic energy one, to form the global energy tensor which is on the
right-hand side of the Einstein equation. In principle, the fields ${\cal E}$ and ${\cal B}$ are not
directly associated with ${\bf E}$ and ${\bf B}$. However, in the natural evolution of electromagnetic
phenomena, the two entities usually coincide.
\par\smallskip

All the examples analyzed in the previous section  satisfy $\rho_{\cal B}=0$ and $\rho_{\bf B}=0$,
hence, they were already well suited to the case $\lambda =1$, corresponding to ${\cal E}={\bf E}$ and
${\cal B}=c{\bf B}$. In particular, the case in spherical coordinates  simulates the real behavior of a
wave generated by an infinitesimal electric dipole ascillating in a vertical direction. Somehow, the
dipole imparts mechanical oscillations to the fluid, in the same direction as the electric field.
Formally, we can now exchange the role of the fields ${\bf E}$ and $c{\bf B}$, by polarizing the
spherical wave by 90 degrees. In this new situation, we  have ${\rm div}{\bf E} =0$, $\rho_{\bf E}=0$
and ${\rm div}{\bf B}\not =0$. By choosing $\lambda =0$, we realize the condition $\rho_{\cal B}=0$ and
the fictitious fields ${\cal E}$ and ${\cal B}$ turn out to be anti-ruotated by 90 degrees. Therefore,
there is no longer coincidence of ${\cal E}$ and ${\cal B}$ with the corresponding ${\bf E}$ and $c{\bf
B}$. Nevertheless, a spherical wave having the second kind of polarization is difficult to observe in
nature, since it should correspond to the one  generated by an infinitesimal magnetic monopole.
\par\smallskip

It is certainly true that our equations are not capable of recognising the polarization of free waves.
This is a property that comes with the initial conditions. However, free waves are created by some
causes inherent to natural events, which have a strong influence in determining  polarization. The
problem resides at the origin, for example in the non existence of magnetic monopoles (we will have a
short discussion about this in section 15). Recall that, in equation (\ref{eq:sfg}), the electric and
 magnetic fields cannot be interchanged. Certainly, this equation influences the creation of a spherical
wave through the mechanical oscillations of an electric charge. The conclusion is that, at least for
free waves, we can expect $\lambda =1$, which implies that the direction of transversal propagation of
the fluid is in accordance with that of the electric field. More precisely, this can be taken as a
definition of electric field. Suppose that an external mechanical perturbation is applied to a free wave
having ${\cal E}={\bf E}$, in a direction not alligned with that of  field ${\bf E}$, in such a way the
direction of ${\cal E}$ changes. Then we may think that the wave reacts by varying its polarization (see
section 7, 8 and 9) in order to correct its posture, bringing  field ${\bf E}$ to once again coincide
with  field ${\cal E}$. In other words, the electric field turns out to be identified with the one that
follows the transversal oscillations of the fluid, and such a definition matches  reality.

\par\bigskip

\setcounter{equation}{0}

\section{Other developments and conclusions}
\smallskip

We start by recalling the primary results obtained by the paper. In section 9, we introduced the
following equations:
\begin{equation}\label{eq:sfe2}
{\partial {\bf E}\over \partial t}~=~ c^2 {\rm curl} {\bf B}~
-~({\rm div}{\bf E}) {\bf V}
\end{equation}
\begin{equation}\label{eq:sfb2}
{\partial {\bf B}\over \partial t}~=~ -{\rm curl} {\bf E}~
-~({\rm div}{\bf B}){\bf V}
\end{equation}
\begin{equation}\label{eq:sfgr2}
{D{\bf V}\over Dt}~=~\mu\big({\bf E}~+~{\bf V}\times{\bf B}\big)
\end{equation}
where ${\bf E}$ is the electric field, ${\bf B}$ is the magnetic field and
${\bf V}$ is a velocity field satisfying:
\begin{equation}\label{eq:normal22}
\vert {\bf V}\vert ~=~c
\end{equation}
The constant $\mu$ is a charge divided by a mass, and $c$ is the speed of light.
\par\smallskip

Then, in sections 11 and 12, we wrote the equations in covariant form.
In the same order they appear above, we have, for $i=0,1,2,3$:
\begin{equation}\label{eq:frm2}
(\nabla_k F^{ik})V^0~=~(\nabla_k F^{0k})V^i
\end{equation}
\begin{equation}\label{eq:frm2h}
(\nabla_m \hat F^{im})V^0~=~(\nabla_m \hat F^{0m})V^i
\end{equation}
\begin{equation}\label{eq:geofor2}
{DV^i\over Dt }~+~\Gamma^i_{jk}V^jV^k~=~-{\mu\over c}~F^{im}V_m
\end{equation}
where $V^0=c$, $F_{ik}$ is the electromagnetic tensor and  $\hat F^{jm}$ its dual.
In the general system of coordinates, the normalizing condition takes the form:
\begin{equation}\label{eq:nor22}
g_{im}V^iV^m~=~0
\end{equation}
\par\smallskip

The metric tensor $g_{ik}$ is not given, but has to be determined through the
Einstein equation:
$$
R_{ik}~-~{\textstyle{1\over 2}}g_{ik}R
$$
\begin{equation}\label{eq:eindue}
~=~{\chi\mu\over c^4}\Big( -\mu g^{mj}F_{im}F_{kj}~+~{\textstyle{1\over 4}}
\mu g_{ik}F_{mj}F^{mj}~+~V_iV_k~\nabla_m F^{0m}\Big)
\end{equation}
 where on the right-hand side we find a suitable
energy tensor, obtainable with the rules provided in section 12 (in the construction of $\rho_{\bf E}=
\nabla_m F^{0m}$ remember to take into account the warnings at the end of section 14). Such a coupling
corresponds to a quite complex system, able to describe space-time geometry in conjunction with
electromagnetic phenomena.
\par\smallskip

Our set of equations contains the embryo of some of the main laws of Physics. In (\ref{eq:sfe2}) and
(\ref{eq:sfb2}), we recognise the equations of electromagnetism, more or less with the same structure as
the Maxwell equations. We discovered that, when ${\bf V}$ is irrotational, then (\ref{eq:normal22}) is
the eikonal equation, so that the Huygens principle is also latent. On the right-hand side of
(\ref{eq:sfe2}) we partly recognize the Amp\`ere law. The equation (\ref{eq:sfgr2}) expresses the
Lorentz law, anticipating the Newton law in the form of momentum equation for the dynamics of fluids. In
fact, we claimed  that the light rays can be assimilated to stream-lines of a certain fluid of density
$\rho_{\bf E}$. Moreover, we  know that a continuity equation holds for $\rho_{\bf E}$.
\par\smallskip

Throughout the paper, we assumed we were in a universe that we could
 call ``pre-Coulombian''. As a matter of fact, we developed a theory of
 electromagnetism without introducing any charges, and we spoke
about fluids without having any masses. The only elements at our disposal were the fields. Here comes
the big question: can we now build matter from these fields respecting the rules that we wrote? In other
terms: can an elementary particle  be ``solution'' to our set of equations?
\par\smallskip

A particle is a quite complicated thing. It has charge, magnetic momentum, spin, mass. It evolves and
interacts with other particles according to the rules of quantum mechanics. Can we contain all these
factors in a solution localized in space? This problem was mentioned in section 10 where we discussed
 possible solutions, consisting of a stable system of two rotating solitons. Although the framework
is still incomplete we collected some pieces of evidence, whose details will shortly be discussed below,
that support the possibility of creating particles from fields. We recall that other authors, through a
qualitative analysis, followed a similar idea of building electrons from photons (see for example
\cite{william} and the references therein).
\par\smallskip

We can give a rough idea of how a ``particle solution''  looks by examining figure 7, that shows,
projected on a plane, the rotation of the fronts around an axis. From a qualitative viewpoint,  field
${\bf E}$ oscillates radially, but, in the average, mainly pointing inward (or outward). This creates
the polarity of the electrical charge. Field ${\bf B}$ is orthogonal to the page. The rays form closed
orbits and their vector curvature ${\bf G}$ points toward the center, producing a non-vanishing
gravitational field. If the sign of ${\bf E}$ is changed,
 then ${\bf G}$ again points toward the inside (gravity has only one
polarity). The displacement of field ${\bf V}$ matches the idea that something is ``spinning'', and the
associated electromagnetic fluid corresponds to a kind of vortex.
\par\smallskip

Still referring to figure 7, let us suppose that the particle is an electron. Then ${\bf E}$ should be
directed toward the center and, using the standard vector product $\times$, ${\bf B}$ points downwards.
Nevertheless, a negative charge rotating clockwise produces a spin angular momentum pointing downward
and a magnetic field pointing upward, which is in contrast to what previously found.  As we remarked in
section 8, this happens because we do not use the suitable vector product $\times$. In fact, the correct
one is the left-handed one. Since the magnetic dipole moment is independent of the sign of $\times$, the
change of parity now confirms that ${\bf B}$ points upward. If we want to maintain the same set of
equations, we can solve the problem just by changing the sign of the electric field, so that the
electron has a chance of existing only if the electric field vectors point outward. We can still call
this particle an ``electron'' and give a negative sign to it, but we have to comply with the new rule
stating that currents flow from a negative pole to a positive one.
\par\smallskip

Nevertheless, one can see that such a situation is still not compatible with equations
(\ref{eq:sfe2})-(\ref{eq:sfb2})-(\ref{eq:sfgr2})-(\ref{eq:normal22}). One of the reasons is that the
outer orbits of the light rays are longer than the inner orbits, and this does not match the condition
(\ref{eq:normal22}), telling us that the information propagates at constant speed. In order to have
chances of finding solutions to the form described above, the use of the general relativity framework is
unavoidable. The modification of space-time geometry  allows for the preservation of the momentum of
inertia (a typical mechanical concept), providing the ``glue'' that keeps the particle together. The
rotating wave follows the geodesics in the new metric. At the same time, the curvature of such geodesics
has to be compatible, through (\ref{eq:geofor2}), with the electromagnetic setting. The geometry alters
the relation between space and time in such a way that the rays, always travelling at speed $c$, can
accomplish paths of different length in the same amount of time. This recalls the problem of the rigid
rotating disk in general relativity. It is clear that the particle solution (if it exists) involves the
use of the whole set of equations. Therefore, its determination, even from the point of view of
numerical computations, is a demanding problem. Finally, by heuristic arguments, one can recognise that
a similar solution, where the magnetic field is exchanged with the electric one, should be forbidden by
equation (\ref{eq:geofor2}). This would imply the impossibility of building magnetic monopoles.
\par\smallskip

We finish the paper with some further speculations, not having  enough theoretical background. One
positive aspect is that particle solutions are expected to be extremely stable (an electron is quite a
difficult object to destroy). Another aspect is that they are in some sense ``unique'' (there is only
one type of electron or proton), and this property raises other questions. The equations (\ref{eq:sfep})
and (\ref{eq:sfbp})  are ``scalable'', by meaning that we can multiply the fields of a free wave by a
constant, oncemore obtaining a solution. Thus, free waves may be of any size and intensity. But, if we
take into account constrained waves, then this property is no longer true, since (\ref{eq:normal22}) is
not a scalable equation. Together with $c$, $\mu$ and $\chi$ in (\ref{eq:eins}), another constant is
hidden in the set of equations (\ref{eq:frm2})-(\ref{eq:frm2h})-(\ref{eq:geofor2})-(\ref{eq:nor22}),
which is related to some ``magnitude'' of the geometry. This result does not penalize our theory.
Actually it may give more strength to it. As a matter of fact, we cannot have electrons of any size! We
have no elements for quantifying the values of the various parameters, unless we find the particle
solution explicitly.
\par\smallskip

The last issue we discuss is the convenience of setting up experiments validating our theory. The
problem is left to the experts. However, we think that many convincing arguments, also based on a
multitude of practical observations, have already been collected, showing that our model is adequate.
The real breakthrough would be in predicting the realization of an electromagnetic device, capable of
producing gravitational field.

\par

\end{document}